\documentclass[12pt]{iopart}

\usepackage[square,sort&compress]{natbib}
\expandafter\let\csname equation*\endcsname=\relax
\expandafter\let\csname endequation*\endcsname=\relax
\usepackage{aas_macros}
\usepackage{amsmath}
\usepackage{amssymb}
\usepackage{amsfonts}
\usepackage[utf8]{inputenc}
\usepackage[T1]{fontenc}
\usepackage[english]{babel}
\usepackage{bm}
\usepackage{empheq}
\usepackage{xcolor}
\usepackage{mathrsfs}
\usepackage{tabularx}
\usepackage{multirow}
\usepackage{tablefootnote}
\usepackage{soul}
\usepackage[breaklinks=true,colorlinks,citecolor=blue,linkcolor=blue,urlcolor=blue]{hyperref}

\def\dd{\mathrm{d}}
\def\pptime{\tau}
\def\obqt{\epsilon}
\def\obqtL{\obqt_\lune}
\def\tcb{t}
\def\tdb{t^{**}}
\def\tcg{T}
\def\tcl{\mathcal{T}}
\def\A{\mathrm{e}}
\def\B{\mathrm{l}}  
\def\ext{\mathrm{A}}
\def\pptimeA{\tau_\A}
\def\tcbA{\tcb_\A}
\def\tdbA{\tdb_\A}
\def\tcgA{\tcg_\A}
\def\tclA{\tcl_\A}
\def\pptimeB{\tau_\B}
\def\tcbB{\tcb_\B}
\def\tdbB{\tdb_\B}
\def\tcgB{\tcg_\B}
\def\tclB{\tcl_\B}
\def\posl{\mathcal{X}}
\def\posb{x}

\def\posg{X}  
\def\dirl{\mathcal{N}}
\def\xl{\vect{\posl}}
\def\xg{\vect{\posg}}
\def\xb{\vect{\posb}}
\def\xbd{\vect{\posb}^{**}}
\def\xlA{\vect{\posl}_\A}
\def\xgA{\vect{\posg}_\A}
\def\xbA{\xb_\A}
\def\xbdA{\xbd_\A}
\def\xlB{\vect{\posl}_\B}
\def\xgB{\vect{\posg}_\B}
\def\xbB{\xb_\B}
\def\xbdB{\xbd_\B}
\def\xlG{\vect{\posl}_\ext}
\def\Nl{\vect{\mathcal{N}}}
\def\NlB{\Nl_{\!\B}}
\def\NlG{\Nl_{\!\ext}}
\def\NlE{\Nl_{\!\terre}}

\def\Ng{\vect{N}}
\def\NgG{\vect{N}_\ext}
\def\NgA{\vect{N}_\A}
\def\dirlG{\dirl_\ext}
\def\vitb{v}
\def\vitg{V}
\def\vitl{\mathcal{V}}

\def\vitlB{\vitl_\B}

\def\vitlG{\vitl_\ext}

\def\vl{\vect{\vitl}}
\def\vg{\vect{\vitg}}
\def\vb{\vect{\vitb}}
\def\vlB{\vect{\vitl}_\B}
\def\vgB{\vect{\vitg}_\B}
\def\vbB{\vect{\vitb}_\B}

\def\vgA{\vect{\vitg}_\A}

\def\accb{\vect{a}}

\def\abB{\accb_\B}
\def\poteff{\mathcal{W}}
\def\potGCRSeff{W}
\def\potM{\poteff_{\mathrm{\lune}}}
\def\potE{\potGCRSeff_{\mathrm{E}}}
\def\potGCRSE{\potGCRSeff_{\mathrm{E}}}
\def\potMnot{\overline w_{\mathrm{\lune}}}
\def\pottide{\poteff_{\mathrm{tidal}}}
\def\potGCRStide{\potGCRSeff_{\mathrm{tidal}}}

\def\pottideS{\poteff_{\mathrm{tidal}}^{\mathrm{M}}}
\def\pottideNS{\poteff_{\mathrm{tidal}}^{\mathrm{NM}}}
\def\potiner{\poteff_{\mathrm{iner}}}
\def\potGCRSiner{\potGCRSeff_{\mathrm{iner}}}
\def\vpoteff{\vect{\poteff}}
\def\vpotM{\vpoteff_{\mathrm{\lune}}}

\def\vpottide{\vpoteff_{\mathrm{tidal}}}
\def\vpotiner{\vpoteff_{\mathrm{iner}}}
\def\scalpot{\Phi}
\def\scalpotmoy{\bar{\Phi}}

\def\potsel{\bar{\Phi}_{\mathrm{sel}}}
\def\terre{\mathrm{E}}
\def\lune{\mathrm{L}}
\def\sun{\mathrm{S}}
\def\csphR{\mathcal{R}}
\def\csphT{\theta}
\def\csphL{\varphi}

\def\csphTA{\theta_\A}
\def\csphLA{\varphi_\A}
\def\csphRB{\csphR_\B}
\def\csphTB{\theta_\B}
\def\csphLB{\varphi_\B}
\def\csphRG{\csphR_\ext}
\def\csphTG{\theta_\ext}
\def\csphLG{\varphi_\ext}

\def\eulphi{\phi}
\def\eultheta{\theta}
\def\eulpsi{\psi}
\def\angvitCass{\vect{\Omega}_{\mathrm{CL}}}

\def\angvitGP{\vect{\Omega}_{\mathrm{GP}}}
\def\angvitLTP{\vect{\Omega}_{\mathrm{LTP}}}
\def\angvitTP{\vect{\Omega}_{\mathrm{TP}}}
\def\angvitiner{\vect{\Omega}_{\mathrm{iner}}}
\def\Rsel{R_{\mathrm{sel}}}
\def\topo{\mathrm{topo}}
\def\Rtopo{R_{\topo}}
\def\polax{s}
\def\epol{\vect{\polax}}
\def\ellnum{\ell_{G}}
\def\ellnumtopo{\ell_{\mathrm{topo}}}
\def\suml{\mathcal{S}}
\def\summ{\mathcal{U}}
\def\nongeoacc{\vect{\nongeoaccnorm}}
\def\nongeoaccnorm{\mathcal{Q}}
\def\spin{\mathcal{S}}

\def\vspinM{\vect{\spin}_\lune}
\def\vspinG{\vect{\spin}_\ext}
\def\spinG{\spin_\ext}

\newcommand{\vect}[1]{\bm{#1}}
\newcommand{\cvect}[2]{#1^{#2}}
\newcommand{\cflin}[2]{#1_{#2}}

\newcommand{\cvl}[1]{\vitl^{#1}}

\newcommand{\metl}[2]{\mathcal{G}_{#1#2}}

\newcommand{\be}[1]{\vect{e}_{#1}}
\newcommand{\DD}[2]{\frac{\dd #1}{\dd #2}}
\newcommand{\cnongeoacc}[1]{\nongeoaccnorm_{#1}}
\newcommand\microsec{\textnormal{~$\mu$s}}

\def\TCL{\mathrm{TCL}}
\def\TCG{\mathrm{TCG}}
\def\TCB{\mathrm{TCB}}
\def\TL{\mathrm{TL}}
\def\TT{\mathrm{TT}}
\def\UTC{\mathrm{UTC}}
\def\UTCk{\mathrm{UTC}(\mathrm{k})}
\def\TDB{\mathrm{TDB}}

\begin{document}

\title{Lunar Reference Timescale}

\ead{adrien.bourgoin@obspm.fr, p.defraigne@oma.be, frederic.meynadier@bipm.org}

\author{A.~Bourgoin$^1$, P.~Defraigne$^2$,  F.~Meynadier$^3$}
\address{$^1$ LTE, Observatoire de Paris, Universit\'e PSL, Sorbonne Universit\'e, Université de Lille, LNE, CNRS, 61 avenue de l'Observatoire, 75014 Paris, France}
\address{$^2$ Royal Observatory of Belgium, 1180 Brussels, Belgium}
\address{$^3$ Time Department, BIPM, Pavillon de Breteuil, 92310 S\`evres, France}

\date{\today}

\begin{abstract}
    Setting up a relativistic lunar reference frame is of a prime importance in the context of future exploration missions to the Moon. If the procedure for building a consistent reference frame within the framework of the general theory of relativity is well established (cf. resolutions B.3 of IAU 2000), there is still some freedom in the choice of the coordinate timescale to be adopted as reference in the lunar region. In this paper, we review the orders of magnitude of the relativistic effects resulting from (i) the gravitational redshift of a clock on the lunar surface and (ii) the time transformations between a clock on the surface of the Moon and a clock on the surface of the Earth. We then discuss possible options for a lunar reference timescale with their advantages and drawbacks, taking note that the solution which is adopted for the Moon shall then be reemployed for Mars and other planets. Finally, we propose possible realizations of the lunar reference timescale as well as its traceability to UTC.
\end{abstract}

\fbox{
\parbox{.9\textwidth}{
Manuscript version: Accepted Manuscript\\
\url{https://doi.org/10.1088/1681-7575/ae2c03}\\
This Accepted Manuscript is © 2025 The Author(s). Published on behalf of BIPM by IOP Publishing Ltd.
As the Version of Record of this article is going to be / has been published on a gold open access basis under a CC BY 4.0 licence, this Accepted
Manuscript is available for reuse under a CC BY 4.0 licence immediately.
Everyone is permitted to use all or part of the original content in this article, provided that they adhere to all the terms of the licence
\url{https://creativecommons.org/licences/by/4.0}\\
Although reasonable endeavours have been taken to obtain all necessary permissions from third parties to include their copyrighted content
within this article, their full citation and copyright line may not be present in this Accepted Manuscript version. Before using any content from this
article, please refer to the Version of Record on IOPscience once published for full citation and copyright details, as permissions may be required.
All third party content is fully copyright protected and is not published on a gold open access basis under a CC BY licence, unless that is
specifically stated in the figure caption in the Version of Record.
View the article online for updates and enhancements.}}

\maketitle


\section*{Introduction}
\label{sec:intro}

The interest in Moon exploration has substantially grown in the latest years. With the gradual launches of large-scale lunar exploration missions, the demand for high-precision navigation and positioning of lunar probes has become increasingly urgent. To support the next generation of institutional and private lunar exploration missions,  space agencies are developing lunar communications and navigation services that would allow positioning accuracy during the final descent of landers on the Moon surface.  In particular, NASA is proposing the LunaNET \citep{Israel2020}, and ESA is proposing the Moonlight program \citep{Moonlight2022}, both aiming to provide communication and Position Navigation and Timing (PNT) services to institutional and commercial lunar missions.  China has also proposed its own ``Queqiao comprehensive constellation for communications, navigation and remote sensing'' \citep{China2019}, which is planned to be completed in three phases to provide relay communications, navigation, positioning and timing services for operation in the vicinity of the Moon, and for lunar surface exploration activities. The lunar  constellation composed of Moon orbiters  can achieve continuous coverage of specific areas of the lunar surface or even the entire lunar surface, and can provide necessary relay communication, navigation, positioning and other services for robotic landers first, then manned missions, to fully carry out lunar exploration in the future. 

Future scientific or more industrial applications on the lunar surface, as well as communication in the lunar environment, will require a standard reference timescale. One particular application requiring a reference timescale is certainly the positioning, navigation and timing as envisaged by the satellite navigation systems like Moonlight, LunaNET or Queqiao. As in classical Earth Global Navigation Satellite System (GNSS), the different satellite clocks should be synchronized between them, or more precisely, their offset with respect to a common reference should be known at user level \citep{Handbook_GNSS}. Furthermore, the lunar explorers should be able to coordinate their activities with Earth-based control centers, meaning that realizations of their timescales need to have a known and predictable difference to realizations of UTC on the Earth---at least up to a certain level of accuracy. 
This implies the urgent need to define a theoretical scale to be used in the lunar environment, as well as to propose some solutions for its realization. 

The requirement that future localization on the lunar surface is expected within a meter precision, which translates to the nanosecond precision for time keeping, implies that general relativity cannot be neglected. Let us remind that on the Earth, the GNSS relativistic transformations imply corrections at the level of several tens of microsecond per day \citep{Ashby2003} which can of course not be ignored when a navigation precision of one meter is targeted. The Moon is orbiting the Earth in its annual revolution around the Sun, and the respective gravitational and relative velocity differences suggest that a common reference timescale cannot be used for activities both on  the Moon and on the Earth, and that a new definition and a new terminology should be fixed and adopted by the whole scientific and industrial communities. 

The  terminology for astronomical reference systems and timescales in the solar system, with the associated relativistic framework, have been provided by the International Astronomical Union (IAU) \citep{IAUGA1991}. They defined two different barycentric coordinate systems: a global one ``with spatial origin at the center of mass of the solar system'', and a local one ``with spatial origin at the center or mass of the Earth'', together with their associated timescales: the Barycentric Coordinate Time (TCB) and the Geocentric Coordinate Time (TCG), respectively. \citet{IAUGA1991} also provided the coordinate transformations between global and local coordinates. These definitions were then refined  in the resolution B1.3 of the 24th IAU General Assembly \citep{IAUGA2000}, which adopted the terms Barycentric Celestial Reference System (BCRS) and Geocentric Celestial Reference System (GCRS) to designate them.

The IAU 2024 resolution II \citep{IAUGA2024} recommends constructing a Lunar Celestial Reference System (LCRS), with its coordinate time designated as Lunar Coordinate Time (TCL), using the same techniques as for constructing the GCRS and TCG from BCRS and TCB, and keeping  the unit of measurement of TCL consistent with the SI second. It has been adopted during the IAU General Assembly and provides the explicit definitions needed to designate this lunar reference frame, in direct continuity to previous recommendations. Recently, studies of the relations between lunar and terrestrial timescales have been proposed by \citet{Ashby2024,Kopeikin2024}, and \citet{2025ApJ...985..140T}. We propose here a more comprehensive development of the time rate at the surface of the Moon, and from there we discuss possible options for the future reference timescale on the Moon.

Section \ref{sec:Earth} of this paper proposes a review of the timing system in the Earth environment, summarizing the transformation between the proper time of clocks and the coordinate time of the geocentric reference system. The definition of the different coordinate times, as recommended by the IAU and adopted by the metrological community, are presented. Section \ref{sec:Moon} describes the relation between the proper time of a clock in lunar environment and the coordinate time of the lunar reference system. The details of the mathematical developments used in this section are presented in \ref{app:comp}.  Section \ref{sec:TCL-TCG} details the relation between the time on Earth and on the Moon, including proper and coordinate time differences, and their link to the light travel time in the context of a one-way time transfer. We then present in section \ref{sec:options} some features of possible scaled versions of TCL that can be used as lunar reference timescale, discussing their possible advantages and drawbacks, and acknowledging that whichever option is eventually adopted for the Moon will likely be applied for Mars and other planets in the near future. Finally, we discuss the realization of the lunar timescale as well as its traceability to UTC in section \ref{sec:realization}.

\section*{Notations}

Throughout the paper, we use different coordinate systems with, sometimes, different timescales associated with the same coordinate system. Although the notations are explained when introduced in the text, they are also resumed in table \ref{tab:summary}.

\begin{table}
    \caption{Table of the notations and definitions used in this paper.}
    \vspace{0.2cm}
    \begin{tabularx}{\textwidth}{l X}
        \hline\hline
        \multicolumn{2}{l}{\textbf{Abbreviations}}\\
        \hline
        BCRS & Barycentric Celestial Reference System \hfill \citep{IAUGA1991,IAUGA2000}\\
        GCRS & Geocentric Celestial Reference System \hfill \citep{IAUGA1991,IAUGA2000}\\
        LCRS & Lunar Celestial Reference System \hfill \citep{IAUGA2024}\\
        TCB & Barycentric Coordinate Time associated with BCRS \hfill \citep{IAUGA1991}\\
        TDB & Barycentric Dynamical Time associated with BCRS; linear scaling of TCB using constant $L_B$ [cf. Eq. \eqref{eq:TDB-TCB}] \hfill \citep{IAUGA2006}\\
        TCG & Geocentric Coordinate Time associated with GCRS \hfill \citep{IAUGA1991}\\
        TT & Terrestrial Time associated with GCRS; linear scaling of TCG using constant $L_G$ [cf. Eq. \eqref{eq:TCG-TT}] \hfill \citep{IAUGA1991,IAUGA2000}\\
        TCL & Lunocentric Coordinate Time associated with LCRS \hfill \citep{IAUGA2024}\\
        TL & Lunocentric Time associated with LCRS; linear scaling of TCL using constant $\Delta f$ [cf. Eq. \eqref{TL}]. Three options are considered in this paper for $\Delta f$ (cf. options \ref{item:1}, \ref{item:2}, \ref{item:3} in Sect. \ref{sec:options}) \hfill [this paper]\\
        \hline
        \multicolumn{2}{l}{\textbf{Coordinate reference systems}}\\
        \hline
        $(\tcb, \xb)$ & TCB-compatible BCRS coordinates\\
        $(\tdb, \xb^{**})$ & TDB-compatible BCRS coordinates; linear scaling w.r.t. TCB-compatible coordinates using constant $L_B$ [cf. Eq. \eqref{eq:TDB-TCB}]\\
        $(\tcg, \xg)$ & TCG-compatible GCRS coordinates\\
        $(\tcg^*,\xg^*)$ & TT-compatible GCRS coordinates; linear scaling w.r.t. TCG-compatible coordinates using constant $L_G$ [cf. Eq. \eqref{eq:TCG-TT}]\\
        $(\tcl, \xl)$ & TCL-compatible LCRS coordinates\\
        $(\tcl^\dag, \xl^\dag)$ & TL-compatible (in the sense of option \ref{item:2}) LCRS coordinates; linear scaling w.r.t. TCL-compatible coordinates using constant $L_L$ [cf. Eq. \eqref{eq:teclscale-tcl}] so that, on average, $\tcl^\dag$ is close to the proper time of a clock on a given selenoid (cf. option \ref{item:2} in Sect. \ref{sec:options})\\
        $(\tcl^\ddag, \xl^\ddag)$ & TL-compatible (options \ref{item:1}, \ref{item:2}, \ref{item:3}) LCRS coordinates; linear scaling w.r.t. TCL-compatible coordinates using constant $\Delta f$ [cf. Eq. \eqref{TL}]. Three options are considered in this paper for $\Delta f$ (cf. options \ref{item:1}, \ref{item:2}, \ref{item:3} in Sect. \ref{sec:options}) \\
        \hline
        \multicolumn{2}{l}{\textbf{Coordinate velocities}}\\
        \hline
        $\vb$ & BCRS coordinate velocity; both TCB- and TDB-compatible\\
        $\vg$ & GCRS coordinate velocity; both TCG- and TT-compatible\\
        $\vl$ & LCRS coordinate velocity; both TCL- and TL-compatible\\
        \hline
        \multicolumn{2}{l}{\textbf{Subscripts}}\\
        \hline
        $\terre$ & label for the Earth\\
        $\lune$ & label for the Moon (`L' as in Lunar)\\
        $\sun$ & label for the Sun\\
        $\A$ & label for the clock at rest on the surface of the Earth\\
        $\B$ & label for the clock at rest on the surface of the Moon\\
        \hline
    \end{tabularx}
    \label{tab:summary}
\end{table}

\section{Timescales in the geocentric reference system}
\label{sec:Earth}

Time and frequency comparisons at the level of current atomic clock performances should be formulated within the post-Newtonian theory of reference systems. This formalism, which is the outcome of decades of intensive research and collaboration, was rigorously established in the early 90s (see, e.g., \cite{1989ASSL..154.....K,1990CeMDA..48...23B,1991ercm.book.....B,1993PhRvD..48.1451K}; see also \citet{Damour1991,Damour1992,1993PhRvD..47.3124D,1994PhRvD..49..618D} and references therein) and was subsequently adopted by the IAU in 2000 \citep{Soffel2003}. This formalism provides a comprehensive description, at the first post-Newtonian level, of (i) the global dynamics (``external problem'') of $N$ spinning extended bodies arbitrarily composed and shaped, (ii) the local gravitational structure of each body (``internal problem''), and (iii) the mutual consistency of these descriptions (``relativistic theory of reference systems'').

The treatment of the external problem requires a \emph{global} reference system (i.e., the BCRS) whereas the local gravitational structure of each body is more conveniently represented in its own \emph{local} reference system (e.g., the GCRS for the Earth or the LCRS for the Moon). In local frames, the influence of the external gravitational field is largely ``effaced'', in the sense that the effective external potential acting locally on the body and its environment reduces primarily to tidal terms. Owing to this ``effacement'' property, local reference frames are particularly suitable for describing the space-time geometry in the vicinity of each body. However, because non-inertial and tidal terms respectively increase linearly and quadratically with distance from the body’s center of mass, the local space-time metric becomes increasingly complex as the spatial domain of interest extends outward. Maintaining a given level of precision therefore requires including higher-order terms in the expansion of the tidal and non-inertial potentials.

Consequently, in order to preserve simplicity in the experimental modeling, one should employ a local frame that will ensure that the spatial extent of the experiment will remain limited to a region where only first-order terms in the expansion of the tidal and non-inertial potentials are actually needed. The GCRS fulfills this condition for experiments conducted in the vicinity of the Earth, typically up to geostationary orbits.

\subsection{Geocentric Celestial Reference System}
\label{subsec:GCRS}

Within the solar system, the global frame is the BCRS and its associated coordinate time is TCB; the origin is at the center of mass of the Solar System  \citep{IAUGA2000}. When neglecting the influence of the gravitational field of the Galaxy, the axes of the BCRS are kinematically and dynamically non-rotating (cf. e.g., \citet{1993A&A...279..273K}) with respect to distant extra-galactic sources. The global frame is used for constructing theories of motion of the Solar System bodies and for constructing catalogues of distant objects (outside the Solar System). In the Earth environment, the local frame is the GCRS and its associated coordinate time is TCG; the origin is at the center of mass of the Earth \citep{IAUGA2000}. The axes of GCRS are kinematically non-rotating with respect to the BCRS ones, namely the GCRS undergoes a relativistic precession with regard to the BCRS. The Earth local frame is used for describing local physical processes such as the rotation of the Earth or its local gravitational field.

The components of the metric tensor in GCRS coordinates $(\cvect{X}{\mu}) = (c\tcg,\xg)$ was recommended by the \citet{IAUGA2000}, with $c$ the speed of light in a vacuum, $\tcg$ the coordinate time TCG, and $\xg$ the GCRS position of any event. Detailed description of the relativistic theory for time and frequency comparisons around the Earth can be found for instance in \citet{Wolf-Petit1995} or \citet{Petit-Wolf2005}. Only a brief summary will be given hereafter.

Now let us consider a clock `$\A$' staying in the immediate region around the Earth so that its motion is conveniently regarded in GCRS. According to general relativity, a clock can be seen as a physical device that follows a timelike worldline $\mathscr{L}$ in space-time and provides a sequence of measurable events ($\mathscr{P}_i$, $\forall i\in\mathbb{Z}$) sampling $\mathscr{L}$. The clock's proper time between two events $\mathscr{P}_i$ and $\mathscr{P}_{i+N}$ is $KN$, where $K$ is a constant affine factor fixing the unit of time measurement and where $N$ is an integer. Note that once $K$ is fixed, the proper time becomes independent of the parametrization of the clock trajectory $\mathscr{L}$. In other words, one is free to describe, in a given reference system, the evolution of the spatial coordinates of a clock as a function of the associated coordinate time. For instance, in the GCRS, the position of clock `$\A$' might be expressed as a function of TCG, namely $\xgA=\xgA(\tcg)$; its coordinate velocity is thus $\vgA=\vgA(\tcg)$ with $\vgA = \dd\xgA/\dd\tcg$. Hence, within the IAU recommended framework, the proper time of that clock is given by
\begin{equation}
\label{eq:dtaudTCG}
\frac{\dd \pptimeA - \dd \tcg}{\dd \tcg}=-\frac{1}{2c^2}\left[\vitg_\A^2(\tcg)+2\potGCRSeff(\tcg,\xgA)\right] + \mathcal{O}(c^{-4}) \, ,
\end{equation}
where $\vitg_\A = \vert \vgA \vert$ and where the effective potential $\potGCRSeff$ is given by
\begin{equation}
  \potGCRSeff(\tcg,\xgA) = \potGCRSE(\tcg,\xgA) + \potGCRStide(\tcg,\xgA) + \potGCRSiner(\tcg,\xgA) \, .
\end{equation}
$\potGCRSE$ is the Newtonian potential of the Earth, $\potGCRStide$ is the tidal contribution from the  Moon, the Sun, and the other planets, and $\potGCRSiner$ is the inertial potential caused by the non-geodesic acceleration of the Earth center of mass (namely the couplings between the Earth non-spherical part of the gravitational potential to external gravitational fields from other bodies). All potentials are evaluated at the clock geocentric position $\xgA$.

Note that for most of the applications, only $c^{-2}$ terms need to be taken into account when considering a precision at the level of $10^{-16}$ (see e.g., \citet{Soffel2003, blanchet2001, Wolf-Petit1995}). This justifies that only scalar potentials are considered in equation \eqref{eq:dtaudTCG}, as the vector potentials contribute at the $c^{-4}$ level.

The three components of the effective potential $W$ should be characterized separately. 
The inertial potential is scaling such as $\vect{Q} \cdot \xgA$ with $\vect{Q}$ being the non-geodesic acceleration of the Earth. According to \citet{blanchet2001}, $\vert \vect{Q} \vert$ is of the order of $10^{-11}\ \mathrm{m}\,\mathrm{s}^{-2}$, with the main contribution coming from the monopole of the Moon interacting with the quadrupole moment of the Earth. For a clock at rest on the surface of the Earth, the inertial potential shall therefore generate a relative frequency variation of the order of $10^{-22}$ which can safely be ignored at the level of performances of existing atomic clocks. Note, however, that the inertial potential grows linearly with the distance to the center of mass of the Earth. We thus expect it to contribute at the level of $10^{-16}$ for radial distances corresponding to several astronomical units which is far from being restrictive in practice.

The tidal potential is due to the gravitational attraction of the solar system bodies (limited here to their monopole term) and reads such as
\begin{equation}
        \potGCRStide(\tcg,\xgA) = \sum_{\ext \neq \terre} \frac{Gm_\ext}{\vert \xg_\ext \vert} \sum_{j=2}^{+\infty} \left(\frac{\vert \xgA \vert}{\vert \xg_\ext \vert}\right)^{\!j}  P_{j}(\NgA \cdot \NgG) \, ,
        \label{eq:tidalGCRS}
\end{equation}
where $G$ is the gravitational constant, $m_\ext$ is the mass of the body $\ext$, $P_j$ are the Legendre polynomials of degree $j$, and  $\Ng_\A=\xgA / \vert \xgA \vert$. The GCRS position of body $\ext$ is seen as a function of TCG: $\xg_\ext = \xg_\ext (\tcg)$ and so is $\Ng_\ext=\xg_\ext / \vert \xg_\ext \vert$. For a clock at the Earth surface, the impact of the Moon and Sun [i.e., $\rm A = \rm L$ and $\rm A = \rm S$ in Eq. \eqref{eq:tidalGCRS}] tidal quadrupole term [i.e., $j=2$ in Eq. \eqref{eq:tidalGCRS}] is periodic with a magnitude at the level of $10^{-17}$ in the relative frequency difference between the proper time and TCG (see e.g., \citet{Wolf-Petit1995,Klioner2008}). Note that the tidal potential grows quadratically with the distance to the center of mass of the Earth. We thus expect the next-to-leading order term [i.e., $j = 3$ and $\rm A = \rm L$ in Eq. \eqref{eq:tidalGCRS}] to be at the level of $10^{-19}$ for that clock on the Earth surface. The $j=3$ tidal term would therefore contribute at the level of $10^{-16}$ for clocks situated at altitudes around $28\,000\ \mathrm{km}$, which roughly corresponds to geostationary orbits.

Finally, for a clock in the vicinity of the Earth, at a distance $\vert \xgA \vert$ from the center of mass of the Earth, the  Newtonian potential of the Earth is given by 
\begin{align}
  \potE (\tcg,\xgA) & = \frac{Gm_\terre}{\vert \xgA \vert} \, \Bigg\{ 1 + \sum_{\ell=2}^{+\infty}\sum_{m=0}^{\ell} \left(\frac{R_\terre}{\vert \xgA \vert}\right)^{\!\ell} \nonumber\\
  & \times P_{\ell m} (\sin \csphTA) \left[ C^\terre_{\ell m} \cos (m \csphLA) + S^\terre_{\ell m} \sin (m \csphLA) \right] \Bigg\} \, ,
  \label{eq:potE}
\end{align}
where $m_\terre$ is the mass of the Earth, $R_\terre$ is its (mean) equatorial radius, $\csphTA$ and $\csphLA$ are the latitude and longitude of the clock (measured from Earth's equator and Greenwich prime meridian, respectively), $P_{\ell m}$ are the associated Legendre polynomials of degree $\ell$ and order $m$, and where $C^\terre_{\ell m}$ and $S^\terre_{\ell m}$ are the Stokes coefficients depending on the mass distribution inside Earth. The most important term in this potential is of course the monopole. When introduced into equation \eqref{eq:dtaudTCG} its contribution to the clock rate with respect to TCG, is  about $7 \times 10^{-10}$ for a clock on the surface of the Earth. The second most important term corresponds to the Earth flattening, $J_2^\terre$ (with $J_2^\terre = - C_{20}^\terre$), and gives a contribution of about $8 \times 10^{-13}$. All other coefficients of the Earth gravitational field are at least three orders of magnitude smaller than $J_2^\terre$'s. Note however that for high precision clocks on the Earth surface, the series expansion in equation \eqref{eq:potE} is not adapted; local gravimetric measurements must be favored instead (see discussion below).

For a clock at rest on the Earth surface, the term proportional to the square of the GCRS velocity of the clock in equation \eqref{eq:dtaudTCG} corresponds to the centrifugal potential caused by Earth's proper rotation. Considering a clock on Earth's equator, this returns a contribution to the clock rate of about $10^{-12}$ with respect to TCG.

\subsection{Terrestrial Time}
\label{sec:TT}

The sum of the averaged centrifugal and effective potentials gives the mean equipotential at any position in the rigidly rotating frame of the Earth. All ideal clocks located at rest on a mean equipotential surface in the Earth vicinity, will tick at the same rate in the GCRS (up to small periodic terms of order $10^{-17}$), namely they will nearly beat the same rate with respect to TCG. Note that of course real clocks will have rate differences due to their own limited accuracies.

One particular equipotential surface is originally the geoid, which coincides with the mean sea level. The \citet{IAUGA1991} defined in 1991 a new coordinate time: the Terrestrial Time (TT), a scaled version of the coordinate time TCG, so that the mean rate of TT agrees as much as possible with that of the proper time of a clock located on the geoid. However, noting the intricacy and temporal changes of the definition and realization of the geoid, the IAU in 2000 \citep{IAUGA2000} re-defined TT by fixing the scaling constant to $L_G = 6.969\,290\,134 \times 10^{-10}$ such that
\begin{equation}
\label{eq:TCG-TT}
 \TT - \TCG = - L_G \times (\mathrm{JD}_{\TT} -T_0) \times 86\,400\ \mathrm{s} \, .
\end{equation}
where $\mathrm{JD}_{\TT}$ is the TT Julian date with $T_0 = 2\,443\,144.500\,372\,5$ being the TT Julian day 1977 January 1st, $0^{\mathrm{h}}\,0^{\mathrm{m}}\,32.184^{\mathrm{s}}$.

Let us note that according to \citet{Klioner2008}, scaling the coordinate time is necessarily accompanied by the corresponding scaling of spatial coordinates: $\xg^* = (1-L_G)\xg$, and mass parameter of celestial bodies: $(Gm)^* = (1-L_G) \, (Gm)$. This is needed to ensure that a range (difference of proper times) satisfies the general covariance principle as discussed in section~\ref{sec:TCL-TCG}. The scaled coordinate times and spatial coordinates can be thought of as defining
a new reference systems: one with coordinates $(c\tcg^*, \xg^{*} )$ where $\tcg^*$ denotes TT. This
new reference system can be characterized by its own metric tensor, different from
the one of the GCRS \citep{Klioner2010}.

Fixing the constant $L_G$ also fixes the value to the reference Earth potential $\potGCRSeff_0$ used for defining TT. 
Clocks located exactly on the geoid $\potGCRSeff_0 = 62\,636\,856\ \mathrm{m}^2\,\mathrm{s}^{-2}$ provide a direct realization of TT up to a precision of one part in $10^{-17}$ \citep{Klioner2008}, with an uncertainty depending on the clock accuracy. 
The proper time of a clock `$\A$' located at an orthometric altitude $H_\A$ (i.e., altitude above the geoid) then deviates from TT by
\begin{equation}
\label{eq:tau-TT}
\frac{\dd \tau_\A - \dd \tcg^*}{\dd \tcg^*}=\frac{1}{c^2}\int_0^{H_\A} g_\A(\tcg^*,h) \, \mathrm{d}h \, ,
\end{equation}
where $g_\A$ is the gravity acceleration on Earth (i.e., the gradient of centrifugal plus effective potentials). 
This rate deviation corresponds to  a frequency shift of about $10^{-16}$ per altitude meter.
A more detailed derivation can be found for instance in \citet{Wolf-Petit1995}.

One realization of TT is the International Atomic Time (TAI) to which is added an offset of 32.184 s. This timescale is calculated a posteriori every month by the Bureau International des Poids et Mesures (BIPM), from a clock ensemble algorithm combining about 450 atomic clocks distributed around the world.  To ensure the best agreement between TAI and the SI second duration, the result of this clock ensemble (called Echelle Atomique Libre, EAL) is then steered using the comparison with a dozen of primary and secondary frequency standards, namely cold atoms atomic fountains or more recently optical clocks \citep{Panfilo2019}.
As a consequence, the primary and secondary frequency standards measurements used to ensure the accuracy of the TAI have to be first corrected for their relativistic redshift, as TAI is a realization of TT which was defined in such a way that its rate is indiscernible from the mean rate of an ideal clock on the geoid defined by the equipotential $W_0$ up to terms of order $10^{-17}$ \citep{Klioner2008}. 
As an example, the frequency standards operating at the US National Institute of Standard and Technology (NIST), located at about $1\,660\ \mathrm{m}$ above the geoid, must be corrected for a relativistic redshift  of  $(-1.798\,7 \pm 0.000\,3) \times 10^{-13}$ \citep{PavlisWeiss2017}. 
This value can be estimated from precise altitude measurements combined with geoid models and gravity measurements, see for instance \cite{PavlisWeiss2017} for more details. 

Finally, the  international reference for time and frequency is UTC, obtained from TAI after applying leap seconds. UTC being computed a posteriori, it is not available for a real time use. All time laboratories
contributing to UTC by submitting their clock data therefore maintain a realization of UTC called UTC(k), where `k' is the acronym of the laboratory. The BIPM publishes each month the ``Circular T'' which presents the differences between the different UTC(k) and UTC \citep{CircularT}. From the Circular T it is seen that some UTC(k) are able to stay close to UTC within a few nanoseconds. In most cases they are available in real time as a physical signal for synchronization and can be used to demonstrate the traceability to UTC. 

\section{Timescales in the selenocentric reference system}
\label{sec:Moon}

In this section, we aim at deriving the rate of proper time of a clock at rest on the surface of the Moon. To this end, we first introduce the LCRS and its associated space-time metric. Then, we derive the expression of the rate of proper time with respect to TCL for any lunar clock. This expression is applied to the clock on the lunar surface. We systematically neglect physical effects smaller than one part in $10^{16}$, which is the long-term stability of UTC \citep{Petit2015}---this threshold is  well beyond the  accuracy of the future clocks that will live in the lunar region in the next decade. Finally, mimicking what was done for the Earth, the rate of proper time of the clock on the Moon is determined with respect to a rescaled version of TCL by invoking a lunar equipotential $\poteff_{0}$ (similar to $\potGCRSeff_0$ in GCRS).

\subsection{Lunar Celestial Reference System}

The IAU 2024 resolution II \citep{IAUGA2024} recommends constructing a LCRS---with TCL as the coordinate time---using the same approach as for constructing the GCRS---with TCG as the coordinate time. It also recommends that the unit of measurement of TCL be consistent with the SI second. 

The local LCRS coordinate system is denoted by $(\cvect{\posl}{\mu}) = (c\tcl , \xl)$; it is centered at the center of mass of the Moon, with the $\mathcal X$-axis directed towards the vernal equinox of J2000.0, the $\mathcal Z$-axis pointing towards the celestial pole of Earth's J2000.0 equator, and the $\mathcal Y$-axis completing the right-handed triad. The spatial axis of LCRS are thus kinematically non-rotating with respect to the BCRS ones, namely the LCRS undergoes a relativistic precession with regard to the BCRS.

According to IAU 2024 resolution II, the LCRS space-time metric, denoted by $\metl{\alpha}{\beta}$, is constructed in a similar fashion as GCRS's, namely based on the resolution B1.3 of the IAU 2000 \citep{Soffel2003}; the Earth related quantities are replaced by Moon's:
\begin{subequations}\label{eq:metl}
  \begin{empheq}[left=\empheqlbrace]{align}
    \metl{0}{0} (\tcl,\xl) & = -1+2c^{-2}\poteff(\tcl,\xl)-2c^{-4}\poteff^2(\tcl,\xl)+\mathcal{O}(c^{-5}) \, ,\\
    \metl{0}{a} (\tcl,\xl) & = -4c^{-3}\cvect{\poteff}{a}(\tcl,\xl)+\mathcal{O}(c^{-5}) \, ,\\
    \metl{a}{b} (\tcl,\xl) & = \cflin{\delta}{ab}\left[1+2c^{-2} \poteff(\tcl,\xl)\right]+\mathcal{O}(c^{-4}) \, .
  \end{empheq}
\end{subequations}
The effective potentials $\poteff$ and $\cvect{\poteff}{a}$ are called the scalar potential and the ($a$-th component of the) vector potential $\vpoteff$, respectively. The scalar potential reads
\begin{equation}
  \poteff(\tcl,\xl) = \potM(\tcl,\xl) + \pottide(\tcl,\xl) + \potiner(\tcl,\xl) \, ,
  \label{eq:Wscal}
\end{equation}
with $\potM$ the lunar gravitational potential, $\pottide$ the tidal potential (mainly from Earth and Sun), and $\potiner$ the inertial potential (caused by the non-geodesic acceleration of the Moon center of mass). The vector potential is given by
\begin{equation}
  \vpoteff(\tcl,\xl) = \vpotM(\tcl,\xl) + \vpottide(\tcl,\xl) + \vpotiner(\tcl,\xl) \, .
  \label{eq:Wvect}
\end{equation}
See for instance \citet{Soffel2003} and references therein (in particular \citet{Damour1991,Damour1992}) for full expressions of the different contributions in $\poteff$ and $\vect{\poteff}$ including terms up to $\mathcal{O}(c^{-2})$.

In LCRS, the position and the (coordinate) velocity of a lunar clock `$\B$' can be seen as functions of TCL, namely $\xlB = \xlB(\tcl)$ and $\vlB = \vlB(\tcl)$ with $\vlB = \dd \xlB / \dd \tcl$. Hence, $\pptimeB$, the proper time of clock `$\B$', is given by
\begin{equation}
  \dd \pptimeB = \dd \tcl \big[ - \metl{0}{0}(\tcl,\xlB) - 2c^{-1} \, \metl{0}{a}(\tcl,\xlB) \, \cvl{a}_\B - c^{-2}\metl{a}{b}(\tcl,\xlB) \, \cvl{a}_\B \cvl{b}_\B \big]^{1/2} \, ,
  \label{eq:pptime}
\end{equation}
where the components of the LCRS space-time metric $\metl{\alpha}{\beta}$ are evaluated at time $\tcl$ along the worldline of clock `$\B$'. $\cvl{a}_\B$ denotes the $a$-th component of $\vlB$. Then, by substituting equations \eqref{eq:metl} for $\metl{\alpha}{\beta}$ into equation \eqref{eq:pptime}, we find, after Taylor expanding the so-obtained expression in power series of $c^{-1}$, the rate of proper time of clock `$\B$' with respect to TCL:
\begin{align}
  & \frac{\dd \pptimeB - \dd \tcl}{\dd \tcl} = -\frac{1}{2c^2} \left[ \vitlB^2 + 2\poteff(\tcl,\xlB) \right] \nonumber\\
  & \qquad - \frac{1}{8c^4} \left[ \vitlB^4 + 12 \vitlB^2 \poteff(\tcl,\xlB) - 4\poteff^2(\tcl,\xlB) - 32 \vlB \cdot \vect{\poteff}(\tcl,\xlB) \right] + \mathcal{O}(c^{-5}) \, ,
  \label{eq:dpptimedtcl}
\end{align}
where $\vitlB$ represents the norm of the LCRS velocity of clock `$\B$', namely $\vitlB = \vert \vlB \vert$. Equation \eqref{eq:dpptimedtcl} is referred to as the relative frequency difference between clock `$\B$'s proper time and TCL. 

The LCRS, analogously to the GCRS, constitutes a local nearly freely falling frame that provides a convenient framework for investigating physical phenomena in the vicinity of the Moon. As previously discussed, the influence of external gravitational fields is largely suppressed within such local frames. Consequently, the effective external potential acting on the Moon and its immediate environment reduces essentially to tidal contributions, which considerably simplifies the local spacetime geometry. In practice, retaining only the leading terms in the series expansion of the tidal potential (fact that we denote as the ``minimal'' version of LCRS) is typically sufficient for describing phenomena within a spatial domain small enough to be encompassed by the LCRS, in its ``minimal'' version. However, note that the spatial extent of the ``minimal'' LCRS is smaller than that of the ``minimal'' GCRS one. Indeed, the next-to-leading-order term beyond the quadrupole contribution in the tidal potential becomes significant---at the level of $10^{-16}$---for clocks situated at altitudes larger than $6\,200\ \mathrm{km}$.

\subsection{Rate of proper time with respect to TCL}
\label{proper time Moon}

Let us now consider the clock `$\B$' to be at rest on the lunar surface. All terms present in equation~\eqref{eq:dpptimedtcl} are discussed and then estimated in \ref{app:comp} in this context. We summarize the discussion hereafter.

The position of clock `$\B$' is given, to a sufficient accuracy, in terms of its (spherical) selenographic coordinates $(\csphRB,\csphTB,\csphLB)$ by the following relation:
\begin{equation}
  \xlB (\tcl) = \csphRB \, \NlB (\tcl) \, ,
\end{equation}
where $\csphRB$ is the (mean) radial distance of clock `$\B$' measured from the center of mass of the Moon and $\NlB$ is the unit vector defined such as
\begin{equation}
  \NlB (\tcl) = \be{A} (\tcl) \cos \csphTB \cos \csphLB + \be{B} (\tcl) \cos \csphTB \sin \csphLB + \be{C} (\tcl) \sin \csphTB \, .
\end{equation}
$\csphTB$ and $\csphLB$ being the (mean) selenographic latitude and longitude of clock `$\B$', respectively (with $-90^\circ \leqslant \csphTB \leqslant 90^\circ$ and $-180^\circ < \csphLB \leqslant 180^\circ$). The unit vectors $\be{A}$, $\be{B}$, and $\be{C}$ form the vectorial basis attached to the principal axis of inertia of the Moon. The $A$ and $C$-axis are directed along the directions of minimal and maximal moments of inertia of the Moon, respectively. The $B$-axis completes the right-handed triad.

The coordinate velocity of clock `$\B$' is given, to a sufficient accuracy, by
\begin{equation}
  \vlB (\tcl) = n_\lune \csphRB \, \be{C} (\tcl) \times \NlB (\tcl) \, ,
\end{equation}
$n_\lune$ being the lunar mean motion; it corresponds to a  sideral period of $27.32\ \mathrm{day}$, namely $n_\lune=2.66\times 10^{-6}\ \mathrm{rad}\,\mathrm{s}^{-1}$. This expression neglects lunar physical librations that shall not exceed $10^{-3} \ \mathrm{rad}$ (see discussion below and in \ref{app:comp}).

The clock trajectory describes a timelike worldline in space-time and its velocity tells us how fast the clock moves along it. From that worldline, we can now deduce the rate of the proper time $\pptime_\B$ of clock `$\B$' with respect to TCL, up to a precision of one part in $10^{16}$. We find the following expression:
\begin{align}
  \frac{\dd \pptime_\B - \dd \tcl}{\dd \tcl} & = - \frac{1}{c^2} \scalpot ( \csphRB,\csphTB,\csphLB ) \, ,
  \label{eq:dpptimedtclsimpfinal}
\end{align}
where the equipotential $\scalpot$ reads as
\begin{equation}
  \scalpot ( \csphRB,\csphTB,\csphLB ) = \frac{\vitlB{}^2(\csphTB)}{2}  + \potM( \csphRB,\csphTB,\csphLB ) + \pottide( \csphRB,\csphTB,\csphLB ) \, .
  \label{eq:equipot}
\end{equation}
According to results derived in \ref{app:comp}, the equipotential is constant in time at that precision and is given by the following relation:
\begin{align}
  \scalpot (\csphRB,\csphTB,\csphLB) & = \frac{Gm_\lune}{R_\lune} \, \Bigg\{ 1 + \left(\frac{R_\lune}{\csphRB} - 1\right) + \frac{J_2^\lune}{2} + \frac{n_\lune^2 R_\lune^3}{2\,Gm_\lune} \nonumber\\
  & - \left( \frac{3J_2^\lune}{2} + \frac{n_\lune^2 R_\lune^3}{2\,Gm_\lune} \right) \sin^2\csphTB + 3 \left( C_{22}^\lune \cos 2 \csphLB + S_{22}^\lune \sin 2 \csphLB \right) \cos^2\csphTB \nonumber\\
  & + \sum_{\ell=3}^{\ellnum}\sum_{m=0}^{\ell} P_{\ell m} (\sin \csphTB) \left[ C^\lune_{\ell m} \cos (m \csphLB) + S^\lune_{\ell m} \sin (m \csphLB) \right] \nonumber\\
  & + \frac{1}{2} \left( \frac{m_\terre}{m_\lune} \right) \left( \frac{R_\lune}{a_\lune} \right)^{\!3} (3\cos^2\csphTB\cos^2\csphLB-1) \Bigg\} \, . \label{eq:equipotPhi}
\end{align}
In this equation, $m_\lune$ and $R_\lune$ denote the lunar mass and (mean) equatorial radius of the Moon, respectively. The term $a_\lune$ represents the semi-major axis of the Moon's orbit about the Earth-Moon barycenter. $J_2^\lune$ is the lunar quadrupole moment of the mass distribution, namely $J_2^\lune = - C_{20}^\lune$, while $C_{\ell m}^\lune$ and $S_{\ell m}^\lune$ are the Stokes coefficients of the Moon gravitational potential. The parameter $\ellnum$ represents the degree to which the partial sum over $\ell$ is stopped; for a precision of one part in $10^{16}$, we choose $\ellnum=150$. Below and in \ref{sec:potLune3}, we further discuss this choice which represents a compromise between computational time and precision.

After integration, relation \eqref{eq:dpptimedtclsimpfinal} simply reads such as
\begin{equation}
    \pptime_\B (\tcl) - \tcl = [\pptime - \tcl]_{(\tcl,\csphRB,\csphTB,\csphLB)} \, ,
    \label{eq:diffpptTCL}
\end{equation}
where the function $[\pptime - \tcl]_{(\tcl,\csphRB,\csphTB,\csphLB)}$ returns the difference between the proper time of a clock at rest on the surface of the Moon---at spatial coordinates $(\csphRB,\csphTB,\csphLB)$---and TCL; it is defined as
\begin{equation}
  [\pptime - \tcl]_{(\tcl,\csphRB,\csphTB,\csphLB)} = -\frac{1}{c^2} \scalpot (\csphRB,\csphTB,\csphLB) \, (\tcl - \tcl_0) + \mathrm{const} \, ,
  \label{eq:diffpptTCLbis}
\end{equation}
with $\tcl_0$ a conventional origin of time for TCL (defined by the \citet{IAUGA2024} such that the reading of TCL be 1977 January 1st, $0^{\mathrm{h}}\,0^{\mathrm{m}}\,32.184^{\mathrm{s}}$ when the reading of TCB be 1977 January 1st, $0^{\mathrm{h}}\,0^{\mathrm{m}}\,32.184^{\mathrm{s}}$) and `$\mathrm{const}$' representing $[\pptime - \tcl]_{(\tcl_0,\csphRB,\csphTB,\csphLB)}$, namely the value of $\pptimeB (\tcl_0) - \tcl_0$, that is the switch-on synchronization offset of the clock.

\begin{figure}[t]
  \begin{center}
    \includegraphics[scale=0.8]{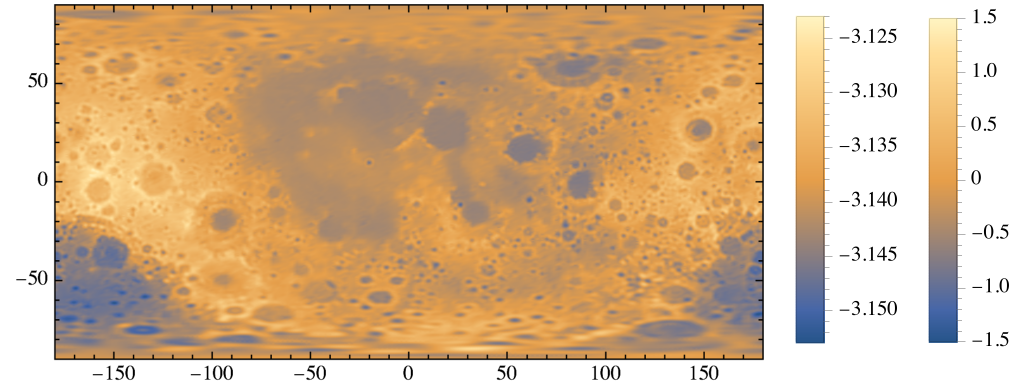}
  \end{center}
  \setlength{\unitlength}{1.0cm}
  \begin{picture}(0,0)(-8,0)
    \put(-2.8,0.25){\rotatebox{0}{longitude $\csphLB$ $[{}^{\circ}]$}}
    \put(-7.6,2.6){\rotatebox{90}{latitude $\csphTB$ $[{}^{\circ}]$}}
    \put(3.6,6.2){\rotatebox{0}{\footnotesize $[\times 10^{-11}]$}}
    \put(5.4,6.2){\rotatebox{0}{\footnotesize $[\times 10^{-13}]$}}
    \put(3.65,1.0){\rotatebox{0}{\footnotesize $(a)$}}
    \put(5.45,1.0){\rotatebox{0}{\footnotesize $(b)$}}
    \put(6.9,5.0){\rotatebox{270}{rel. freq. diff.}}
  \end{picture}
  \vspace{0.0cm}
  \caption{Map of the relative frequency difference between the proper time of a clock at rest on the lunar surface and $(a)$ TCL [i.e., $\tcl$ in Eq. \eqref{eq:dpptimedtclsimpfinal}], $(b)$ a rescaled TCL [i.e., $\tcl^\dag$ in Eq.~\eqref{eq:tau-TL}]. The clock is assumed at radial position $\csphRB=\Rtopo(\csphLB,\csphTB)$ [cf. Eq.~\eqref{eq:Rtopo}]. The origin of the longitude represents lunar prime meridian.}
  \label{fig:lunar_total}
\end{figure}

Figure \ref{fig:lunar_total} depicts the dependence to $(\csphTB,\csphLB)$ of the relative frequency difference between proper time and TCL [cf. Eq. \eqref{eq:dpptimedtclsimpfinal}]  for a clock `$\B$' at rest on the lunar surface, namely a clock forced to follow the lunar topography: $\csphRB = \Rtopo(\csphTB,\csphLB)$ (cf. explanations in \ref{sec:topo}).

It is possible to trace back in equations \eqref{eq:dpptimedtclsimpfinal} or \eqref{eq:diffpptTCL} the different contributions in \eqref{eq:equipotPhi}. For instance, the first term in curly brackets in \eqref{eq:equipotPhi} (i.e., one) represents the monopole contribution of the lunar gravitational potential $\potM$. This represents the main effect in equation~\eqref{eq:dpptimedtclsimpfinal}; it is at the level of $3.14 \times 10^{-11}$ [cf. Eq. \eqref{eq:potMest}].

The second term in equation \eqref{eq:equipotPhi} (i.e., $R_\lune/\csphRB-1$) corresponds to the effect of the altitude of clock `$\B$' on the lunar surface (w.r.t. the equatorial radius). It comes from the lunar gravitational potential $\potM$. According to the discussion in \ref{sec:topopot}, this term shall induce in equation \eqref{eq:dpptimedtclsimpfinal} an effect of the order $1.62 \times 10^{-13}$ due to maximum altitude variation of $+7\ \mathrm{km}$ to $-8\ \mathrm{km}$ with respect to the equatorial radius [cf. Eq.~\eqref{eq:RtopoRM}; see also figure~\ref{fig:topo_sel}, and discussion in \ref{sec:selenoid} for a comparison with respect to the selenoid]. As depicted in figure \ref{fig:lunar_gravity_field_topo}, this term highly depends on the location of the clock on the lunar surface.

The terms proportional to $n_\lune^2 R_\lune^3/(G m_\lune)$ in equation \eqref{eq:equipotPhi} represent the contribution from the centrifugal potential evaluated at the level of clock `$\B$'. This term contributes only to the level of $1.19 \times 10^{-16}$ in equation \eqref{eq:dpptimedtclsimpfinal} [cf. Eq. \eqref{eq:est_centr_pot}] and depends on the selenographic latitude of the clock. The small value of the centrifugal term is due to the lunar spin orbit resonance 1:1 which makes the clock at the lunar
surface experiences a complete period in
one Moon’s sidereal orbital period. This also justifies us not including the physical lunar librations in this analysis.

All together, the contributions from degree two lunar gravitational potential $\potM$, namely terms proportional to $J_2^\lune$, $C_{22}^\lune$ and $S_{22}^\lune$, induce a maximum effect at the level of $7.92 \times 10^{-15}$ which depends on the longitude and latitude on the Moon (cf. Fig. \ref{fig:lunar_gravity_field_2}).

In equation \eqref{eq:equipotPhi}, each one of the $\ell \geqslant 3$ terms do not contribute significantly to equation~\eqref{eq:dpptimedtclsimpfinal} when they are evaluated independently from each other. However, the sum of them all might still be significant if the sum over the $m$'s in the $\ell$-th degree is reasonably close to the sum over the $m$'s in the $(\ell-1)$-th degree. In other words, the contributions of the terms with $\ell \geqslant 3$ in equation \eqref{eq:equipotPhi} depends critically on the convergence properties of the series. However, it is well known that this series expansion converges poorly when the field point lies close to the surface of the non-spherical body. For this reason, the discussion below should not be regarded as suitable for future high-precision applications---it merely provides an order-of-magnitude estimate of the proper time rate for a clock located on the lunar surface. In the next section, a more general framework based on local gravimetric measurements is introduced; this latter approach should be preferred for future experiments.

As shown in figure~\ref{fig:POTM_loca} (cf. the \emph{blue curve} representing the summation over $\ell$ from $\ell \geqslant 3$ to 300 and the \emph{red curve} representing the summation over $m$ from $0 \leqslant m \leqslant \ell$ for each degree $\ell$), the series expansion $\suml_{\ellnum} (\csphTB,\csphLB)$ [cf. Eq. \eqref{eq:sumSlG}] would not converge after considering only the first tenth degrees for a clock located at the intersection between the lunar prime meridian and lunar equator, namely $(\csphTB,\csphLB) = (0^\circ,0^\circ)$. As a matter of fact $\vert \suml_{10} (0^\circ,0^\circ) \vert = 2.26 \times 10^{-5}$ while $\vert \suml_{300} (0^\circ,0^\circ) \vert = 3.25 \times 10^{-7}$, which returns
\begin{equation}
  \left( \frac{Gm_\lune}{c^2R_\lune} \right) \vert \suml_{10} (0^\circ,0^\circ) \vert = 7.10 \times 10^{-16} \, , \qquad \left( \frac{Gm_\lune}{c^2R_\lune} \right) \vert \suml_{300} (0^\circ,0^\circ) \vert = 1.02 \times 10^{-17} \, ,
  \label{eq:potLS3deg300}
\end{equation}
after being inserted into equation \eqref{eq:dpptimedtclsimpfinal}. In other words, a series expansion up to the order $\ellnum=10$ cannot be considered accurate when considering a precision at the level $10^{-16}$ since it has not converged yet as shown in equation \eqref{eq:potLS3deg300}. In every cases we investigated, the series expansion reaches rough convergence for $\ellnum \gtrsim 150$ which justifies our choice for the numerical value of $\ellnum = 150$. With this, the maximum effect induced by $\ell \geqslant 3$ terms from the non-spherical part of the lunar gravitational potential in \eqref{eq:dpptimedtclsimpfinal} is of the order $6.34 \times 10^{-15}$.  Figure \ref{fig:lunar_gravity_field_3_TO_100} depicts the dependency of this effect to the location of the clock on the lunar surface. Note that it is different from the topography contribution which is shown in figure \ref{fig:lunar_gravity_field_topo}.

Finally, the last line in equation \eqref{eq:equipotPhi} represents the permanent tidal potential raised on the Moon by the Earth. Once inserted into equation \eqref{eq:dpptimedtclsimpfinal}, it contributes at the level of $2.37 \times 10^{-16}$ as discussed in \ref{sec:tides}. Note that this relative frequency shift is due to the Moon spin orbit resonance 1:1 which renders the tides raised by the Earth on the Moon permanent, unlike tides raised by the Moon and Sun on the Earth which are only periodic. We emphasize that the tidal permanent effect depends on the location of the clock on the lunar surface as depicted in figure \ref{fig:lunar_tides}.

\subsection{Rate of proper time with respect to a rescaled TCL}
\label{sec:refsel}

The expression \eqref{eq:dpptimedtclsimpfinal} suggests that, similar to what has been done for GCRS, we could introduce an equipotential surface close to the physical surface of the Moon. As discussed in \ref{sec:selenoid}, the time averaging of the equipotential $\scalpot$ in equation~\eqref{eq:equipot} can be used to impose a numerical value for the equipotential of the selenoid. For instance, we can choose to define the selenoid such as $\potsel=2.82\times10^{6}\ \mathrm{m}^2\,\mathrm{s}^{-2}$ (see \ref{sec:selenoid} for more details); let us call this equipotential $\poteff_{0}=\potsel$ in order to match GCRS's notations.

Then, by expanding $\scalpot(\xlB)$ around $\poteff_{0}$ in equation \eqref{eq:dpptimedtclsimpfinal}, the rate of proper time with respect to TCL, for a clock `$\B$' located at an orthometric altitude $H_\B$ relative to the equipotential surface of reference $\poteff_{0}$, will then be given by
\begin{equation}
\label{eq:tau-TCL}
 \frac{\dd \tau_\B - \dd \tcl}{\dd \tcl}=- \frac{1}{c^2}\left[\poteff
_{0} - \int_0^{H_\B} g_\B (\tcl,h) \, \mathrm{d}h \right] \, ,
\end{equation}
where $g_\B$ is the lunar gravity acceleration (i.e., the gradient of centrifugal plus effective potentials) at the clock location. In the context of future high-precision applications, the expression \eqref{eq:tau-TCL} is more suitable than \eqref{eq:dpptimedtclsimpfinal} [with $\scalpot$ given by \eqref{eq:equipotPhi}]. It indeed relies on local gravimetric measurements rather than on the series expansion of the gravitational potential which is poorly converging as discussed previously.

The integral in equation \eqref{eq:tau-TCL} corresponds to  a frequency shift of about $2 \times 10^{-17}$ per altitude meter. While this factor is about 5 times smaller than on the Earth, the topography of the Moon surface is much more contrasted than on the Earth, with minima and maxima reaching 7.5 km with respect to the equipotential surface $\poteff
_{0}$ (see Fig. \ref{fig:topo_sel} and \citet{Topo}). The frequency shift between the proper time of two clocks at the lowest and highest altitudes on the Moon is therefore about $3.0 \times 10^{-13}$ [cf. $(b)$-label in figure~\ref{fig:lunar_total}].

The reader can notice that the term $\poteff_{0}$ has been kept in equation \eqref{eq:tau-TCL} while no $W_0$ was present in the corresponding equation \eqref{eq:tau-TT} for a clock on the Earth. The reason is that equation~\eqref{eq:tau-TT} is not written in term of TCG but rather in term of TT, a rescaled coordinate time which was precisely defined to remove $W_0$ from the averaged rate of proper time of a clock on the geoid. An equivalent approach for the Moon (if needed) would require us to introduce a new coordinate time being a scaled version of TCL, in a similar way as TT is a rescaled coordinate time with respect to TCG. Such a scaled coordinate time (denoted $\tcl^\dag$ hereafter and called ``option \ref{item:2}'' in Sect. \ref{sec:options}), if it is constructed with the same purpose as TT in the geocentric context, would be given by $\tcl^\dag = \tcl + [\tcl^\dag - \tcl]_{(\tcl)}$, where the function $[\tcl^\dag - \tcl]_{(\tcl)}$ is 
\begin{equation}
    [\tcl^\dag - \tcl]_{(\tcl)} = - L_L ( \tcl - \tcl_0 ) \, ,
    \label{eq:teclscale-tcl}
\end{equation}
with $L_L = \poteff_{0}/c^2$ a defining constant and $\tcl_0$ introduced back in equation \eqref{eq:diffpptTCLbis}. As expected, making use of this new coordinate timescale rather than TCL allows one to remove the constant rate $\poteff_{0}$ from the proper time equation which shall now reduce (at leading order in $c^{-1}$) to
\begin{equation}
\label{eq:tau-TL}
 \frac{\dd \tau_\B - \dd \tcl^\dag}{\dd \tcl^\dag}= \frac{1}{c^2} \int_0^{H_\B} g_\B (\tcl^\dag,h) \, \mathrm{d}h \, .
\end{equation}
This result can alternatively be written such as
\begin{equation}
    \pptime_\B(\tcl^\dag) - \tcl^\dag = [\pptime - \tcl^\dag]_{(\tcl^\dag,\xlB^\dag)}
\end{equation}
where the function $[\pptime - \tcl^\dag]_{(\tcl^\dag,\xlB^\dag)}$ returns the difference between the proper time of the clock---situated at spatial coordinates $\xlB^\dag=(\csphRB^\dag,\csphTB^\dag,\csphLB^\dag)$---and $\tcl^\dag$; it is given by
\begin{equation}
  [\tau - \tcl^\dag]_{(\tcl^\dag,\xlB^\dag)} = -\frac{1}{c^2} \left[ \scalpot (\csphRB^\dag,\csphTB^\dag,\csphLB^\dag) - \poteff_{0} \right] (\tcl^\dag - \tcl_0^\dag) + \mathrm{const}^\dag \, .
\end{equation}
In this expression $\tcl_0^\dag$ represents the time origin of $\tcl^\dag$ (not defined by the IAU) and `$\mathrm{const}^\dag$' is the difference $\pptime_\B(\tcl_0^\dag) - \tcl_0^\dag$, namely the clock switch-on synchronization offset. Relying on $\tcl^\dag$ rather than $\tcl$ for computing the proper time of a clock on the lunar surface will remove the constant rate of $3.14 \times 10^{-11}$ due to the monopole contribution [cf. $(a)$-label in Fig. \ref{fig:lunar_total}] and leave behind a peak-to-peak amplitude of $3.0 \times 10^{-13}$ mainly due to the lunar topography as depicted in the $(b)$-label of figure~\ref{fig:lunar_total}.

Let us emphasize that the scaling of the lunar coordinate time will force similar scaling into distances and mass parameters \citep{Klioner2008}: $\xl^\dag = (1-L_L) \xl$ and $(Gm)_\dag = (1-L_L) \, (Gm)$, respectively. Hence, $\scalpot(\csphRB^\dag,\csphTB^\dag,\csphLB^\dag)$ shall return the exact same numerical value as $\scalpot(\csphRB,\csphTB,\csphLB)$. Note also that the scaled $(Gm)_\dag$ of the current section would be different from the one introduced in section \ref{sec:TT} [i.e., $(Gm)_*$] as expected from the occurrence of parameters $L_L$ and $L_G$ in the scaling relations. 

It is important to note that $\tcl^\dag$, the scaled counterpart of TCL, has neither been established nor adopted in any IAU resolution to date. Hence, introducing $\tcl^\dag$ for the lunar system---in a similar way as TT is introduced in the geocentric context---is a possible way (see e.g., \citet{Ashby2024,Kopeikin2024}, and \citet{2025ApJ...985..140T}), but not a required one. Indeed, before the adoption of a fully relativistic framework by the IAU at the end of the 20th century, timescales similar to TT and Barycentric Dynamical Time (TDB) were already in widespread operational use. In other words, TT and TDB existed (with different definitions) long before the coordinate timescales TCB and TCG were introduced by the IAU in 1991. Abandoning completely the pre-existing timescales was not feasible due to their long-standing practical utility. It was thus preferable to equip them with consistent definitions within the new relativistic formalism. In 2000 and 2006, TT and TDB were thus defined from TCG and TCB, respectively.

For the lunar system, there is no historical precedent of $\tcl^\dag$ being employed independently of any Earth-based timescales. Therefore, the necessity of defining and adopting $\tcl^\dag$, as introduced in equation \eqref{eq:teclscale-tcl}, remains open to question. This point will be further discussed in sections \ref{sec:options} and \ref{sec:Discussion}. 

\section{Geocentric and selenocentric timescale difference}
\label{sec:TCL-TCG}

In previous sections, we derived the rates of proper times with respect to local coordinate timescales in GCRS and LCRS for clocks on the Earth and on the Moon, respectively. In view of the future communication between the Earth and the Moon it is also needed to establish the relation between the local coordinate times employed around the Moon and around the Earth. As it will be shown in this section, this relation depends on the intermediate reference system which is used for the light travel time computation, as well as the position of the clocks. 

Hereafter we consider a clock `$\B$' on or around the Moon  and a clock `$\A$' on or around the Earth,   
and a one-way synchronisation between them. We consider that the synchronizing link is an electromagnetic signal (encoding the proper time of clock `$\A$') emitted by clock `$\A$' at geocentric coordinate time $\tcgA$ and then received by clock `$\B$' at selenocentric coordinate time $\tclB$. The light travel time is computed in an intermediate reference system using its associated coordinate time (rescaled or not). We consider three possible intermediate reference systems: (i) BCRS with TCB or TDB, (ii) GCRS with TCG or TT, and finally (iii) LCRS with TCL or a rescaled TCL. Each of these cases is discussed in turn with their advantages and drawbacks. 

\subsection{Using BCRS with TCB or TDB}

Let us start investigating the range equation (i.e., the difference between proper times) in BCRS while using TCB as intermediate timescale. The range between clock `$\B$' and clock `$\A$' obtained through a one-way signal reads as follows:
\begin{equation}
  \pptimeB - \pptimeA = [\pptime - \tcl]_{(\tclB,\xlB)} + [\tcl - \tcb]_{(\tcbB,\xbB)} + [\tcbB - \tcbA]_{(\tcbA,\xbA,\xbB)} - [\tcg - \tcb]_{(\tcbA,\xbA)} - [\pptime - \tcg]_{(\tcgA,\xgA)} \, ,
  \label{eq:range}
\end{equation}
where $[\pptime - \tcl]_{(\tclB,\xlB)}$ and $[\pptime - \tcg]_{(\tcgA,\xgA)}$ are the transformations between the proper time of clock `$\B$' and TCL, and the proper time of clock `$\A$' and TCG, respectively. These two functions have been presented in equations \eqref{eq:diffpptTCL} and \eqref{eq:dtaudTCG}, respectively considering clock `$\B$' on the lunar surface. Note that, according to equations~\eqref{eq:TCG-TT} and \eqref{eq:tau-TT}, the function $[\pptime - \tcg]_{(\tcgA,\xgA)}$ is usually separated into two pieces [cf. Eq. \eqref{eq:pptimeTT} below]: (i) the transformation from proper time of clock `$\A$' to TT, namely $[\pptime - \tcg^*]_{(\tcgA^*,\xgA^*)}$, and (ii) the transformation from TT to TCG, that is $[\tcg^*-\tcg]_{(\tcgA)}$.

Then, the term $[\tcbB - \tcbA]_{(\tcbA,\xbA,\xbB)}$ in equation \eqref{eq:range} corresponds to the light travel time of the signal expressed in the global coordinate system: the BCRS with TCB. For the one-way experiment under consideration, the time of reception of the signal is unknown. Therefore, the light travel time is a function of the barycentric time of emission $\tcbA$, the barycentric position of emitter $\xbA=\xbA(\tcbA)$, and the barycentric position of receiver $\xbB=\xbB(\tcbB)$. The function $[\tcbB - \tcbA]_{(\tcbA,\xbA,\xbB)}$ is thus called an emission time transfer function \citep{2008CQGra..25n5020T}. It returns the difference between the (coordinate) time of reception and the (coordinate) time of emission:
\begin{equation}
  \tcbB - \tcbA = [\tcbB - \tcbA]_{(\tcbA,\xbA,\xbB)} \, .
  \label{eq:ttf}
\end{equation}
In most applications (except lensing effect), the time transfer function is the one corresponding to the so-called quasi-Minkowskian light path \citep{PhysRevD.93.044028}, and is thus expressible such as
\begin{equation}
  [\tcbB - \tcbA]_{(\tcbA,\xbA(\tcbA),\xbB(\tcbB))} = \frac{\vert \xbB(\tcbB) - \xbA(\tcbA) \vert}{c} + \frac{\Delta(\tcbA,\xbA(\tcbA),\xbB(\tcbB))}{c} \, ,
  \label{eq:lighttime}
\end{equation}
with the following associated condition: $\vert \xbB(\tcbB) - \xbA(\tcbA) \vert \gg \Delta(\tcbA,\xbA(\tcbA),\xbB(\tcbB))$. In these expressions, $\Delta/c$ is called a delay function \citep{2008CQGra..25n5020T}; it usually contains a gravitational delay, called the Shapiro delay, and an atmospheric delay due to the presence of the Earth atmosphere---see e.g., \citet{PhysRevD.101.064035} for an expression of the delay function containing simultaneously gravitational and atmospheric delays and see \ref{sec:lighttt} for estimates of the delays in the one-way transfer considered here. Note that the light-time equation~\eqref{eq:ttf} [together with Eq. \eqref{eq:lighttime}] is implicit in $\tcbB$, it must therefore be solved iteratively in order to eventually determine $\xbB(\tcbB)$. In the \ref{sec:lighttt}, we present analytical expressions of $\tcbB$ including $\mathcal{O}(c^{-3})$ terms.

Finally, the functions $[\tcl - \tcb]_{(\tcb,\xb)}$ and $[\tcg - \tcb]_{(\tcb,\xb)}$ in equation \eqref{eq:range} are part of 4D relativistic transformations between (i) LCRS and BCRS, and (ii) GCRS and BCRS, respectively. Note that  both transformations are expressed at the point-event with BCRS coordinates $(c\tcb,\xb)$. Hereafter, we shall discuss each function in turn.

\subsubsection{Transformation from TCB to TCG.}

The one-before-last term in equation \eqref{eq:range}, namely $[\tcg - \tcb]_{(\tcb,\xb)}$, is the difference between geocentric and barycentric coordinate times, expressed at barycentric coordinate time $\tcb$ and at barycentric position $\xb$. This difference is given by the IAU Resolution B1.5 (2000) (considering here only $c^{-2}$ terms):
\begin{equation}
[\tcg -\tcb]_{(\tcb,\xb)} = - \frac{1}{c^{2}} \left\{\int_{t_0}^{t}\left[\frac{\vitb^2_\terre(\tcb')}{2}+\overline w_{\terre}(\xb_\terre(\tcb'))\right] \dd \tcb' + \vb_\terre (\tcb) \cdot [\xb-\xb_\terre(\tcb)] \right\}+\mathcal{O}(c^{-4}) \, ,
  \label{eq:TCB-TCG}
\end{equation}
where $\xb_\terre$ and $\vb_\terre$ are the position and velocity of the geocenter in BCRS, and $\overline w_{\terre}$ is the Newtonian potential (evaluated at the geocenter) due to all bodies in the solar system besides Earth. The reference epoch $t_0$ is 1977 January 1st, $0^{\mathrm{h}}\,0^{\mathrm{m}}\,32.184^{\mathrm{s}}$ in TCB, or 1977 January 1st, $0^{\mathrm{h}}\,0^{\mathrm{m}}\,0^{\mathrm{s}}$ TAI.

\subsubsection{Transformation from TCB to TCL.}

The second term of equation (\ref{eq:range}), namely $[\tcl - \tcb]_{(\tcb,\xb)}$, is the equivalent for TCL of the term just described for TCG; it gives the difference between selenocentric and barycentric coordinate times, expressed at barycentric coordinate time $\tcb$ and at barycentric position $\xb$. IAU Resolution II (2024) recommends that the transformation between TCL and TCB be given by a relation similar to the one described in IAU Resolution
B1.5  (2000), with Earth related quantities replaced by those of the Moon. The relationship between TCB and TCL is therefore  given by the time part of the full 4D relativistic transformation between the barycentric and selenocentric reference systems: 
\begin{equation}
[\tcl -\tcb]_{(\tcb,\xb)} = -\frac{1}{c^2} \left\{\int_{t_0}^{\tcb}\left[\frac{\vitb^2_\lune(\tcb')}{2}+\overline w_{\lune}(\xb_\lune(\tcb'))\right] \dd \tcb' + \vb_\lune (\tcb) \cdot [\xb-\xb_\lune(\tcb)] \right\}+\mathcal{O}(c^{-4}) \, ,
  \label{eq:TCB-TCL}
\end{equation}
where $\xb_\lune$ and $\vb_\lune$ are the position and velocity of the center of mass of the Moon  in BCRS, and $\overline w_{\lune}(\xb_\lune)$ is the Newtonian potential (evaluated at the Moon center of mass) due to all the bodies in the solar system besides Moon. The reference epoch $t_0$ is the same as for the difference $\TCG-\TCB$. Currently, the difference between TCG and TCB [cf. Eq.~\eqref{eq:TCB-TCG}] can be computed using the IAU service establishing and maintaining an accessible and authoritative set of algorithms and procedures that implement models used in Standards Of Fundamental Astronomy (SOFA) \citep{SOFA}. The difference is then obtained for a given coordinate time and position $(\tcb,\xb)$. The lunar-equivalent is not available to date in the SOFA but could, in principle, be provided in a similar way. 

\subsubsection{Timescale transformation in BCRS.}

\begin{figure}[t]
  \begin{center}
    \includegraphics[scale=0.7]{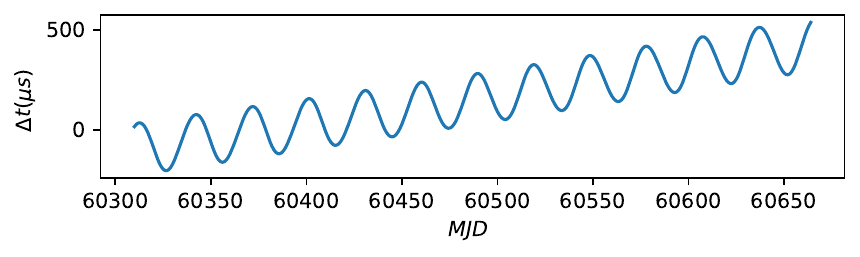}
  \end{center}
  \caption{Coordinate time differences $[\tcl - \tcb]_{(\tcbB,\xbB)}-[\tcg - \tcb]_{(\tcbA,\xbA)}$ over the year 2024 with $\xbB$ at a lunar latitude and longitude ($0^\circ$,$0^\circ$) and $\xbA$ at a Earth latitude and longitude ($0^\circ$,$0^\circ$).}
  \label{fig:TCL-TCG}
\end{figure}

\begin{figure}[t]
  \begin{center}
    \includegraphics[scale=0.7]{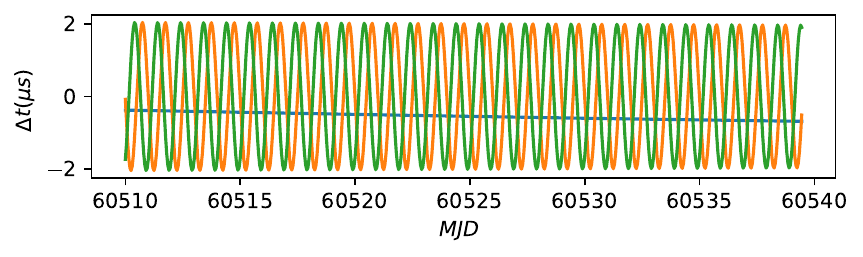}
    \includegraphics[scale=0.7]{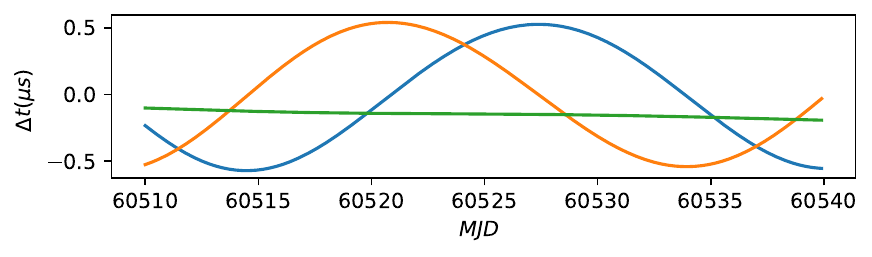}
  \end{center}
  \caption{\emph{Top}: scalar product  $\vb_\terre \cdot (\xb-\xb_\terre)$ from Eq. \eqref{eq:TCB-TCG}  over one month in  2024, evaluated at Earth latitudes and longitudes ($0^\circ$,$0^\circ$) \emph{in green}, ($90^\circ$,$0^\circ$) \emph{in blue} and ($0^\circ$,$120^\circ$) \emph{in orange}. \emph{Bottom}: scalar product  $\vb_\lune \cdot (\xb-\xb_\lune)$ from Eq. \eqref{eq:TCB-TCL}  over one month in 2024, evaluated at Moon latitudes and longitudes ($0^\circ$,$0^\circ$) \emph{in green}, ($90^\circ$,$0^\circ$) \emph{in blue} and ($0^\circ$,$90^\circ$) \emph{in orange}.}
  \label{fig:TCL-TCG_2pos}
\end{figure}

From these expressions, it appears that in BRCS, it is not possible to provide a unique relation for $\TCG-\TCB$ and $\TCL-\TCB$ as those depend on the location of the point-events where the transformations are to be performed. 
As an example, we show in figure \ref{fig:TCL-TCG} the difference $[\tcl - \tcb]_{(\tcbB,\xbB)}-[\tcg - \tcb]_{(\tcbA,\xbA)}$ appearing in equation \eqref{eq:range}, using relations \eqref{eq:TCB-TCG} and \eqref{eq:TCB-TCL}. Equation \eqref{eq:TCB-TCG} is computed at point-event $(\tcbA,\xbA)$ with $\xbA$ corresponding to an Earth station located at a $0^\circ$ longitude and a $0^\circ$ latitude in geographic coordinates. Equation \eqref{eq:TCB-TCL} is computed at point-event $(\tcbB,\xbB)$ with $\xbB$ corresponding to a lunar station at a $0^\circ$ longitude and a $0^\circ$ latitude in selenographic coordinates. 
The computation is realized using Astropy's default ephemerides, namely DE430 \citep{Folkner2014}, and considering only the gravitation potential of the Sun and the Earth for $\TCL-\TCB$, and the Sun and the Moon for $\TCG-\TCB$. As seen in figure \ref{fig:TCL-TCG}, we obtain a difference  $[\tcl - \tcb]_{(\tcbB,\xbB)} -[\tcg - \tcb]_{(\tcbA,\xbA)}$ mainly characterized by a linear drift (also called secular term) of about $1.5\microsec /  \mathrm{day}$ (i.e., a relative frequency offset of $1.7 \times 10^{-11}$) plus a monthly periodicity with an amplitude of $127\microsec$. Note that the linear drift only contains the contribution from the two coordinate time transformations being used. The total secular term in the range between a clock at rest on the surface of the Earth and a clock at rest on the surface of the Moon also contains the relative frequency difference between proper times and local coordinate times in the GCRS (about $6.97 \times 10^{-10}$) and in the LCRS (about $3.14 \times 10^{-11}$) [cf. the first and last terms in Eq.~\eqref{eq:range}]. Summing together all these contributions, we finally find a range rate difference of about $6.58 \times 10^{-10}$, or about $56\microsec / \mathrm{day}$ (see also \citet{Kopeikin2024,Ashby2024}, and \citet{2025ApJ...985..140T}). 
The periodic variation in figure~\ref{fig:TCL-TCG} is mainly due to the integrated terms in equations~\eqref{eq:TCB-TCG} and \eqref{eq:TCB-TCL}. Those are independent of the location and time where the time transformations are performed. The location-dependent terms give rise to periodic variations with an amplitude limited to a few microseconds (not visible on Fig. \ref{fig:TCL-TCG}) and depending mainly on the latitude of the clocks on the surface of the Earth and on the surface of the Moon. The longitude of clocks only changes the phase of periodic variations as depicted in figure \ref{fig:TCL-TCG_2pos}. In the top panel, the location-dependent terms in $\TCG-\TCB$ produces daily variations due to the motion of the clock on the Earth which rotates at the Earth sidereal period as seen from the BCRS. In bottom panel, the location-dependent terms in $\TCL-\TCB$ produces monthly variations due to the motion on the lunar clock which rotates at the lunar sidereal period as seen from the BCRS. Hence, only seasonal variations are present for a clock at the lunar poles. For clocks located on the surface of Earth and Moon, the amplitude variations of position-depend terms remain at the level a few microseconds. However, amplitude variations are expected to be larger for clock's in orbit due to larger geocentric and selenocentric distances.

\subsubsection{Using TDB instead of TCB.} Let us emphasize that the numerical value of the range in equation~\eqref{eq:range} is independent of the barycentric timescale being employed for the computation of the light travel time. Either TCB or its rescaled version, TDB, could have been used indifferently. TDB was defined by the IAU Resolution B3 \citep{IAUGA2006} as a scaled version of TCB with a scaling factor chosen in order to have the same average rate than TT at the geocenter:
\begin{equation}
\TDB - \TCB = - L_B \times (\mathrm{JD}_{\TCB} - T_0) \times 86\,400\ \mathrm{s} + \TDB_0 \, ,
\label{eq:TDB-TCB}
\end{equation}
where $\mathrm{JD}_{\TCB}$ is the TCB Julian date; $T_0$ was given earlier in equation \eqref{eq:TCG-TT}. $L_B$ and $\TDB_0$ are two defining constants: $L_B = 1.550\,519\,768 \times 10^{-8}$ and $\TDB_0 = -6.55 \times 10^{-5}\ \mathrm{s}$, respectively. For the range to be independent of using either $\TDB$ or TCB in equation~\eqref{eq:range}, some cancellations of coordinate-dependent terms must occur. For this to happen, the scaling of TDB-compatible quantities such as distances and mass parameters has to be carried out closely as discussed in \ref{sec:TDBVSTCB}. There, we indeed show that the range is generally covariant only if the scaling of TDB-compatible quantities is properly accounted for in the light travel time expression. Ignoring this scaling would induce relative errors of the order of $L_B$ in the computation of proper time difference $\pptimeB - \pptimeA$; this is because the scaling is affecting the leading order term in the light travel time equation [i.e., the Minkowskian term in Eq. \eqref{eq:lighttime}]. Considering that planetary ephemerides are computed and distributed in a TDB-compatible frame (see e.g., INPOP19 \citep{2019NSTIM.109.....F} and DE440/441 \citep{2021AJ....161..105P}), it is more advantageous and safe to compute the light-time in TDB rather than in TCB, and this means inserting equation \eqref{eq:TCBlt2TDBlt} into \eqref{eq:range}, so that we eventually obtain the following symmetrical expression for the range:
\begin{align}
  \pptimeB - \pptimeA & = [\pptime - \tcl]_{(\tclB,\xlB)} + [\tcl - \tcb]_{(\tcbB,\xbB)} - [\tdb - \tcb]_{(\tcbB)} + [\tdbB - \tdbA]_{(\tdbA,\xbdA,\xbdB)} \nonumber\\
  & + [\tdb - \tcb]_{(\tcbA)} - [\tcg - \tcb]_{(\tcbA,\xbA)} - [\pptime - \tcg]_{(\tcgA,\xgA)} \, ,
  \label{eq:rangeTDB}
\end{align}
where $\tdb$ is TDB and $\xb^{**}$ denotes TDB-compatible coordinates. The function $[\tdb - \tcb]_{(\tcb)}$ thus denotes the right-hand side of equation \eqref{eq:TDB-TCB}.

\subsection{Using GCRS with TCG or TT}

The clock comparison presented in the previous section involved two coordinate time transformations computed at two different positions when using BCRS as intermediate system for the computation of the light travel time (relying either on TCB or TDB). An alternative path would be to use instead GCRS as the intermediate system. In this case, a single coordinate time transformation is required at the level of Moon's clock with the light travel time computed in GCRS (relying either on TCG or TT). Indeed, the range between clock `$\A$' and clock `$\B$' through the one-way signal then reads
\begin{equation}
  \pptimeB - \pptimeA = [\pptime - \tcl]_{(\tclB,\xlB)} + [\tcl - \tcg]_{(\tcgB,\xgB)} + [\tcgB - \tcgA]_{(\tcgA,\xgA,\xgB)} - [\pptime - \tcg]_{(\tcgA,\xgA)} \, ,
  \label{eq:rangeKop}
\end{equation}
where the light travel time is expressed in the GCRS with  emission and reception coordinate times in TCG. Let us emphasize that the difference $[\tcl - \tcg]_{(\tcgB,\xgB)}$, should be known at $\xgB$, namely the geocentric position of the lunar clock; the position of the clock on Earth does not show up in the coordinate time transformation.

\subsubsection{Timescales transformation in GCRS.}

This is the approach followed by \citet{Kopeikin2024} to derive the coordinate time transformation between TCL and TCG. Reworking equation \eqref{eq:TCB-TCL} in order to reveal geocentric quantities, and then subtracting equation~\eqref{eq:TCB-TCG} to it, allows one to reduce the influence of external bodies to the Earth-Moon system to tidal terms only. As emphasized by \citet{Kopeikin2024}, this is a consequence of the Einstein equivalence principle. The difference $\TCL-\TCG$ reads
\begin{align}
    [\tcl - \tcg]_{(\tcgB,\xgB)} = & - \frac{1}{c^{2}} \Bigg\{ \int_{\tcg_0}^{\tcgB}\left[\frac{\vitg^2_\lune (\tcg) }{2}+\frac{G(m_\terre-2m_\lune)}{\vert \xg_\lune (\tcg) \vert} + \potGCRStide (\tcg,\xg_\lune(\tcg))\right] \dd \tcg \nonumber\\
    & + \vg_\lune (\tcgB) \cdot [\xgB (\tcgB) - \xg_\lune (\tcgB) ] \Bigg\} \, , 
    \label{eq:Kop}
\end{align}
with $\xg_\lune$ and $ \vg_\lune $  the position and velocity of the Moon center of mass in the GCRS, and $\potGCRStide$ the quadrupole of the tidal gravitational potential of the Sun:
\begin{align}
    \potGCRStide(\tcg,\xg_\lune) =\frac{1}{2}\frac{G m_\sun}{\vert \xg_\sun \vert} & \Bigg\{ \left( \frac{\vert \xg_\lune \vert}{\vert \xg_\sun \vert}\right)^{\!2} \left[ 3\left( \vect{N}_\sun \cdot \vect{N}_\lune\right)^2 -1 \right] \nonumber\\
    & \!+\left( \frac{\vert \xg_\lune \vert}{\vert \xg_\sun \vert}\right)^{\!3} 
    \left[ 5\left( \vect{N}_\sun \cdot \vect{N}_\lune\right)^3 - 3 \left( \vect{N}_\sun \cdot \vect{N}_\lune\right)\right]\Bigg\}\, ,
    \label{eq:tidesintime}
\end{align}
with the unit vectors $\vect{N}_\sun=\xg_\sun /\vert \xg_\sun \vert $ and $\vect{N}_\lune= \xg_\lune/ \vert \xg_\lune \vert$; $\xg_\sun$ being the geocentric position of the Sun. Note that equation \eqref{eq:Kop} is similar to equation (12) of \citet{Kopeikin2024} except that (i) the integration time is TCG rather than TCB, and (ii) the tidal potential [cf. Eq. \eqref{eq:tidesintime}] contains the next-to-leading order beyond the quadrupole term since it is of the order of $1.65 \times 10^{-16}$. Note however that our expression \eqref{eq:Kop} cannot be retrieved from equation (20) of \citet{Fienga2024} which seems to contain a misprint in its velocity-dependent term. Also note that equation \eqref{eq:Kop} is limited to post-Newtonian terms only since post-post-Newtonian effects are expected to be in the order of $10^{-17}$ \citep{Kopeikin2024}.

\begin{figure}[t]
  \begin{center}
    \includegraphics[scale=0.7]{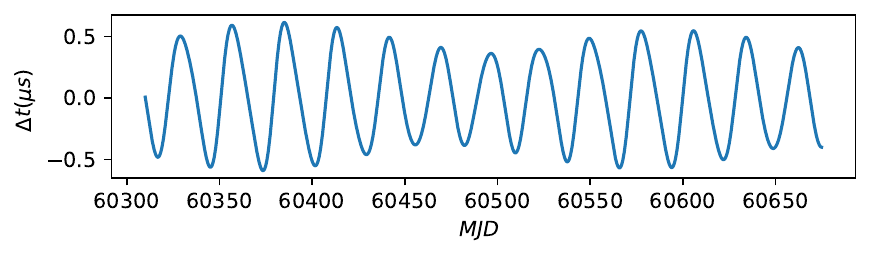}
    \includegraphics[scale=0.7]{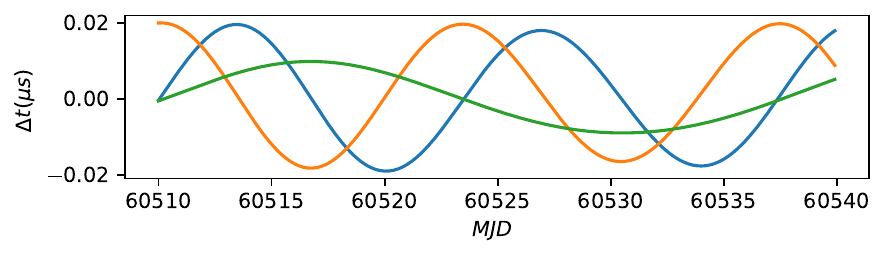}
  \end{center}
  \caption{\emph{Top}: Time differences $[\tcg - \tcl]_{(\tclB,\xlB)}$ over the year 2024, with $\xlB$ at a lunar latitude and longitude ($0^\circ$,$0^\circ$), after removing the secular term of $1.5 \microsec / \mathrm{day}$. \emph{Bottom}: Scalar product  $\vg_\lune \cdot (\xgB-\xg_\lune)$ from Eq. \eqref{eq:Kop}  over two months in   2024, evaluated at Moon latitudes and longitudes ($0^\circ$,$0^\circ$) \emph{in green}, ($90^\circ$,$0^\circ$) \emph{in blue}, and ($0^\circ$,$90^\circ$) \emph{in orange}. 
  }
  \label{fig:TCG-TCL_Kop}
\end{figure}

The so-obtained difference $\TCL-\TCG$ shows the same $1.6 \times 10^{-11}$ (or $1.5 \microsec / \mathrm{day}$) secular term as in the BCRS case, but with periodic variations not larger than $0.5\microsec$ in magnitude. An example is given in the top panel of figure \ref{fig:TCG-TCL_Kop}
for a clock located at $0^\circ$ longitude and $0^\circ$ latitude on the Moon (the secular term is removed for clarity). The lower amplitude of periodic terms when relying on GCRS rather than BCRS can be explained by the fact that 
 the Earth and the Moon forming a 2-body system in free fall  in the gravitational
field of the Sun (and other planets of the solar system)  appears only in the form of tidal terms \citep{Kopeikin2024}. The timescale difference $\TCL-\TCG$ is compatible with a light travel time computed in GCRS. 
In this case, the numerical value of  $[\tcl - \tcg]_{(\tcgB,\xgB)} $ depends on the lunar clock position via the location-dependent term in equation \eqref{eq:Kop}: $\vg_\lune \cdot (\xgB-\xg_\lune)$. It generates differences up to 20 ns between separate locations as shown in the bottom panel of figure~\ref{fig:TCG-TCL_Kop}.

An alternative to the GCRS as intermediate reference system is discussed in the appendix of \cite{Kopeikin2024} where they rely on IAU 2000 resolutions to build a Earth-Moon local coordinate system with the origin at the Earth-Moon barycenter. In doing so, they recover, and enlarge, previous results by \cite{Ashby2024} and then perform some comparison with the GCRS. The magnitudes of the periodic terms they obtain are very similar than in GCRS as the barycenter of the Earth-Moon system is lying inside the Earth. 

Using BCRS or GCRS to estimate the numerical value of the range between the clock on the Earth and the clock on the Moon shall not affect the outcome $\pptimeB - \pptimeA$. We thus expect some cancellations of coordinate-dependent terms between the light travel time equation and the timescales transformation when 4D relativistic transformations between BCRS and GCRS are properly applied. For this to happen, a quick comparison of equations \eqref{eq:range} and \eqref{eq:rangeKop}, reveals that the following relation:
\begin{equation}
  [\tcl - \tcg]_{(\tcgB,\xgB)} + [\tcgB - \tcgA]_{(\tcgA,\xgA,\xgB)} = [\tcl - \tcb]_{(\tcbB,\xbB)} + [\tcbB - \tcbA]_{(\tcbA,\xbA,\xbB)} - [\tcg - \tcb]_{(\tcbA,\xbA)} \, ,
  \label{eq:temp1}
\end{equation}
shall be verified for the range to remain generally covariant. In the \ref{sec:GCRSVSBCRS}, we demonstrate that the left-hand side of the previous expression does indeed reduce to the right-hand side when 4D relativistic transformations between BCRS and GCRS are applied consistently.

\subsubsection{Using TT instead of TCG.} TCG is not the only coordinate time which might be employed to compute the light travel time in GCRS coordinates; TT could be used too, and hence, the light-time would split into the three following contributions:
\begin{equation}
  [\tcgB - \tcgA]_{(\tcgA,\xgA,\xgB)} = [\tcg^*-\tcg]_{(\tcgA)}+[\tcgB^*-\tcgA^*]_{(\tcgA^*,\xgA^*,\xgB^*)}-[\tcg^*-\tcg]_{(\tcgB)} \, ,   
\end{equation}
where $[\tcgB^*-\tcgA^*]_{(\tcgA^*,\xgA^*,\xgB^*)}$ is the TT-compatible light travel time, and where $[\tcg^*-\tcg]_{(\tcg)}$ is given in equation \eqref{eq:TCG-TT}. To ensure the general covariance while computing the difference of proper times $\pptimeB - \pptimeA$, it is important to employ TT-compatible distances and mass parameters in the light travel time expression (this statement can be demonstrated in a similar way as in \ref{sec:TDBVSTCB} for BCRS with TDB or TCB). However, let us emphasize that mass parameters and distances which are distributed by planetary ephemerides are usually TDB-compatible. This means that two scalings have to be employed to go from TDB-compatible quantities to TT-compatible ones. For instance, to pass from a TDB-compatible mass parameter $(Gm)_{**}$ to the TT-compatible one, $(Gm)_*$, a first scaling to go from TDB to TCB/TCG-compatible framework must be applied, and then, a second scaling to go from TCG to the TT-compatible mass parameter. The relation between $(Gm)_{*}$ and $(Gm)_{**}$ thus eventually reads such as
\begin{equation}
 (Gm)_{*} = \left(\frac{1-L_G}{1-L_B} \right) (Gm)_{**} \, .
\end{equation}
Considering that planetary ephemerides are computed and distributed in a TDB-compatible frame, it is more advantageous and safe to compute the light-time in TCG rather than in TT---in TCG-compatible frame, only one scaling shall be applied to TDB-compatible parameters whereas two scalings are needed in a TT-compatible frame. However, note that if TT is used somewhere else in the range calculation, the double scaling issue will persist. Indeed, TT is usually not employed at the level of the light-time computation but rather at the level of the proper time of the clock on Earth as stated earlier below equation~\eqref{eq:range}. Therefore, the rate of clock `$\A$'s proper time with respect to TCG in equation \eqref{eq:rangeKop} is usually replaced by
\begin{equation}
  [\pptime - \tcg]_{(\tcgA,\xgA)} = [\pptime - \tcg^*]_{(\tcgA^*,\xgA^*)} + [\tcg^* - \tcg]_{(\tcgA)} \, ,
  \label{eq:pptimeTT}
\end{equation}
while the light-time computation is maintained in TCG. Note however that, from a theoretical point of view, this does not change the double scaling issue since the proper time must now be computed with TT-compatible quantities as it is clearly visible from the presence of TT-compatible arguments in the function $[\pptime - \tcg^*]_{(\tcgA^*,\xgA^*)}$.

\subsection{Using LCRS with TCL or rescaled TCL}

Finally, we can also consider working in the LCRS, where the corresponding range equation reads
\begin{equation}
  \pptimeB - \pptimeA = [\pptime - \tcl]_{(\tclB,\xlB)} + [\tclB - \tclA]_{(\tclA,\xlA,\xlB)} - [\tcg - \tcl]_{(\tclA,\xlA)} - [\pptime - \tcg]_{(\tcgA,\xgA)} \, .
  \label{eq:rangeLCRS}
\end{equation}
In this expression, the light travel time is now computed in LCRS coordinates and a unique coordinate time transformation is actually needed: $[\tcg - \tcl]_{(\tclA,\xlA)}$. It is expressed at the level of the clock on the Earth unlike what was done in the previous subsection where GCRS was employed as intermediate reference system. There, the timescale transformation was expressed at the level of the clock on the Moon.

\subsubsection{Timescale transformation in LCRS.}

\begin{figure}[t]
  \begin{center}
    \includegraphics[scale=0.7]{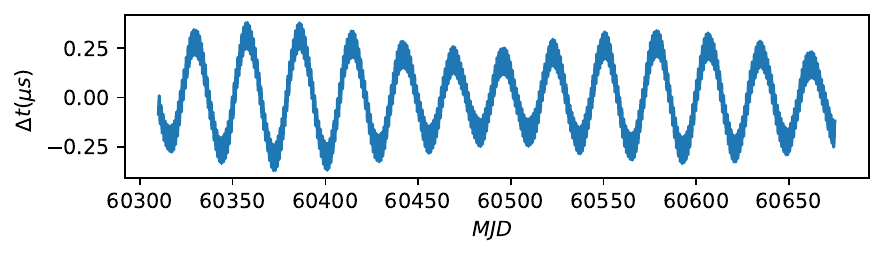}
   \includegraphics[scale=0.7]{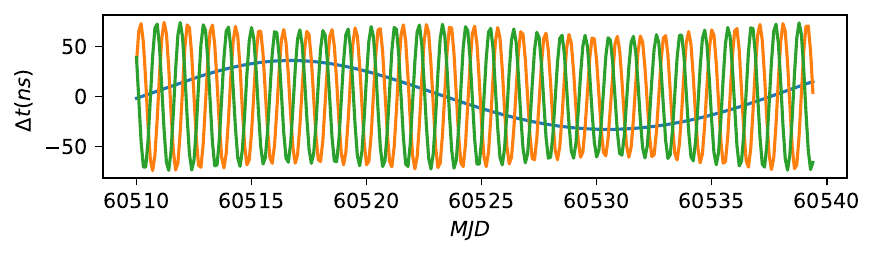}
   \end{center}
  \caption{\emph{Top}: Time differences $[\tcg - \tcl]_{(\tclA,\xlA)}$ over the year 2024, with the Earth clock $\xlA$ at latitude and longitude ($0^\circ$,$0^\circ$), after removing the secular term of $1.5 \microsec / \mathrm{day}$. \emph{Bottom}: Scalar product  $\vl_\terre \cdot (\xlA-\xl_\terre)$ from Eq. \eqref{eq:Kop}  over one month in 2024, evaluated at Earth latitudes and longitudes ($0^\circ$,$0^\circ$) \emph{in green}, ($90^\circ$,$0^\circ$) \emph{in blue} and ($0^\circ$,$120^\circ$) \emph{in orange}.
  }
  \label{fig:KopLune}
\end{figure}

Using the same analytical development as in \citet{Kopeikin2024}, but in the LCRS, we obtain the equation for the relative rate of TCG with respect to TCL:
\begin{align}
    [\tcg - \tcl]_{(\tclA,\xlA)} & = \frac{1}{c^2} \Bigg\{ \int_{\tcl_0}^{\tclA}
    \left[\frac{\vitl^2_\terre(\tcl)}{2}+\frac{G(m_\lune-2m_\terre)}{\vert \xl_\terre (\tcl) \vert} + \pottide (\tcl,\xl_\terre(\tcl)) \right] \dd \tcl \nonumber\\
    & + \vl_\terre (\tclA) \cdot [\xlA(\tclA)-\xl_\terre(\tclA)] \Bigg\} \, ,
    \label{eq:KopLune}
\end{align}
where $\xl_\terre$ and $\vl_\terre$ are the position and the velocity of the Earth in the LCRS, and $\pottide$ is the quadrupole of the tidal potential of the Sun:
\begin{align}
   \pottide (\tcl,\xl_\terre) = \frac{1}{2} \frac{G m_{\sun}}{\vert \xl_\sun \vert} 
    & \Bigg\{\left( \frac{\vert \xl_\terre \vert}{\vert \xl_\sun \vert}\right)^2 
    \left[ 3 \left( \Nl_\sun \cdot \Nl_\terre \right)^2 -1 \right] \nonumber\\
    &\!+\left( \frac{\vert \xl_\terre \vert}{\vert \xl_\sun \vert}\right)^{\!3} 
    \left[ 5 \left( \Nl_\sun \cdot \Nl_\terre \right)^3 - 3\left( \Nl_\sun \cdot \Nl_\terre \right) \right]\Bigg\}\, ,
\end{align}
with $\Nl_\sun=\xl_\sun/\vert \xl_\sun  \vert $ and $\Nl_\terre=\xl_\terre/\vert \xl_\terre  \vert$; $\xl_\sun$ being the lunocentric position of the Sun. The difference $ [\tcg - \tcl]_{(\tclA,\xlA)}$  shows  the same secular term of $1.5 \microsec / \mathrm{day}$ as when relying on BCRS or GCRS as intermediate systems, plus monthly oscillations with a magnitude of about $0.25 \microsec$ as shown in the top panel of figure \ref{fig:KopLune}. On top of these are superimposed smaller daily variations whose magnitude and phase depend on the Earth clock position  via the location-dependent term in equation~\eqref{eq:KopLune}: $\vl_\terre \cdot (\xlA-\xl_\terre)$ . The magnitude of these smaller variations can reach more than 50 ns as shown in the bottom panel of figure \ref{fig:KopLune}. Using LCRS instead of BCRS or GCRS to perform the computation of the difference of proper times $\pptimeB - \pptimeA$ does not change the outcome of the range numerical value. The proof of this statement is exactly similar to the demonstration in the \ref{sec:GCRSVSBCRS} where the range expression rely on GCRS instead of BCRS. In other words, according to the general covariance principle, we expect the following relation [comparison between range Eqs. \eqref{eq:rangeLCRS} and \eqref{eq:range}]:
\begin{equation}
  [\tclB - \tclA]_{(\tclA,\xlA,\xlB)} - [\tcg - \tcl]_{(\tclA,\xlA)} = [\tcl - \tcb]_{(\tcbB,\xbB)} + [\tcbB - \tcbA]_{(\tcbA,\xbA,\xbB)} - [\tcg - \tcb]_{(\tcbA,\xbA)} \, .
\end{equation}
to be always satisfied.

\subsubsection{Using a rescaled TCL instead of TCL.} In section \ref{sec:refsel}, we introduced $\tcl^\dag$, a rescaled coordinate time in LCRS, such that the average rate of the proper time of a clock located on the selenoid is suppressed when using $\tcl^\dag$ instead of $\tcl$. If such a timescale is needed and adopted by the IAU, which is not the case yet, one might wish to employ it for synchronizing procedure in lunar environment. This can be done at two different levels: (i) in the light travel time computation, or (ii) in the proper time computation. When invoking a rescaled TCL-compatible light-time $[\tclB^\dag - \tclA^\dag]_{(\tclA^\dag,\xlA^\dag,\xlB^\dag)}$, the TCL-compatible light-time in equation~\eqref{eq:rangeLCRS} shall be replaced by
\begin{equation}
  [\tclB - \tclA]_{(\tclA,\xlA,\xlB)} = [\tcl^\dag - \tcl]_{(\tclA)} + [\tclB^\dag - \tclA^\dag]_{(\tclA^\dag,\xlA^\dag,\xlB^\dag)} - [\tcl^\dag - \tcl]_{(\tclB)} \, ,
  \label{eq:TCLVSrescaledTCL}
\end{equation}
where the function $[\tcl^\dag - \tcl]_{(\tcl)}$ has been introduced previously in equation \eqref{eq:teclscale-tcl}. Let us emphasize that in order to ensure a perfect equality in equation \eqref{eq:TCLVSrescaledTCL}, it is required that rescaled TCL-compatible quantities be employed in the light travel time expression, that is to say rescaling distances and mass parameters. As discussed above in the case of TT and TCG, two scalings are actually needed to pass from TDB-compatible quantities to rescaled TCL-compatible ones. For instance, if a TDB-compatible mass parameter $(Gm)_{**}$ is provided by planetary ephemerides, after applying the double scaling, the rescaled mass parameter $(Gm)_{\dag}$ is given by
\begin{equation}
 (Gm)_{\dag} = \left(\frac{1-L_L}{1-L_B} \right) (Gm)_{**} \, ,
\end{equation}
with $L_L$ introduced back in equation \eqref{eq:teclscale-tcl}. From a practical point of view, avoiding multiple scaling is preferable. Hence, considering the light travel time in TCL rather than in its rescaled version is more advantageous since only one scaling is needed to pass from TDB-compatible quantities to TCB/TCL-compatible ones. However, note that if the rescaled TCL is used somewhere else in the range calculation, as suggested by option (ii), the double scaling issue will still persist. Indeed, if the rate of clock `$\B$'s proper time relative to TCL in equation \eqref{eq:rangeLCRS} is replaced by
\begin{equation}
  [\pptime - \tcl]_{(\tclB,\xlB)} = [\pptime - \tcl^\dag]_{(\tclB^\dag,\xlB^\dag)} + [\tcl^\dag - \tcl]_{(\tclB)} \, ,
\end{equation}
while keeping the light-time computation in TCL, we note that rescaled TCL-compatible quantities are still in use in the function $[\pptime - \tcl^\dag]_{(\tclB^\dag,\xlB^\dag)}$.

\subsection{Summary}

\subsubsection{General covariance.} As already discussed in this section, the expression of the range between two clocks depends on the intermediate reference system and its associated coordinate timescale being adopted for the computation of the light travel time. However, as imposed by the general covariance principle, the numerical value of the range has to remain unchanged no matter the intermediate reference system being employed. To ensure a full covariance, the structure of the equation relating the two proper times invariably relies on three contributions: (i) a function relating observer's proper time to a local coordinate time, (ii) a transformation between a local coordinate time and an intermediate coordinate time (used for expressing the light travel time), and (iii) a time transfer function expressed in the intermediate coordinate time. In a one-way experiment, when the intermediate reference system is BCRS, the transformation (ii) involves two timescale transformations applied at the two space-time events $(\tcbB,\xbB)$ and $(\tcbA,\xbA)$. In GCRS (resp. LCRS) only one timescale transformation is required; it is applied to the space-time event $(\tcgB,\xgB)$ [resp. $(\tclA,\xlA)$]. However, let us emphasize that this greater simplicity is only apparent, it is indeed counterbalanced by the fact that the light travel time must be expressed in GCRS (or LCRS) whereas the positions and velocities are usually distributed in BCRS by planetary ephemerides. Thus, a spatial relativistic transformation from BCRS to GCRS (or LCRS) is generally required into the light-time expression. Moreover, if the intermediate coordinate time is a rescaled coordinate time, the general covariance imposes one to use rescaled distances and mass parameters in the light-time expression or in the proper time computation, and this might be a source of confusion and errors (we recall that the scaling of distances impacts the leading order term in the time transfer function). This is even more critical if the intermediate coordinate time is TT or a rescaled version of TCL since two scalings are required for a complete invariance: a first scaling from TDB to TCB/TCG/TCL and a second one to TT or to the rescaled version of TCL. Therefore, in a given application, if planetary ephemerides must be called to compute a light-time and a proper time, we would advise to rely on expression \eqref{eq:rangeTDB} (without rescaled coordinate times) in order to avoid possible confusion at the level of coordinate transformations and scaling factors. 

\subsubsection{Practical use.} As we have seen, the difference between TCG and TCL is showing a secular term of about $1.5 \microsec / \mathrm{day}$ plus periodic variations with magnitude depending on the coordinate system and on the clock positions where this difference is computed. 
Users on Earth willing to communicate with users on the Moon (or inversely), will have to use timescale transformations which are consistent with the intermediate reference system being adopted for the computation of the light travel time. As discussed earlier, these transformations being part of 4D relativistic transformations, they depend on a location. For users on the surface of the Earth and on the surface of the Moon, the location-dependent terms seem to have straightforward dependency to latitudes and longitudes. It would therefore be possible to derive analytical expressions, such that the users would just need to input their local coordinates into the model broadcast by the lunar communication and navigation services. In the difference between a lunar reference coordinate time and UTC, only the secular term does not depend on the clock positions and the intermediate coordinate system. The exact value of this frequency shift will slightly vary over long time scales due to the secular relative motion of the Moon and the Earth. However, a standard value could be used by the different space agencies as a convention, avoiding the need to broadcast it continuously. The variations with respect to the conventional value, could then be broadcast as a polynomial. 

\section{Three options for the lunar reference timescale}
\label{sec:options}

In order to allow interoperability between different PNT systems  on the Moon, it is important to define a common reference timescale, exactly as TT (and its realization TAI, or further UTC)  has been chosen and commonly adopted for all activities in the vicinity of the Earth. We suggest to call the reference timescale in the Moon environment the Lunar Time (TL)\footnote{We do not recommend using LT since this abbreviation is used for ``Local Time''.}. Several criteria have to be considered for this lunar timescale to be adopted as a reference.

\begin{itemize}
    \item The lunar reference timescale should be defined as a linear function of TCL.

    \item A physical realization should be available so that all actors could steer their clocks and internal timescales on this reference. Note that this physical realization should be based on Earth's clocks as long as there is no accurate clock on the Moon.

    \item  This reference should have a clear and documented relationship with UTC. UTC produced by the BIPM, based on TAI, is indeed the only recommended timescale for international reference as stated in the Resolution 2 of the 26th CGPM in 2018 \citep{CGPM2018res2}.

    \item The  approach  chosen for defining the Lunar reference time should be applicable to Mars, and possibly other planets, in the future.
\end{itemize}

The first criteria implies that TL is an affine function of TCL, namely scaled by some constant frequency offset:
\begin{equation}
    \label{TL}
    \TL=\TCL + \Delta f \, (\TCL-\TCL_0) + \mathrm{const}_0 \, .
\end{equation}
We consider here three options that could make sense for lunar operations:
\begin{enumerate}
    \item $\Delta f$ is such that the rate of TL is exactly the same as TCL, namely $\Delta f=0$; \label{item:1}
    \item $\Delta f$ is such that the rate of TL corresponds, on average, to the rate of proper time of a clock on a given selenoid; \label{item:2}
    \item $\Delta f$ is such that there are only periodic variations between TL and TT. \label{item:3}
\end{enumerate}

As these three options are based on frequency offsets, before discussing them in turn, we give in table \ref{tab:magnitude} the order of magnitude of the mean rate differences between TCL and either the proper time of clocks on the lunar surface [which would be the scaling for option \eqref{item:2}] or TT [which would be the scaling for option \eqref{item:3}]. The first value has been computed as in section \ref{proper time Moon}. The  value of $\langle\mathrm{TCL} - \mathrm{TT}\rangle_{\mathrm{mean\; rate}}$ has been estimated from the difference between equations \eqref{eq:TCB-TCG} at the geocenter and \eqref{eq:TCB-TCL} at the Moon center of mass, then adding the rate difference $\TCG-\TT$, namely $L_G$. For comparison, we also provide the corresponding values for the TCG, which both correspond to $\pm L_G$. Furthermore, in order to consider the fourth criteria for choosing the lunar reference time mentioned here above, we also provide an estimation of the corresponding values for Mars. While Mars Coordinate Time (TCM) has not yet been formally defined by the IAU, its relations to TCB must  be equivalent to the one for the Earth and the Moon [Eqs. \eqref{eq:TCB-TCG} and \eqref{eq:TCB-TCL}]. We estimate here $\langle\mathrm{TCM} - \mathrm{TT}\rangle_{\mathrm{mean\; rate}}$  in the same way as explained here above for the Moon. The difference $\langle\tau_{\mathrm{surface}}-\mathrm{TCM}\rangle_{\mathrm{mean\;rate}}$ is computed from the gravitational potential on Mars' surface, using the IAU recommended values for the Mars radius, rotation speed and gravitational parameter.  

Let us recall some consequences of scaling TCL as in options \eqref{item:2} and \eqref{item:3}. As mentioned in sections \ref{sec:Earth} and \ref{sec:Moon},  a scaling of a coordinate time is necessarily accompanied by the corresponding scaling of spatial coordinates 
and mass parameters of celestial bodies so that the form of the equations of motion and light propagation are preserved \citep{Klioner2008}.
The scaled coordinate time and spatial coordinates can then be seen as defining a
new reference system  characterized by its own metric tensor. Presently, two different scalings are recommended by the IAU: the one defining TDB from TCB [cf. Eq.~\eqref{eq:TDB-TCB}] and the one defining TT from TCG [cf. Eq.~\eqref{eq:TCG-TT}]. Defining a reference coordinate time based on a scaling of TCL in the lunar environment would thus introduce an additional set of rescaled mass parameters and distances. Applying a similar approach to Mars in the future would again imply another scaling factor and an another set of rescaled masses and distances, leading to possible increasing confusion as more scaled coordinate times would be introduced for other planets. 

\begin{table}[t]
    \centering
       \caption{Magnitude of the mean relative frequency offsets between proper times on the surface and local coordinate times TCX, with TCX meaning TCG for the Earth, 
TCL for the Moon and TCM for Mars.}
       \vspace{0.2cm}
       \begin{tabular}{l |c r| c r}
        \hline\hline
        \multicolumn{1}{c}{} & \multicolumn{2}{c}{$\langle\tau_{\mathrm{surface}}-\mathrm{TCX}\rangle_{\mathrm{mean\;rate}}$} & \multicolumn{2}{c}{$\langle\mathrm{TCX} - \mathrm{TT}\rangle_{\mathrm{mean\; rate}}$}\\
         \hline
         Earth & $-L_G$ & $-60.2 \microsec / \mathrm{day} $ & $L_G$ & $ 60.2 \microsec / \mathrm{day}$\\
         Moon &  $ -3.1 \times 10^{-11}$ & $-2.7 \microsec / \mathrm{day}$ & $6.8 \times 10^{-10}$ & $ 58.7 \microsec / \mathrm{day }$ \\
         Mars & $ -1.4  \times 10^{-10}$ & $ -14.7 \microsec / \mathrm{day}$ & $ 5.8\times 10^{-9}$ & $501 \microsec / \mathrm{day}$ \\
    \hline
    \end{tabular}
   \label{tab:magnitude}
\end{table}

Hereafter, we denote by $\tcl^\ddag$ the TL timescale, whatever the TL option which is selected---according to our previous notation in section \ref{sec:Moon}, option \eqref{item:2} reads as $\tcl^\ddag = \tcl^\dag$ with $\Delta f = -L_L$ [cf. Eq. \eqref{eq:teclscale-tcl}]. For the following discussion, we assume that TL is introduced in the range computation at the level of the proper time of clock `$\B$'. Therefore, we replace $[\pptime-\tcl]_{(\tclB,\xlB)}$ in equation~\eqref{eq:rangeTDB} by
\begin{equation}
  [\pptime-\tcl]_{(\tclB,\xlB)} = [\pptime-\tcl^\ddag]_{(\tclB^\ddag,\xlB^\ddag)} + [\tcl^\ddag - \tcl]_{(\tclB)} \, ,
\end{equation}
where $[\tcl^\ddag - \tcl]_{(\tcl)}$ is given in equation \eqref{TL}. By invoking equation \eqref{eq:pptimeTT} in order to rely on TT rather than TCG at the level of the computation of proper time of clock `$\A$', the symmetrical expression for the range now becomes
\begin{align}
  \pptimeB - \pptimeA & = [\pptime-\tcl^\ddag]_{(\tclB^\ddag,\xlB^\ddag)} + [\tcl^\ddag - \tdb]_{(\tdbB,\xbdB)} + [\tdbB - \tdbA]_{(\tdbA,\xbdA,\xbdB)} \nonumber\\
  & - [\tcg^* - \tdb]_{(\tdbA,\xbdA)} - [\pptime - \tcg^*]_{(\tcgA^*,\xgA^*)} \, .
  \label{eq:rangeTDBTL}
\end{align}
The different steps in that equation are the following: proper time of clock `$\B$' to TL, TL to TDB evaluated at clock `$\B$', TDB-compatible light-time between clocks `$\A$' to `$\B$', then, TDB to TT evaluated at clock `$\A$', and finally TT to proper time of clock `$\A$'. 

\subsection{TL equal to TCL}

This option implies that an ideal clock located on the surface of the Moon would have a frequency shift about $3.14 \times 10^{-11}$, that is to say $2.7 \microsec / \mathrm{day}$, with respect to the reference TCL. However, this frequency offset would  be visible only if the clock accuracy is sufficient, namely better than the $10^{-11}$ level. Note that to date only Caesium clocks, H-masers and optical clocks reach that level on the Earth. Nevertheless, in order to avoid possible large differences between the reference and the time indicated by an accurate clock on the lunar surface, its frequency could be adjusted by a synthesizer from which the output signal would tick at a rate corresponding to the TCL. This is similar to what is currently done  on the Earth for the UTC(k) which are steered on UTC. These UTC(k) are driven from clocks at different altitudes, hence with different rates with respect to TT, and their frequency is therefore adjusted to compensate the gravitational redshift and the clock instability in order to provide a timescale aligned on UTC. It is also similar to what is done for some GNSS satellite clocks which are pre-tuned to tick UTC while their proper time would deviate from UTC by about $38 \microsec / \mathrm{day}$. 

Depending on user needs, any clock in lunar environment could be adjusted to the reference time  TCL  broadcast for instance by the lunar navigation systems, or some ground station---on Earth or even on the Moon in the future. For Mars, the frequency offset is about $1.4 \times 10^{-10}$ between a clock on the surface and TCM; the exact same approach as just explained for the Moon, namely introducing frequency corrections, could be employed, depending on user needs.

The main advantage of this option with respect to options \eqref{item:2}  and \eqref{item:3} below is that no additional coordinate time besides TCL has to be introduced, which avoids new definition and avoids the need to introduce new scaling of mass parameters and distances. If this choice is also adopted for Mars and other planets in a near future, it would drastically simplify future transformations between reference systems and timescales. There would be only one time scale per local reference system except for the geocentric system which would still involve TCG and TT for historical reasons.

\subsection{TL rate aligned on the average clock rate on a selenoid}

This option consists in mimicking what has been done for the Earth with TT (see e.g., \citet{Kopeikin2024,Ashby2024}, and \citet{2025ApJ...985..140T} for similar approach). The TL would be defined as a scaled version of TCL, so that the rate difference between TL and the average rate of a clock on a lunar surface of reference is as small as possible. This would require that a selenoid is defined with a reference potential $\poteff_{0}$ and an associated scaling factor $L_L$ by analogy with the Earth $W_{0}$ and $L_G$ scaling. As a consequence, any clock in free-running mode on the defined selenoid would be realizing TL with an accuracy equal to the clock accuracy (and at a maximal level of $10^{-18}$ due to periodic terms from the tidal potential). Such  definitions  of $\poteff_{0}$ and $L_L$ however do not exist to date. 

Moreover, as seen in section \ref{sec:Moon} and in figure~\ref{fig:lunar_total}, due to the high variations of the surface topography, a clock at rest on the lunar surface can experience rate differences up to $1.6 \times 10^{-13}$ (i.e., up to $20\ \mathrm{ns}/\mathrm{day}$) with respect to the reference depending on its selenographic coordinates. Therefore, if the user needs are surpassing that level, the clocks should anyway be steered accordingly on the reference TL, or corrected from local gravity measurements.

Note that, unlike option \eqref{item:1}, option \eqref{item:2} requires implicitly the introduction of a new space-time reference system associated with a scaled version of TCL. This adds possible confusion in the numerical values of mass parameters to be used. The scaling of $3.14 \times 10^{-11}$ would also apply to distances, that is to say about $1\ \mathrm{cm}$ on the Earth-Moon distance, which should be taken into account within a centimeter requirement when expressing for instance the light travel time between the Earth and Moon in the LCRS using TL as the coordinate time.  This is even larger for Mars where the scaling amounts to $1.4 \times 10^{-10}$ which would imply a difference of hundreds of meters on the Earth-Mars distance when working in the Mars Celestial Reference System (MCRS) using a rescaled coordinate time rather than TCM. In both cases, the scaling would also apply on mass parameters distributed by planetary ephemerides. 

\subsection{TL aligned on TT}

In this option, the difference between TL and TT would reduce to periodic terms only (see e.g., \cite{Fienga2024} for similar approach). There would be no long-term drift between TL and TT/TDB, potentially facilitating communication with the Earth.

This option could be particularly interesting when using Earth GNSS on the Moon such as in the LUGRE mission \citep{Parker2022}. Because Earth GNSS reference time is aligned on UTC, the GNSS signals would enable to get a rough timing in the lunar region. It seems however unlikely that simultaneous use of signals from Earth's constellation and lunar PNT satellites could easily be done due to the difference of coordinate reference systems used by the two constellations (we recall that the geocentric position is encoded in the GNSS signal from Earth's constellation).

On a more technical level, using TL aligned in average on TT would permit to use TL as a good first estimate of TDB (i.e., $\tcbB^{**} = \tclB^\ddag$) while solving iteratively for the implicit time transformation equation: $\tcbB^{**} = \tclB^\ddag - [\tcl^\ddag - \tcb^{**}]_{(\tcbB^{**},\xbB^{**})}$ with $[\tcl^\ddag - \tcb^{**}]_{(\tcbB^{**},\xbB^{**})}$ containing only small periodic terms for option \eqref{item:3}. Instead, for option \eqref{item:1} (i.e., $\tclB^\ddag = \tclB$), one would have to start with
\begin{equation}
  \tcbB^{**} = \tclB - \left( L_B - L_C\right) ( \tclB - T_0 ) + \TDB_0 \, ,
\end{equation}
(where $T_0$ and $\TDB_0$ have been introduced in equations \eqref{eq:TCG-TT} and \eqref{eq:TDB-TCB}; $L_B$ and $L_C$ were introduced in \citet{IAUGA1991} and later redefined in \citet{IAUGA2000}) to ensure convergence of the iterative process for any value of $\tclB$. For option \eqref{item:2} (i.e., $\tcl^\ddag = \tcl^\dag$), one should start the iterative resolution with
\begin{equation}
  \tdbB = \tclB^\dag + \left(L_L-L_B+L_C\right) (\tclB^\dag - T_0 - \mathrm{const}_0) - \TDB_0 - \mathrm{const}_0 \, ,
\end{equation}
[where $\mathrm{const}_0$ was introduced in Eq. \eqref{TL}] to ensure convergence for all values of $\tclB^\dag$. Note however that $\tcbB^{**} = \tclB^\ddag$ will also be a good first approximation (better than $1\%$ relative error) for both options \eqref{item:1} and \eqref{item:2} as long as $\vert \tclB^\ddag - T_0 \vert > 5\ \mathrm{day}$ with $T_0$ being 1977 January 1st, $0^{\mathrm{h}}\,0^{\mathrm{m}}\,0^{\mathrm{s}}$ TAI. Therefore, what seemed to be at first glance a privilege advantage of option \eqref{item:3} with respect to the two other options is actually not for current and future data well beyond 1977 January 1st, $0^{\mathrm{h}}\,0^{\mathrm{m}}\,0^{\mathrm{s}}$ TAI. 

The first drawback of option \eqref{item:3} would be that an ideal clock located on the Moon surface would have a significant frequency offset of about $6.6 \times 10^{-10}$ (i.e., $56\microsec / \mathrm{day}$) with respect to the reference. This, however, would be only visible for accurate clocks such as Rubidium/Caesium clocks or better. Some corresponding frequency correction should then be eventually applied to steer these clocks on the reference TL, which is, in the end, not that different than what was proposed in options \eqref{item:1} and \eqref{item:2}.

The main drawback, as for option \eqref{item:2}, is the introduction of a scaling factor between TCL and TL which changes the numerical values of distances and mass parameters. This scaling would be $7 \times 10^{-10}$ for the Moon, that is to say about $30\ \mathrm{cm}$ on the Earth-Moon distance when expressing it in the LCRS using TL as coordinate time rather than TCL.  If the same approach is later used for Mars, the scaling would be much larger as seen in table \ref{tab:magnitude}; it would generate a difference of $1.5\ \mathrm{cm}$ on the Mars radius and of the order of some kilometer on the Earth-Mars distance when working in the MCRS using the rescaled coordinate time rather than TCM.  

Finally, it must be noted that if the scaling is defined to have no mean drift with respect to TT, the value will
be determined with a limited accuracy and would be subject to changes in the future. Fixing a conventional value now would solve for that issue, but a small frequency offset could still be observed in the future between TL and TT due to the limited accuracy of the current determination possibilities. 

\subsection{On the need for clock frequency steering}

To conclude this section, we  summarize in table \ref{tab:steering} the impact of the TL definition on the need for clock steering, as a function of the user requirement. The thresholds where chosen as follows: $10^{-9}$ corresponding roughly to the scaling of option \eqref{item:3}, $10^{-11}$ corresponding roughly to the scaling of option \eqref{item:2}, and $10^{-13}$ corresponding to the proper time rate variations along the Moon topography. In table  \ref{tab:steering}, we observe a difference between the three options only in two cases: (a) steering would be needed only for option \eqref{item:3} when the clock accuracy is higher than the $10^{-9}$ level, and  the  user needs are between $10^{-9}$ and $10^{-11}$ and (b) option \eqref{item:2} only leaves the need of steering when the clock accuracy is better than $10^{-11}$ and the user requirement between $10^{-11}$ and $10^{-13}$. In all other cases, the need for steering does not depend on the chosen option for the reference timescale TL.

\begin{table}[t]
    \centering
    \caption{Options that require a clock steering, as a function of the user need}
    \vspace{0.2cm}
    \begin{tabular}{c | c c c c }
    \hline
        clock accuracy   & \multicolumn{4}{c}{user need (in $10^{-Y}$)}\\
      \multicolumn{1}{c|}{(in $10^{-X}$) }& $Y< 9$ & $Y = \{9-11\}$ & $Y=\{11-13\}$ & $Y>13$ \\
     \hline
     $X< 9$ & \eqref{item:1}, \eqref{item:2}, \eqref{item:3} & \eqref{item:1}, \eqref{item:2}, \eqref{item:3} & \eqref{item:1}, \eqref{item:2}, \eqref{item:3} & \eqref{item:1}, \eqref{item:2}, \eqref{item:3} \\
        $X = \{9-11\}$ & --- , --- , --- \;\, & --- , --- , \eqref{item:3} & \eqref{item:1}, \eqref{item:2}, \eqref{item:3} & \eqref{item:1}, \eqref{item:2}, \eqref{item:3} \\
       $X > 11$ & --- , --- , --- \;\,  & --- , --- , \eqref{item:3} & \eqref{item:1}, --- , \eqref{item:3} & \eqref{item:1}, \eqref{item:2}, \eqref{item:3} \\
       \hline
    \end{tabular}
\label{tab:steering}
\end{table}

\section{Realization of TL and traceability to UTC}
\label{sec:realization}

In the first years, having no atomic clock on the Moon, there will be no local realization of TL. The reference used for the lunar timing will necessarily be based on Earth clocks and some time transfer between lunar  clocks and these Earth clocks. 
The best  references available in real-time on the Earth are the realizations of UTC, named UTC(k). 
From what we have seen before, there is no unique relation between TT and TL, as it  depends on the reference system in which the observer is, and on the clock locations. However,  it is possible to monitor the difference between a lunar clock and the lunar reference timescale using time and frequency transfer with a clock realizing UTC on the Earth. The lunar clock error can  be determined after removing the position-dependent relativistic terms from the clock comparison.  
Let us consider UTC(k) at a given location on the Earth.

The lunar clock with proper time $\pptimeB$, which should be steered on a reference TL, will be compared to the UTC(k) using equation~\eqref{eq:range}, [or its equivalent \eqref{eq:rangeKop} in the GCRS or \eqref{eq:rangeLCRS} in the LCRS], detailed in section \ref{sec:TCL-TCG}. The time transfer measurements will then be corrected for the known relativistic effects as well as the light-time to finally get $[\zeta-\tcl^\ddag]_{(\tclB^\ddag,\xlB^\ddag)}$, where $\tcl^\ddag$ denotes TL (whatever the definition of TL) and $\zeta_\B = \pptimeB + \varepsilon_\B$ with $\varepsilon_\B$ being the clock error. We thus have the following relation:
\begin{align}
\label{eq:trac}
    \varepsilon_\B + [\pptime - \tcl^\ddag]_{(\tclB^\ddag,\xlB^\ddag)} & = [\zeta_\B - \UTCk]_{\mathrm{meas}} - [\tcl^\ddag - \tcb]_{(\tcbB,\xbB)} - [\tcbB - \tcbA]_{(\tcbA,\xbA,\xbB)} \nonumber\\
    & + [\tcg^* - \tcb]_{(\tcbA,\xbA)} + [\UTCk - \tcg^*]_{(\tcgA^*)} \, ,
\end{align}
where $[\pptime - \tcl^\ddag]_{(\tclB^\ddag,\xlB^\ddag)}$ is the difference between the proper time of clock `$\B$' and TL, with $\xlB^\ddag$ the TL-compatible position of the clock in LCRS, whatever TL is. $[\zeta_\B - \UTCk]_{\mathrm{meas}}$ is the measured difference between the lunar clock and $\UTCk$, and $[\UTCk - \tcg^*]_{(\tcgA^*)} $ is the difference between UTC(k) and TT (or $\tcg^*$). After correction for leap second and the 32.184 s,  as UTC(k) is a realization of UTC and hence of TT, this term is always very small.  If the time transfer measurements from the Moon to the Earth, as well as the relativistic models and travel time modeling are realized with given uncertainties, the traceability of the lunar clock to UTC can then be demonstrated by using the difference $\UTC-\UTCk$ from BIPM Circular T for the last term of equation~\eqref{eq:trac}.

Note that if two clocks `$\B_1$' and `$\B_2$' located in different positions around the Moon are synchronous in TL, of course they are not synchronous in TT, 
but if we compare them  with two clocks `$\A_1$' and `$\A_2$' synchronized in TT, also located at different locations on the Earth, and remove all the relativistic corrections, we get the same result for the two comparisons $(\varepsilon_{\B_1} - \varepsilon_{\A_1})$ and $(\varepsilon_{\B_2} - \varepsilon_{\A_2})$. Hence, all clocks around the Moon that will be steered on different UTC(k) maintained on the Earth, using  time links as in equation \eqref{eq:trac}, will be synchronous with respect to TL. It is obvious that for a strict traceability, the uncertainties associated with the relativistic corrections, and the ephemerides used for them, should be considered. Whether their magnitude will be significant with respect to the other sources of uncertainties in the clock comparisons is out of the scope of this paper. 

It is foreseen that in the future there will be autonomous clocks on the Moon. For example, ESA should launch in early thirties the NovaMoon  lunar in-orbit demonstrator and reference station which will provide a lunar-based local differential, geodetic and timing stations to enhance the accuracy of lunar PNT services to sub-meter levels across the South Pole \citep{NovaMoon}.  Such kind of clocks could then be used as local realizations of TL and serve as reference for lunar applications if their signal is made available. To be completely autonomous, these clocks should have a frequency accuracy sufficient for the user needs, and located on a place where the gravitational potential is known with a sufficient accuracy so that the gravitational redshift with respect to the reference can be taken into account. These local realizations of the lunar reference time TL could be called TL(k), and  if a network of clock is then used in an algorithm ensemble like is done for UTC on the Earth, we propose then to call this ensemble timescale TL(MOON). The traceability to UTC should however still be ensured via regular time transfer links between TL(k) and UTC(k). 

\section{Discussion and conclusion}
\label{sec:Discussion}

Since prehistoric times, the time on the Earth was based on the Earth rotation. The duration of a day was defined as $86\,400\ \mathrm{s}$. As soon as clocks were built, they were designed to tick this
time on the surface of the Earth. When special and general relativity entered into the game, with the need to define a coordinate time associated to the GCRS, it was of course natural to continue with the time rate as it was in use since millennia. For the moon, the situation is completely different as there is currently no clock in use on the Moon, a reference timescale has thus to be defined from scratch. The only obligation is to define a lunar reference timescale based on the TCL which is the coordinate time associated with the LCRS. 

Three options were proposed in the previous section: \eqref{item:1} using directly TCL, or using a scaled version of TCL so that \eqref{item:2} the lunar reference timescale corresponds, on average, to the proper time of a clock on the surface of the Moon, or, \eqref{item:3} the lunar reference timescale differs from TT only by periodic variations.

Options \eqref{item:2} and \eqref{item:3} would require an associated scaling of mass parameters and distances, which would unavoidably introduce complexity in every computation, especially while computing Earth-Moon time and frequency transfers. In addition, because we consider that a similar approach should be used later for Mars and maybe for other planets, this would again ask for additional scalings, with, for each of them, a different numerical value for mass parameters and distances in the solar system. In particular, using option \eqref{item:3} in the case of Mars would imply a scaling of the order of $10^{-9}$ which, if not correctly implemented, would lead to errors of the order of few kilometers on the Earth-Mars distance. In the option \eqref{item:3}, in addition, an accurate clock placed on the Moon surface (or later Mars surface) would not tick the reference time but would have a frequency offset of about $7 \times 10^{-10}$ for the Moon and $6 \times 10^{-9}$ for Mars; this frequency offset should be compensated for all applications requiring a higher accuracy. We therefore consider this option as the less convenient one. Aside of the problem of scaling, one advantage of option \eqref{item:2} would be that an accurate clock on the Moon surface would tick the reference time TL. However this argument is valid only at the condition that the clock accuracy is better than $10^{-11}$ and that the user need is less than $10^{-13}$ which is the variation of proper time along the Moon surface due to the topography. For less accurate clocks or more precise user needs, the clocks should be steered on the reference, exactly as if the reference is not defined on the selenoid. In view of this, and considering in addition that option \eqref{item:2} would require the definition of a lunar $\poteff_{0}$ which does not exist to date, we consider that the less constraining and natural solution is to use option \eqref{item:1}, that is to say $\TL=\TCL$, for which only one scaling is required to transform TDB-compatible quantities into TCL/TCB-compatible quantities.

The steering of clocks in the lunar environment should be done from time and frequency transfers with either the Earth clocks, or with lunar navigation satellites. 
 And the adjusted frequency will be maintained by the clocks for a certain period of time, depending on their intrinsic stability,  until the next connection with either the Earth or the satellites. For very accurate clocks, the steering could also be done based on very accurate determination of the orthometric altitude.

For the link between the lunar time and UTC, we have shown that the relationships between the different proper times are complex and can not be uniquely modeled, at accuracy levels well within the expected requirements for applications such as positioning and navigation. The secular trend between Moon time and Earth time, whatever the reference Lunar Time,  could however be sufficient to  time-tag events in UTC at the sub-microsecond level, which is the magnitude of  the periodic terms between TL and TT when working in the Earth-based or Moon-based coordinate systems. We expect UTC to remain the common operational timescale for humans on the Moon, as the contact and coordination with Earth-based operators will be crucial and frequent, and in many cases the synchronization error will be completely negligible  for these contacts. For example, it will be less error-prone to synchronize launches for orbital rendez-vous with the control center in UTC than having to use different timescales for the different spacecrafts involved. In such a scenario, errors committed by neglecting the periodic differences of proper time over the short time span of the manoeuvre are negligible with respect to the trajectory corrections that will be necessary anyway for other reasons. 

For more precise applications related to communications between the Earth and the Moon, the user will have to consider both the periodic terms of $\TL-\TT$ and the travel time of its communication link.  We therefore suggest that the periodic terms from the integrals of equations \eqref{eq:TCB-TCG} and \eqref{eq:TCB-TCL} are  broadcast to the users in a standard way, plus some expression of the scalar products as a function of the latitude and longitude for both the Earth and the Moon. So that it is easy for the timing device to communicate with an Earth station working in UTC, while still keeping its internal timescale reference aligned on TL. These informations could take the form of polynomials, similar to how ($\UTC - \mathrm{GNSS}$ timescale) are currently broadcast in navigation messages. The level of complexity of this scheme (degree of the polynomials, their validity period, sensitivity to ephemerides, expected accuracy) has to be studied more in details.

Finally, the different options for a lunar reference timescale proposed here should be reviewed by the space agencies and analysed in the frame of their operational constraints. Defining the coordinate timescale to be used in the lunar region will be a task for the IAU. As long as there is no accurate clock on the Moon, the UTC(k), being realization of TT, will be used together with the modeling of relativistic effects based on lunar ephemerides, to serve as realization of the lunar timescale reference TL. 

\section*{Acknowledgments}

The authors are grateful to the members of the CCTF on Moon timing for interesting discussions that motivated this work. The authors warmly thank the two anonymous referees for their comments that really helped improving the quality of the manuscript.

\begin{center}
------------------
\end{center}

\appendix

\section{Proper time of a clock on Moon's surface}
\label{app:comp}

Integrating equation \eqref{eq:dpptimedtcl} for an observer at rest on the lunar surface, for instance clock `$\B$', will allow us to determine the function $[\pptime - \tcl]_{(\tclB,\xlB)}$. In what follows, we explicit the different terms of \eqref{eq:dpptimedtcl}, where $\poteff$ and $\vpoteff$ are given in equations \eqref{eq:Wscal} and \eqref{eq:Wvect}, respectively. We keep in the final expression only the terms which are significant at the level of $10^{-16}$ which is the current accuracy of the SI second.

First, we derive the expressions for $\xlB(\tcl)$ and $\vlB(\tcl)$, namely the trajectory and velocity of clock `$\B$' in the LCRS. From the velocity, we derive the centrifugal potential and evaluate its contribution in equation \eqref{eq:dpptimedtcl}. Then, we study successively the lunar gravitational potential and the tidal potential which appear at $\mathcal{O}(c^{-2})$ in equation \eqref{eq:dpptimedtcl}. Finally, we show that the scalar inertial potential and the vector potential are negligible when considering relative frequency differences up to $10^{-16}$ for a clock situated on the lunar surface.

\subsection{Selenographic frame}

We shall introduce a selenographic frame in which clock `$\B$' is at rest. This frame is rigidly attached to the principal axis of inertia of Moon's figure and is thus referred to as the ``lunar principal axes frame''. The $A$ and $C$-axis are directed along the direction of minimal and maximal moments of inertia of the Moon, respectively. The $B$-axis completes the right-handed triad. We shall denote by $(\be{A},\be{B},\be{C})$ the three unit-vectors of the lunar principal axis frame. In this frame, the selenocentric position of clock `$\B$' reads as (see also \citet{Kopeikin2024} for a similar expression without the tidal contribution though)
\begin{equation}
  \xlB (\tcl) = \csphRB \, \NlB (\tcl) + \Delta \xlB (\tcl) \, ,
  \label{eq:posB}
\end{equation}
where the unit vector $\NlB$ is defined by
\begin{equation}
  \NlB (\tcl) = \be{A} (\tcl) \cos \csphTB \cos \csphLB + \be{B} (\tcl) \cos \csphTB \sin \csphLB + \be{C} (\tcl) \sin \csphTB \, .
\end{equation}
$\csphRB$ is the (mean) radial distance of clock `$\B$' from the center of mass of the Moon, $\csphTB$ and $\csphLB$ being the (mean) selenographic latitude and longitude of clock `$\B$', respectively (with $-90^\circ \leqslant \csphTB \leqslant 90^\circ$ and $-180^\circ < \csphLB \leqslant 180^\circ$). The term $\Delta \xlB$ represents the degree 2 lunar surface displacement due to tides raised on Moon by bodies in the solar system (mainly Earth and Sun). This term can be modelled such as (cf. \citet{2010ITN....36....1P})
\begin{align}
  \frac{\Delta \xlB (\tcl)}{R_\lune} & = \sum_{\ext\neq\lune} \left( \frac{m_\ext}{m_\lune} \right) \left(\frac{R_\lune}{\csphR_\ext}\right)^{\!3} \left(\frac{\csphRB}{R_\lune}\right)^{\!4} \nonumber\\
  & \times \Bigg\{ h_2^\lune \left[P_2\left(\NlB (\tcl) \cdot \NlG (\tcl)\right) - \frac{\delta_{\ext \terre}}{2} (3 \cos^2 \csphTB \cos^2 \csphLB - 1 ) \right] \NlB (\tcl) \nonumber\\
  & + 3l_2^\lune (\NlB (\tcl) \cdot \NlG (\tcl) ) \big[\NlG (\tcl) - ( \NlB (\tcl) \cdot \NlG (\tcl) ) \NlB (\tcl) \big] \Bigg\} \, , \label{eq:DeltaxlB}
\end{align}
where  $h_2^\lune$ (resp. $l_2^\lune$) is the nominal degree 2 Love (resp. Shida) number and $P_2$ is the Legendre polynomial of degree 2. $\csphRG$ and $\NlG$ are respectively the norm and direction of the selenocentric position vector of body $\ext$, namely $\xlG = \csphRG \, \NlG$ with $\csphRG = \vert \xlG \vert$. Let us emphasize that we have removed the constant radial part of the tidal displacement due to Earth tides since it shall be included in the lunar topography. This was done by introducing $\delta_{\ext \terre}$ which is $\delta_{\ext \terre}=1$ if $\ext = \terre$ (i.e, if the external body is Earth) and $\delta_{\ext \terre}=0$ otherwise. We can approximate the radial displacement $\Delta \csphRB$ as
\begin{equation}
  \Delta \csphRB \simeq \Delta \xlB \cdot \NlB \, ,
  \label{eq:DeltacsphRB}
\end{equation}
at linear order in $\vert \Delta \xlB \vert/\csphRB$. Using, equation \eqref{eq:DeltaxlB}, we thus arrive to
\begin{align}
  \frac{\Delta \csphRB (\tcl)}{R_\lune} & = h_2^\lune \sum_{\ext\neq\lune} \left( \frac{m_\ext}{m_\lune} \right) \left(\frac{R_\lune}{\csphR_\ext}\right)^{\!3} \left(\frac{\csphRB}{R_\lune}\right)^{\!4} \nonumber\\
  & \times \left[P_2\left(\NlB (\tcl) \cdot \NlG (\tcl)\right) - \frac{\delta_{\ext \terre}}{2} (3 \cos^2 \csphTB \cos^2 \csphLB - 1 ) \right] \, .
  \label{eq:DeltacsphRBES}
\end{align}

Let us now focus on tidal displacement raised by the Earth. According to equation~\eqref{eq:FourSerLeg2} below, we have
\begin{align}
  P_{2}(\NlB\cdot\NlE) & = \frac{1}{2} (3\cos^2\csphTB\cos^2\csphLB-1)-\frac{3\obqtL}{2} \sin 2\csphTB \cos\csphLB \sin (M_\lune + \omega_\lune)  \, ,
  \label{eq:FourSerLeg2approx}
\end{align}
at linear order in $\mathcal{O}(\obqtL)$ and neglecting $\mathcal{O}(e_\lune)$. In this expression, $\obqtL$ is the lunar obliquity ($\obqtL=0.12\ \mathrm{rad}$), $e_\lune$ is the Moon eccentricity ($e_\lune = 0.055$), $M_\lune$ is the mean anomaly of the Moon, and $\omega_\lune$ is its argument of perigee. Because the constant contribution is removed from $P_2(\NlB \cdot \NlG)$ in equation \eqref{eq:DeltacsphRBES}, the radial displacement of the lunar crust is varying with a mean synodic month period and an amplitude of the order of
\begin{equation}
  \frac{\Delta \csphRB (\tcl)}{R_\lune} \propto \frac{3 \obqtL h_2^\lune}{2} \left( \frac{m_\terre}{m_\lune} \right) \left(\frac{R_\lune}{a_\lune}\right)^{\!3} = 4.95 \times 10^{-8} \, ,
  \label{eq:displest}
\end{equation}
where $a_\lune$ denotes the semi-major axis of the Moon; we considered $a_\lune = 3.85\times 10^{5}\ \mathrm{km}$ and $h_2^\lune = 0.03786$ \citep{Folkner2014}. The lunar crust displacement thus amounts to $\pm 8.6\ \mathrm{cm}$. On the other hand, the displacement due to solar tides scales such as
\begin{equation}
  h_2^\lune \left( \frac{m_\sun}{m_\lune} \right) \left(\frac{R_\lune}{a_\terre}\right)^{\!3} R_\lune = 4.2 \ \mathrm{mm} \, ,
  \label{eq:displestSun}
\end{equation}
where $a_\terre$ denotes the semi-major axis of the Earth. The tidal displacement by Sun is thus one order of magnitude smaller than Earth's so it will be systematically neglected. Hereafter, the influence of the tidal displacement of the lunar surface in the time difference $[\pptime - \tcl]_{(\tclB,\xlB)}$ will be further investigated through the centrifugal potential and lunar gravitational potential.

For later convenience, we shall make use of the ecliptic J2000.0 spatial vector basis, that we shall denote by $(\be{x},\be{y},\be{z})$. This basis, such as the (equator) LCRS basis $(\be{\mathcal{X}},\be{\mathcal{Y}},\be{\mathcal{Z}})$, is non-rotating. The $x$-axis is directed toward the vernal equinox of J2000.0, the $z$-axis points toward the celestial pole of the ecliptic J2000.0, and the $y$-axis completes the right-handed triad. The unit-vectors $(\be{x},\be{y},\be{z})$ and $(\be{\mathcal{X}},\be{\mathcal{Y}},\be{\mathcal{Z}})$ are related by a rotation of angle $\obqt$ about the direction of the vernal equinox J2000.0:
\begin{subequations}
  \begin{empheq}[left=\empheqlbrace]{align}
    \be{x} & = \be{\mathcal{X}}\, , \\
    \be{y} & = \be{\mathcal{Y}} \cos\obqt + \be{\mathcal{Z}} \sin\obqt \, , \\
    \be{z} & = -\be{\mathcal{Y}} \sin\obqt + \be{\mathcal{Z}} \cos\obqt \, ,
  \end{empheq}
\end{subequations}
where $\obqt$ is the obliquity of the J2000.0 ecliptic, $\obqt=23^\circ26'$ \citep{2010ITN....36....1P}.

The orientation of the lunar principal axis frame is modelled according to the empirical laws of Cassini, namely neglecting the lunar physical librations that shall not exceed $10^{-3}\ \mathrm{rad}$ \citep{Newhall1996} (see also \citet{Kopeikin2024} for similar approximation) and would only be responsible for introducing periodic terms within the time difference computation. The empirical laws of Cassini state that \citep{Eckhardt1981}:
\begin{enumerate}
  \item[1. ] The Moon rotates uniformly about its polar axis $\be{C}$ with a rotational period equal to $2\pi/n_\lune$, with $n_\lune$ the mean sideral period of its orbit about the Earth.
  \item[2. ] The descending node of the lunar equator on the ecliptic precesses in coincidence with $\Omega_\lune$, the ascending node of the lunar orbit on the ecliptic.
  \item[3. ] The inclination of the lunar equator on the ecliptic is constant; it is $I_\lune=0.027\ \mathrm{rad}$.
\end{enumerate}
The second laws implies that the Moon polar axis $\be{C}$, its orbit normal, and the celestial pole of the ecliptic $\be{z}$ are coplanar. Furthermore, in order to maintain a locally-minimum gravitational energy such as is required for stability, the longest axis of the Moon, which coincides with the lowest moment of inertia, have to point Earthward. Among the two unit-vectors that lie within the lunar equator, it is thus $\be{A}$ which points Earthward whereas $\be{B}=\be{C}\times\be{A}$. The orientation of the principal axes frame with respect to the ecliptic J2000.0, is described by the three following Euler angles: $\eulphi$, $\eultheta$, and $\eulpsi$; $\eulphi$ is the longitude of the descending node of the lunar equator on the ecliptic of J2000.0, $\eultheta$ is the inclination of the lunar equator to the ecliptic J2000.0, and $\eulpsi$ is the angle between the descending node and $A$-axis. Cassini's laws thus imply
\begin{equation}
  \eulphi = \Omega_\lune \, , \qquad \eultheta = -I_\lune \, , \qquad \eulpsi+\phi = L_\lune+\pi \, ,
\end{equation}
where $L_\lune$ is the mean longitude of the Moon. It follows that
\begin{subequations}
  \begin{empheq}[left=\empheqlbrace]{align}
    \be{A} & = \be{x} \big[ \cos I_\lune\sin\Omega_\lune \sin(M_\lune+\omega_\lune)-\cos\Omega_\lune\cos(M_\lune+\omega_\lune)\big]\nonumber\\
    &-\be{y} \big[ \cos I_\lune\cos\Omega_\lune \sin(M_\lune+\omega_\lune)+\sin\Omega_\lune\cos(M_\lune+\omega_\lune)\big] \nonumber\\
    &+\be{z}\sin I_\lune\sin(M_\lune+\omega_\lune)\, , \\
    \be{B} & = \be{x} \big[ \cos I_\lune\sin\Omega_\lune \cos(M_\lune+\omega_\lune)+\cos\Omega_\lune\sin(M_\lune+\omega_\lune)\big]\nonumber\\
    &-\be{y} \big[ \cos I_\lune\cos\Omega_\lune \cos(M_\lune+\omega_\lune)-\sin\Omega_\lune\sin(M_\lune+\omega_\lune)\big] \nonumber\\
    &+\be{z}\sin I_\lune\cos(M_\lune+\omega_\lune)\, ,\\
    \be{C} & = -\be{x} \sin I_\lune \sin\Omega_\lune + \be{y} \sin I_\lune\cos\Omega_\lune + \be{z} \cos I_\lune \, .
  \end{empheq}
\end{subequations}

Let us recall that because of perturbations from the Sun and other planets, the lunar orbit does a retrograde precession along the ecliptic in a time period of $18.6\ \mathrm{yr}$, and a prograde precession of the perigee in $8.85\ \mathrm{yr}$ (cf. e.g., \citet{Simon1994}), which corresponds to mean rates of $\langle\dot\Omega_\lune\rangle=-1.07\times10^{-8}\ \mathrm{rad}\,\mathrm{s}^{-1}$ and $\langle\dot\omega_\lune\rangle=2.25\times10^{-8}\ \mathrm{rad}\,\mathrm{s}^{-1}$, respectively. Therefore, the time derivatives of the unit-vectors of the principal axis frame are given by
\begin{subequations}
  \begin{empheq}[left=\empheqlbrace]{align}
    \DD{\be{A}}{\tcl} & = \angvitCass (\tcl) \times\be{A} (\tcl)\, , \\
    \DD{\be{B}}{\tcl} & = \angvitCass (\tcl) \times\be{B} (\tcl)\, ,\\
    \DD{\be{C}}{\tcl} & = \langle\dot\Omega_\lune\rangle \, \be{z}\times\be{C} (\tcl)\, .
  \end{empheq}
\end{subequations}
where $\angvitCass$ is the angular velocity of the Moon as predicted by Cassini's laws:
\begin{equation}
  \angvitCass (\tcl) = n_\lune \, \be{C} (\tcl) + \langle\dot\omega_\lune\rangle \, \be{C} (\tcl) + \langle\dot\Omega_\lune\rangle \, \be{z} \, .
  \label{eq:angvitCass}
\end{equation}

\subsection{Centrifugal potential}
\label{subsec:potcentrifuge}

The LCRS velocity of clock `$\B$' is given by [time differentiation of Eq. \eqref{eq:posB}]
\begin{equation}
  \vlB (\tcl) = \csphRB \, \angvitCass(\tcl) \times \NlB (\tcl) + \angvitCass(\tcl) \times \Delta \xlB (\tcl) + \DD{\Delta \xlB}{\tcl}\, .
  \label{eq:vlB}
\end{equation}
The second and last terms in the right hand side are due to solid tides which are responsible for displacing the crust of the Moon and hence clock `$\B$'. These two terms are expected to be (at most) of the order of $\vert \angvitCass \vert \Delta \csphRB$ and $n_\lune \Delta \csphRB$, respectively. From equation \eqref{eq:angvitCass}, we see that they are in fact both proportional to $n_\lune \Delta \csphRB$ due to the lunar spin orbit resonance 1:1. Let us show that the tidal displacements lead to negligible contributions in the centrifugal potential (i.e., the square of the LCRS velocity of clock `$\B$'). Indeed, the first term in the right hand side of equation \eqref{eq:vlB} is the dominant one; it returns $\vitlB^2 \sim n_\lune^2R_\lune^2$, that is to say
\begin{equation}
  \left( \frac{n_\lune R_\lune}{c} \right)^{\!2} = 2.38 \times 10^{-16} \, ,
  \label{eq:est_centr_pot}
\end{equation}
after being inserted into equation \eqref{eq:dpptimedtcl}. Therefore, the contribution from tidal displacement generated by Earth [Sun's is smaller than Earth's, see discussion below Eq.~\eqref{eq:displest}] to the relative frequency difference (through the centrifugal potential) is scaling such as
\begin{equation}
  \left( \frac{n_\lune R_\lune}{c} \right)^{\!2} \frac{\Delta \csphRB}{R_\lune} \sim 10^{-24} \, .
\end{equation}
It can thus safely be neglected. Only the first term in the right hand side of equation~\eqref{eq:vlB} is considered. Then, considering that
\begin{equation}
  \frac{\langle\dot\omega_\lune\rangle}{n_\lune} = 8.46\times10^{-3} \, , \qquad \frac{\langle\dot\Omega_\lune\rangle}{n_\lune} = -4.02\times10^{-3} \, ,
  \label{eq:doOlune}
\end{equation}
we keep in equation \eqref{eq:vlB} only first order terms in $\langle\dot\omega_\lune\rangle/n_\lune$ and $\langle\dot\Omega_\lune\rangle/n_\lune$. With these simplifications, the expression of the centrifugal potential evaluated at clock `$\B$' reduces to
\begin{align}
  \vitlB^2 = n_\lune^2 R_\lune^2\cos^2\csphTB\Bigg\{1&+\left[\left(\frac{\csphRB}{R_\lune}\right)^2-1\right]+\frac{2\langle\dot\omega_\lune\rangle}{n_\lune}\left(\frac{\csphRB}{R_\lune}\right)^2+\frac{2\langle\dot\Omega_\lune\rangle}{n_\lune}\left(\frac{\csphRB}{R_\lune}\right)^2\cos I_\lune\nonumber\\
  &-\frac{2\langle\dot\Omega_\lune\rangle}{n_\lune}\left(\frac{\csphRB}{R_\lune}\right)^2\tan \csphTB\sin I_\lune\sin \big(M_\lune +\omega_\lune +\csphLB\big)\Bigg\} \, .
  \label{eq:vitB2}
\end{align}
This relation, once inserted into equation \eqref{eq:dpptimedtcl}, shows indeed that the main contribution from the centrifugal potential is of the order of \eqref{eq:est_centr_pot} at the lunar equator. This corresponds to a rate of $-20.5 \ \mathrm{ps}/\mathrm{day}$ in the time difference $[\pptime - \tcl]_{(\tclB,\xlB)}$. Let us emphasize that this contribution depends on the selenographic latitude of clock `$\B$' through the trigonometric function $\cos^2\csphTB$. It is thus null at the lunar poles, namely on the axis of proper rotation as one could expect for a centrifugal potential term.

The second term in the expression of the centrifugal potential depends on the radial position of clock `$\B$', it thus changes with the lunar topography. However, as shown in equation \eqref{eq:RtopoRM}, we expect
\begin{equation}
  \mathrm{max}\left\vert\left( \frac{\csphRB}{R_\lune}\right)^2 - 1\right\vert\sim 10^{-2} \, ,
\end{equation}
when $\csphRB = \Rtopo(\csphTB,\csphLB)$ with $\Rtopo$ the radial topography of the Moon [cf. Eq. \eqref{eq:Rtopo}]. This means that the contribution from the topography is two orders of magnitude smaller than the dominant one which scales such as in equation \eqref{eq:est_centr_pot}; it is thus negligible.

Moreover, since terms proportional $\langle\dot\omega_\lune\rangle/n_\lune$ and $\langle\dot\Omega_\lune\rangle/n_\lune$ represent few percent of the dominant one [cf. Eqs. \eqref{eq:doOlune}], all the remaining terms in the centrifugal potential can be neglected. Therefore, at the precision we are working, it is sufficient to model the centrifugal potential at the level of clock `$\B$' such as
\begin{align}
  \vitlB^2 = n_\lune^2 R_\lune^2 \cos^2\csphTB\, .
  \label{eq:vitB2b}
\end{align}

The estimate in equation \eqref{eq:est_centr_pot} shows that the term $\vitl^4/c^4$ in equation \eqref{eq:dpptimedtcl} is completely negligible when evaluated at clock `$\B$' (15 orders of magnitude off requirement).

\subsection{Lunar gravitational potential}

According to \citet{Soffel2003}, the lunar gravitational potential at clock `$\B$' is given by the following expression (neglecting $c^{-2}$ terms):
\begin{align}
  \potM (\tcl,\xlB) & = \frac{Gm_\lune}{\vert \xlB (\tcl) \vert} \Bigg\{ 1 + \sum_{\ell=2}^{+\infty}\sum_{m=0}^{\ell} \left(\frac{R_\lune}{\vert \xlB (\tcl) \vert}\right)^{\!\ell} \nonumber\\
  & \times P_{\ell m} (\sin \csphTB (\tcl)) \left[ C^\lune_{\ell m} (\tcl) \cos (m \csphLB (\tcl)) + S^\lune_{\ell m} (\tcl) \sin (m \csphLB (\tcl)) \right] \Bigg\} \, ,
  \label{eq:potM0}
\end{align}
where $C^\lune_{\ell m} (\tcl)$ and $S^\lune_{\ell m} (\tcl)$ are the lunar Stokes coefficients. The time dependency of Stokes coefficients is due to tides raised on the Moon by bodies in the solar system (mainly Earth and Sun) and to lunar angular velocity variations. The latter will be neglected according to discussion and assumptions in \ref{subsec:potcentrifuge}. However, we consider time variations due to tides raised by body $\ext$ on lunar degree 2 coefficients. Then, neglecting the non-elastic response of the Moon to tidal perturbations, we can write (cf. e.g., \citet{2010ITN....36....1P})
\begin{subequations}
  \begin{align}
    C^\lune_{2 m} (\tcl) & = C^\lune_{2 m} + k_{2}^\lune (2-\delta_{0m}) \frac{(2-m)!}{(2+m)!} \nonumber\\
    & \times \sum_{\ext \neq \lune} \left( \frac{m_\ext}{m_\lune} \right)\left[\frac{R_\lune}{\csphRG(\tcl)}\right]^{3}  P_{2 m} (\sin \csphTG(\tcl)) \cos (m \csphLG(\tcl)) \, , \\
    S^\lune_{2 m} (\tcl) & = S^\lune_{\ell m} + k_{2}^\lune (2-\delta_{0m}) \frac{(2-m)!}{(2+m)!} \nonumber\\
    & \times \sum_{\ext \neq \lune} \left( \frac{m_\ext}{m_\lune} \right) \left[\frac{R_\lune}{\csphRG(\tcl)}\right]^{3} \, P_{2 m} (\sin \csphTG(\tcl)) \sin (m \csphLG(\tcl)) \, ,
  \end{align}
\end{subequations}
where $k_{2}^\lune$ is the degree 2 potential Love number of the Moon, $C^\lune_{2 m}$ and $S^\lune_{2 m}$ (without time dependency) denotes the rigid part of the lunar potential coefficients, and $\xlG(\tcl)=(\csphRG(\tcl),\csphTG(\tcl),\csphLG(\tcl))$ are the selenographic coordinates of tides raising body $\ext$. 

The Moon being highly distorted Earthward, it is convenient to separate in equation~\eqref{eq:potM0} the dominant degree 2 from higher order terms, namely $\ell \geqslant 3$. The lunar gravitational potential thus reads as follows:
\begin{align}
  \potM (\tcl,\xlB) = \frac{Gm_\lune}{R_\lune} \, \Bigg\{ 1 & + \left(\frac{R_\lune}{\csphRB} - 1\right) - \frac{\Delta \csphRB (\tcl)}{R_\lune} \left( \frac{R_\lune}{\csphRB} \right)^{\!2} + \frac{J_2^\lune}{2} \! \left(\frac{R_\lune}{\csphRB}\right)^{\!3} \nonumber\\
  & + \left(\frac{R_\lune}{\csphRB}\right)^{\!3} \left[ 3\left( C_{22}^\lune \cos 2 \csphLB + S_{22}^\lune \sin 2 \csphLB \right) \cos^2\csphTB - \frac{3J_2^\lune}{2} \sin^2 \csphTB \right] \nonumber\\
  & + \sum_{\ell=3}^{+\infty}\sum_{m=0}^{\ell} \left(\frac{R_\lune}{\csphRB}\right)^{\!\ell+1} P_{\ell m} (\sin \csphTB) \left[ C^\lune_{\ell m} \cos (m \csphLB) + S^\lune_{\ell m} \sin (m \csphLB) \right] \nonumber\\
  & + \sum_{\ext \neq \lune} \sum_{m=0}^2 k_{2}^\lune (2-\delta_{0m}) \left( \frac{m_\ext}{m_\lune} \right) \left(\frac{R_\lune}{\csphRB}\right)^{\!3} \left[\frac{R_\lune}{\csphRG(\tcl)}\right]^{3} \frac{(2-m)!}{(2+m)!} \nonumber\\
  &\times P_{2 m} (\sin \csphTB) P_{2 m} (\sin \csphTG(\tcl)) \cos\big[m(\csphLG(\tcl)-\csphLB)\big] \Bigg\} \, ,
  \label{eq:potM}
\end{align}
where we included $\Delta \csphRB$, the effect of the radial displacement of the lunar crust caused by Earth and Sun tides [cf. Eqs. \eqref{eq:DeltacsphRB}]. We neglected longitudinal and latitudinal displacements by considering that $\csphTB$ and $\csphLB$ are constants.

Let us now discuss each contribution in equation \eqref{eq:potM} in turn, namely $\ell=0$ (i.e., the monopole), the topography (i.e., the second term in curly brackets), the tidal displacement of clock `$\B$' (i.e., the third term in curly brackets), $\ell=2$, $\ell\geqslant 3$, and finally the degree 2 tidal perturbations of the lunar gravitational potential.

\subsubsection{Monopole contribution.}

The monopole term in $\potM$ is constant in time, we thus expect it to contribute at the level of
\begin{equation}
  \left(\frac{Gm_\lune}{c^2 R_\lune}\right) = 3.14 \times 10^{-11} \, ,
  \label{eq:potMest}
\end{equation}
after being inserted into equation \eqref{eq:dpptimedtcl}. This corresponds to a rate of about $-2.71 \microsec/\mathrm{day}$  in the time difference $[\pptime - \tcl]_{(\tclB,\xlB)}$. 

As we shall see below, the monopole term in $\potM$ is the dominant contribution in the scalar potential $\poteff$. We conclude that the term $\poteff^2/c^4$ in equation \eqref{eq:dpptimedtcl} is thus negligible (it is in fact 6 orders of magnitude smaller than our $10^{-16}$ limit). Furthermore, considering that $\poteff$ is five orders of magnitude larger than $\vitl^2$ for a clock on the lunar surface, the term $\vitl^2\poteff/c^4$ in equation \eqref{eq:dpptimedtcl} is negligible too, being at the level of $10^{-27}$.

\subsubsection{Altitude contribution.}
\label{sec:topopot}

\begin{figure}[t]
  \begin{center}
    \includegraphics[scale=0.8]{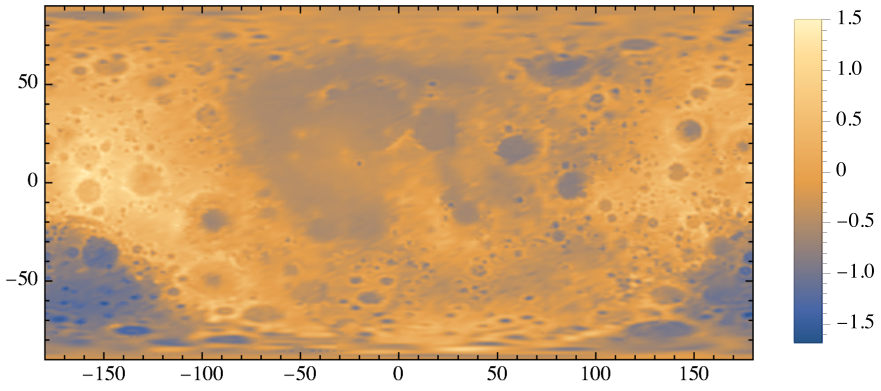}
  \end{center}
  \setlength{\unitlength}{1.0cm}
  \begin{picture}(0,0)(-8,0)
    \put(-2.0,0.25){\rotatebox{0}{longitude $\csphLB$ $[{}^{\circ}]$}}
    \put(-6.8,2.6){\rotatebox{90}{latitude $\csphTB$ $[{}^{\circ}]$}}
    \put(6.0,5.05){\rotatebox{270}{rel. freq. diff.}}
    \put(4.6,6.3){\rotatebox{0}{\footnotesize $[\times 10^{-13}]$}}
  \end{picture}
  \vspace{0.0cm}
  \caption{Map of the contribution to the relative frequency difference in Eq.~\eqref{eq:dpptimedtcl} due to the topography of the Moon. The origin of the longitude represents lunar prime meridian (directed Earthward).}
  \label{fig:lunar_gravity_field_topo}
\end{figure}

Let us now investigate the contribution from the topography of the Moon, namely the second term in equation \eqref{eq:potM}. This term depends on the location of the clock on the Moon through the ratio $R_\lune/\csphRB$. Indeed, for a clock at rest on the lunar surface, we have $\csphRB=\Rtopo(\csphLB,\csphTB)$, where $\Rtopo(\csphLB,\csphTB)$ is given in equation~\eqref{eq:Rtopo}. The contribution of the lunar topography to the relative frequency difference is represented in figure \ref{fig:lunar_gravity_field_topo}. It shows that the clock rate highly depends on its longitude and latitude on the Moon. The amplitude variation reaches $\pm 1.60 \times 10^{-13}$ (i.e., $\pm 13.8\ \mathrm{ns}/\mathrm{day}$) from place-to-place so it cannot be ignored. 

\subsubsection{Tidal displacement contribution.}

Tidal displacement caused by Earth are responsible for (at most) $8.6\ \mathrm{cm}$ amplitude variations of the radial distance of clock `$\B$' [Sun's effects are neglected, see discussion below Eq. \eqref{eq:displest}]. We now want to determine how these impact the time difference $[\pptime - \tcl]_{(\tclB,\xlB)}$ through the lunar potential contribution. Applying $\ext=\terre$ into equation \eqref{eq:DeltacsphRBES} returns 
\begin{align}
  \frac{\Delta \csphRB (\tcl)}{R_\lune} & = h_2^\lune \left( \frac{m_\terre}{m_\lune} \right) \left(\frac{\csphRB}{R_\lune}\right)^{\!4} \left(\frac{R_\lune}{a_\lune}\right)^{\!3} \left(\frac{a_\lune}{\csphR_\terre}\right)^{\!3} \nonumber\\
  & \times \left[P_2\left(\NlB \cdot \Nl_\terre \right) - \frac{1}{2} (3 \cos^2 \csphTB \cos^2 \csphLB - 1 ) \right] \, .
\end{align}
Due to the lunar spin orbit resonance 1:1 and small lunar obliquity, the term in square brackets is proportional to lunar obliquity (and lunar eccentricity) [cf. Eq. \eqref{eq:FourSerLeg2approx}] and is varying with a synodic month period. Therefore, after inserting this into equation \eqref{eq:potM} and then into \eqref{eq:dpptimedtcl}, we infer that the contribution of the tidal displacement to $[\pptime - \tcl]_{(\tclB,\xlB)}$ is a monthly variation with amplitude of the order of
\begin{equation}
  \obqtL h_2^\lune \left(\frac{Gm_\terre}{c^2 n_\lune R_\lune} \right) \left(\frac{R_\lune}{a_\lune}\right)^{\!3} = 3.90 \times 10^{-1} \ \mathrm{ps} \, .
\end{equation}
This oscillating contribution is negligible.

\subsubsection{Degree 2 contribution.}

\begin{figure}[t]
  \begin{center}
    \hspace{0.15cm}
    \includegraphics[scale=0.8]{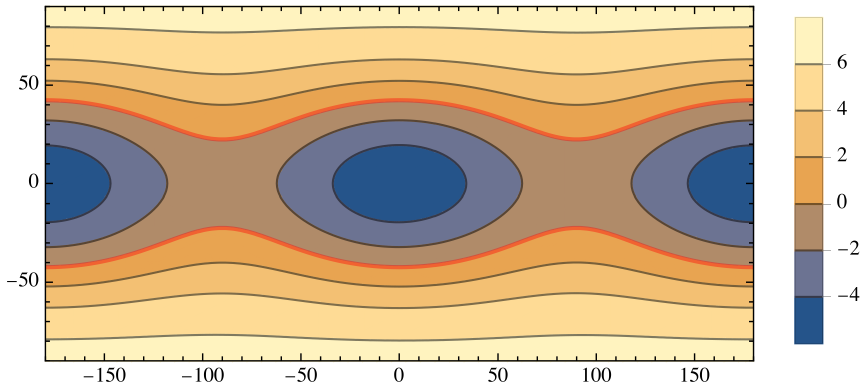}
  \end{center}
  \setlength{\unitlength}{1.0cm}
  \begin{picture}(0,0)(-8,0)
    \put(-1.8,0.25){\rotatebox{0}{longitude $\csphLB$ $[{}^{\circ}]$}}
    \put(-6.6,2.6){\rotatebox{90}{latitude $\csphTB$ $[{}^{\circ}]$}}
    \put(6.3,5.0){\rotatebox{270}{rel. freq. diff.}}
    \put(4.8,6.25){\rotatebox{0}{\footnotesize $[\times 10^{-15}]$}}
    \linethickness{0.4mm}
    \put(4.87,3.43){{\color{red}\line(1,0){0.37}}}
  \end{picture}
  \vspace{0.0cm}
  \caption{Contour map of the contribution to the relative frequency difference in Eq. \eqref{eq:dpptimedtcl} from $\ell = 2$ terms in the lunar potential. The \emph{red contour} corresponds to null frequency rate. The clock is assumed at radial position $\csphRB=\Rtopo(\csphLB,\csphTB)$ [cf. Eq.~\eqref{eq:Rtopo}]. The origin of the longitude represents lunar prime meridian.}
  \label{fig:lunar_gravity_field_2}
\end{figure}

The lunar gravitational field is peculiar because of the Moon spin-orbit resonance 1:1 that makes Earth's tides permanent which elongates the Moon Earthward. This leads the numerical value of $C_{22}^\lune$ to be only one order of magnitude smaller than quadrupole moment $J_2^\lune$ [see numerical values in Eq. \eqref{eq:deg2} below]. The contributions of Moon degree 2 gravitational field to the relative frequency difference is represented in figure \ref{fig:lunar_gravity_field_2}. The maximum amplitude is of the order of $10^{-14}$.

\subsubsection{Degree 3 and beyond contribution.}
\label{sec:potLune3}

Let us now investigate the non-spherical terms $\ell\geqslant 3$ in equation \eqref{eq:potM}. For a clock on the lunar surface, we saw that the monople produces a linear drift of approximately  $3.14 \times 10^{-11}$. This suggests that the double sum over $\ell$ and $m$ in equation \eqref{eq:potM} must reach at least one part in $10^6$  to produce a sensitive effect at the threshold we fixed. Let us, in light of this, determine the appropriate degree $\ellnum$ to which one can stop the summation without loss of precision.

For that purpose, let $\summ_\ell$ be the sum obtained by summing in equation \eqref{eq:potM} each of the non-spherical terms $\ell\geqslant 3$ over $m$ from $m=0$ to $\ell$:
\begin{equation}
  \summ_\ell (\csphTB,\csphLB) = \sum_{m=0}^{\ell} \left(\frac{R_\lune}{\csphRB}\right)^{\!\ell+1} P_{\ell m} (\sin \csphTB) \left[ C^\lune_{\ell m} \cos (m \csphLB) + S^\lune_{\ell m} \sin (m \csphLB) \right] \, ,
\end{equation}
with $\csphRB=\Rtopo(\csphTB,\csphLB)$. Then, let $\suml_{\ellnum}$ be the partial sum obtained by summing $\summ_\ell$'s over $\ell$ from $\ell=3$ to $\ellnum$, that is
\begin{equation}
  \suml_{\ellnum} (\csphTB,\csphLB) = \sum_{\ell=3}^{\ellnum} \summ_\ell (\csphTB,\csphLB) \, .
  \label{eq:sumSlG}
\end{equation}

\begin{figure}[t]
  \begin{center}
    \includegraphics[scale=0.8]{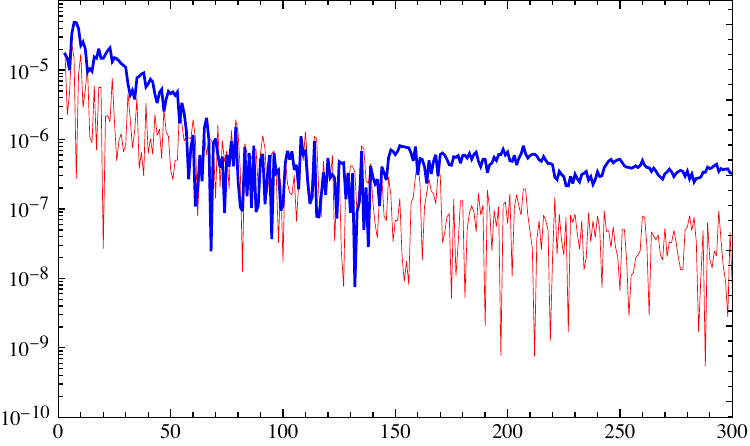}
  \end{center}
  \setlength{\unitlength}{1.0cm}
  \begin{picture}(0,0)(-8,0)
    \put(-0.6,0.3){\rotatebox{0}{degree $\ellnum$}}
    \put(-6.0,2.7){\rotatebox{90}{$\vert\suml_{\ellnum}\vert$ and $\vert\summ_{\ellnum}\vert$}}
  \end{picture}
  \vspace{0.0cm}
  \caption{Evolution of the absolute values of partial sums $\suml_{\ellnum}(0^\circ,0^\circ)$ (\emph{thick blue}) and sum $\summ_{\ellnum}(0^\circ,0^\circ)$ (\emph{light red}) with degree $\ellnum$. Both sums are computed for a clock located at the intersection between the lunar prime meridian and equator, namely $\xlB=(\csphRB,0^\circ,0^\circ)$ with $\csphRB = \Rtopo(0^\circ,0^\circ)$ [cf. Eq. \eqref{eq:Rtopo}].}
  \label{fig:POTM}
\end{figure}

The evolution of $\suml_{\ellnum}(0^\circ,0^\circ)$ and $\summ_{\ellnum}(0^\circ,0^\circ)$ with degree $\ellnum$ is shown in figure \ref{fig:POTM} considering the lunar gravitational potential derived in \href{https://pds-geosciences.wustl.edu/grail/grail-l-lgrs-5-rdr-v1/grail_1001/shadr/gggrx_1200a_sha.tab}{GRGM1200A} modeling (cf. e.g., \citet{Lemoine2014}, \citet{2016LPI....47.1484G}; see also \citet{Konopliv2014}). There, the \emph{red curve} shows that the trend of $\summ_{\ellnum}$'s is a decrease with $\ellnum$ (at least up to $\ellnum=300$), and hence, $\suml_{\ellnum}$ is slowly converging (because we are considering a field point on the lunar surface). Indeed, from the \emph{blue curve} in figure~\ref{fig:POTM}, we infer that $\vert\suml_{\ellnum}(0^\circ,0^\circ) \vert$ reaches convergence around $\ellnum=150$.

\begin{figure}[t]
  \begin{center}
    \includegraphics[scale=0.8]{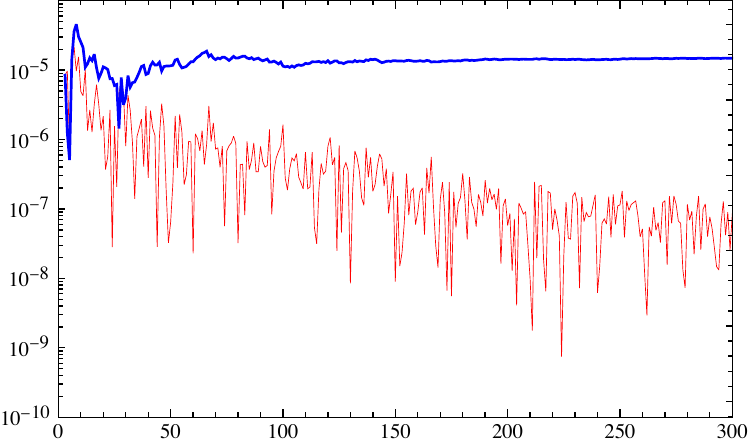}
  \end{center}
  \setlength{\unitlength}{1.0cm}
  \begin{picture}(0,0)(-8,0)
    \put(-0.6,0.3){\rotatebox{0}{degree $\ellnum$}}
    \put(-6.0,2.7){\rotatebox{90}{$\vert\suml_{\ellnum}\vert$ and  $\vert\summ_{\ellnum}\vert$}}
  \end{picture}
  \vspace{0.0cm}
  \caption{Evolution of the absolute values of partial sums $\suml_{\ellnum}(90^\circ,0^\circ)$ (\emph{thick blue}) and sum  $\summ_{\ellnum}(90^\circ,0^\circ)$ (\emph{light red}) with degree $\ellnum$. Both sums are computed for a clock located at the lunar north pole, namely $\xlB=(\csphRB,90^\circ,0^\circ)$ with $\csphRB = \Rtopo(90^\circ,0^\circ)$ [cf. Eq. \eqref{eq:Rtopo}].}
  \label{fig:POTM_loca}
\end{figure}

In addition, let us emphasize that
\begin{equation}
  \vert\suml_{\ellnum>150}(0^\circ,0^\circ)\vert \sim 10^{-7} \, ,
\end{equation}
which, according to our requirement on precision ($\suml_{\ellnum}>10^{-6}$), leads in the computation of $[\pptime - \tcl]_{(\tclB,\xlB)}$, to a negligible contribution from $\ell\geqslant 3$ terms, when clock `$\B$' is at the intersection between the lunar prime meridian and equator. In figure \ref{fig:POTM_loca}, we show the evolution with degree $\ellnum$ of partial sums $\suml_{\ellnum}(90^{\circ},0^\circ)$ and $\summ_{\ellnum}(90^\circ,0^\circ)$, namely for a clock located at the North pole of the Moon. There, the situation is different than in figure~\ref{fig:POTM} for two reasons. Firstly, because convergence is reached (roughly speaking) around $\ellnum=100$ instead of $\ellnum=150$. Secondly, because
\begin{equation}
  \vert\suml_{\ellnum>150}(90^\circ,0^\circ)\vert \sim 10^{-5} \, ,
\end{equation}
which produces a contribution to $[\pptime - \tcl]_{(\tclB,\xlB)}$ which is one order of magnitude above the threshold we fixed. Therefore, for an arbitrary position of clock `$\B$' on the lunar surface, we cannot neglect $\suml_{\ellnum} (\csphTB,\csphLB)$ in general. Furthermore, has shown in figure \ref{fig:POTM} and explained in the text above, we cannot consider only the first tenth degrees, so we have to consider all degrees up to at least $\ellnum = 150$ in order to ensure convergence of the partial summation.

\begin{figure}[t]
  \begin{center}
    \includegraphics[scale=0.8]{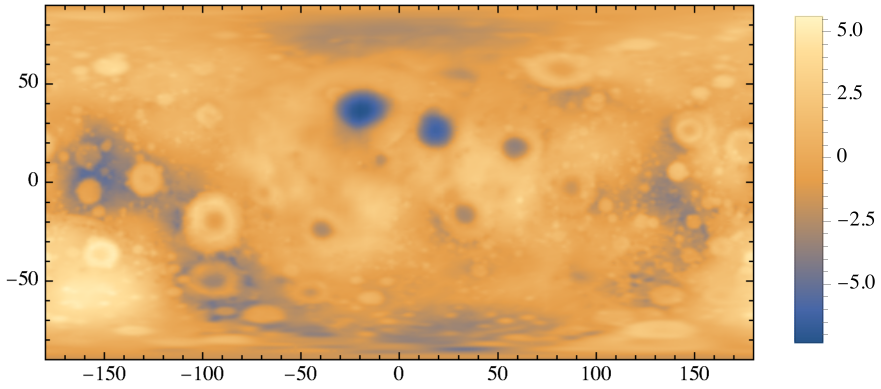}
  \end{center}
  \setlength{\unitlength}{1.0cm}
  \begin{picture}(0,0)(-8,0)
    \put(-2.0,0.25){\rotatebox{0}{longitude $\csphLB$ $[{}^{\circ}]$}}
    \put(-6.9,2.56){\rotatebox{90}{latitude $\csphTB$ $[{}^{\circ}]$}}
    \put(6.2,5.2){\rotatebox{270}{rel. freq. diff.}}
    \put(4.6,6.22){\rotatebox{0}{\footnotesize $[\times 10^{-15}]$}}
  \end{picture}
  \vspace{0.0cm}
  \caption{Map of the contribution to the relative frequency difference in Eq.~\eqref{eq:dpptimedtcl} from the $\ell = \{3,\ldots,150\}$ terms in the lunar potential. The clock is assumed at radial position $\csphRB=\Rtopo(\csphLB,\csphTB)$ [cf. Eq.~\eqref{eq:Rtopo}]. The origin of the longitude represents lunar prime meridian.}
  \label{fig:lunar_gravity_field_3_TO_100}
\end{figure}

Hereafter, we will keep the sum over $\ell$ to go to $\ellnum$ within formal expressions and consider summation up to $\ellnum=150$ for the next-coming numerical application. The contribution of $\ell = \{3,\ldots,150\}$ terms to the relative frequency difference is shown in figure \ref{fig:lunar_gravity_field_3_TO_100}. It is approximately the same amplitude as the $\ell = 2$ terms discussed earlier.

\subsubsection{Tidal contribution.}

As mentioned earlier, the selenographic coordinates of the Earth are almost constant in time due to spin orbit resonance 1:1 and small lunar obliquity. Hence, the tidal perturbation by Earth, through the variation of the lunar gravitational potential coefficients, will induce a constant contribution to the relative frequency difference at the level of
\begin{equation}
  k_{2}^\lune \left(\frac{Gm_\terre}{c^2 R_\lune}\right) \left(\frac{R_\lune}{a_\lune}\right)^{\!3} = 5.63 \times 10^{-18} \, ,
\end{equation}
once plugged into equation \eqref{eq:dpptimedtcl}; the numerical value of $k_{2}^\lune$ is $0.024$ \citep{Folkner2014}. This contribution  can therefore be neglected in view of our $10^{-16}$ threshold. Concerning the Sun, its selenographic coordinates are not constant as for the case of the Earth. Therefore, when plugging the tidal perturbation by Sun into equation \eqref{eq:dpptimedtcl}, oscillating signature is expected in the relative frequency difference with an amplitude of the order of
\begin{equation}
  k_{2}^\lune \left(\frac{Gm_\sun}{c^2 n_\lune R_\lune}\right) \left(\frac{R_\lune}{a_\terre}\right)^{\!3} = 1.19 \times 10^{-2} \ \mathrm{ps} \, ,
\end{equation}
and a period which is equal to the synodic period of the Moon $(n_\lune-n_\terre)$, that we approximated to $n_\lune$. This oscillating contribution being below the picosecond, it can safely be neglected. Therefore, both Earth and Sun tidal distortions of the lunar potential can be omitted.

\subsubsection{Summary.}

To sufficient approximation, the expression of the contribution of the total lunar gravitational potential to $[\pptime - \tcl]_{(\tclB,\xlB)}$ is eventually given by
\begin{align}
  -\frac{1}{c^2} \int_{\tcl_0}^{\tclB} \dd\tcl & \, \potM(\tcl,\xlB) = - \left( \frac{Gm_\lune}{c^2R_\lune} \right) \Bigg\{ 1 + \left(\frac{R_\lune}{\csphRB} - 1\right) + \frac{J_2^\lune}{2} \nonumber\\
  & + 3 \left( C_{22}^\lune \cos 2 \csphLB + S_{22}^\lune \sin 2 \csphLB \right) \cos^2\csphTB - \frac{3J_2^\lune}{2} \sin^2 \csphTB \nonumber\\
  & + \sum_{\ell=3}^{\ellnum}\sum_{m=0}^{\ell} P_{\ell m} (\sin \csphTB) \left[ C^\lune_{\ell m} \cos (m \csphLB) + S^\lune_{\ell m} \sin (m \csphLB) \right] \Bigg\} (\tclB-\tcl_0) \, ,
\end{align}
with $\ellnum=150$. In this expression, we omitted the topography contribution in terms proportional to the non-spherical part of the lunar gravitational potential.

\subsection{Tidal potential}
\label{sec:tides}

In this section and the next one (cf. \ref{sec:neglpot}), we adopt the Einstein summation convention on repeated indices whatever the position of the repeated indices. When dealing with sequences of spatial indices (latin indices), we shall adopt the following notation: a spatial multi-index containing $j$ indices is simply denoted $J$ (always a capital-letter), namely $J\equiv i_1 \ldots i_j$. We also make use of Symmetric and TraceFree tensors (denoted STF tensors). STF tensors are denoted via the use of angular
brackets around capital-letter, such as $n^{\langle J \rangle} \equiv n^{\langle i_1 \ldots i_j \rangle}$. The reader is refereed to \citet{1994CeMDA..60..139H} or \citet{Poisson2014} for a detailed description of the definition and the use of STF tensors in celestial mechanics or in general relativity, respectively. 

According to \citet{Soffel2003}, at Newtonian order, the tidal potential is given by the following relationship:
\begin{equation}
  \pottide(\tcl,\xl) = \potMnot (\tcb,\xb_\lune+\xl) - \potMnot (\tcb,\xb_\lune) - \delta_{ai}\cvect{\posl}{a}\partial_i \potMnot (\tcb,\xb_\lune) \, ,
  \label{eq:pottideSoffel}
\end{equation}
where the arbitrary function of time has been fixed to $\potMnot (\tcb,\xb_\lune)$ for later convenience. In this expression and below, $\xb_\ext$, the BCRS position of a body $\ext$, must be understood as its BCRS trajectory, namely $\xb_\ext=\xb_\ext(\tcb)$. In equation \eqref{eq:pottideSoffel}, because $\pottide$ is already multiplied by $c^{-2}$ in equation \eqref{eq:dpptimedtcl}, the 4D relations which are used to pass from BCRS space-time coordinates to LCRS's, are limited to the Newtonian terms only, that is
\begin{equation}
  \tcb = \tcl \, , \qquad \xb = \xb_\lune + \xl \, .
  \label{eq:pass_BCRS_LCRS}
\end{equation}

The (Newtonian) external gravitational potential to the Moon is given by
\begin{equation}
  \potMnot (\tcb,\xb) = G \sum_{\ext \neq \lune} \int_{\ext} \frac{\rho(\tcb,\xb_\ext+\xg')}{\vert \xb-\xb_\ext-\xg'\vert} \, \dd^3 \posg' \, ,
\end{equation}
with $\xg'$ the position vector of the source point taken from the center of mass of body $\ext$. Typically, $\xg'$ is spanning a volume enclosing body $\ext$. When the external potential is evaluated at the lunar center of mass, namely $\xb=\xb_\lune$, the distance $\vert\xb_\lune-\xb_\ext\vert$ (with $\ext \neq \lune$) is much larger than $\vert\xg'\vert$, it is thus appropriate to expand the denominator in a Taylor series. Hence, the expression of the Moon external potential reduces to
\begin{equation}
  \potMnot (\tcb,\xb) = G \sum_{\ext \neq \lune} \sum_{j=0}^{+\infty} \frac{(-1)^{j}}{j!} I_{\ext}^{\langle J \rangle} (\tcb) \, \partial_{J} \vert \xb-\xb_\ext \vert^{-1} \, ,
  \label{eq:potMnot}
\end{equation}
where $I_{\ext}^{\langle J \rangle}$ denotes the (Newtonian) STF mass multipole moment of body $\ext$:
\begin{equation}
  I_\ext^{\langle J\rangle} (\tcb) = \int_\ext \rho(\tcb,\xb) (\xb-\xb_\ext)^{\langle J \rangle} \, \dd^3 \posb \, .
\end{equation}

Expanding, in equation \eqref{eq:pottideSoffel}, the external potential $\potMnot (\tcb,\xb_\lune+\xl)$ in Taylor series around $\xl = 0$, and then inserting equation \eqref{eq:potMnot} evaluated at $\potMnot(t,\xb_\lune)$, leads to the following expression:
\begin{equation}
  \pottide(\tcl,\xl) = \pottideS(\tcl,\xl) + \pottideNS(\tcl,\xl) \, ,
\end{equation}
where the tides by the monopole and non-spherical components of body $\ext$ gravitational fields are given by
\begin{subequations}\label{eq:pottidal2}
  \begin{empheq}[left=\empheqlbrace]{align}
    \pottideS(\tcl,\xl) & = G \sum_{\ext \neq \lune} m_{\ext} \sum_{j=2}^{+\infty} \frac{(-1)^j}{j!} \cvect{\posl}{\langle J \rangle} \partial_{J} \csphR_{\ext}^{-1}(\tcl) \, ,\label{eq:pottidal2S}\\
    \pottideNS(\tcl,\xl) & = G \sum_{\ext \neq \lune} \sum_{j=2}^{+\infty} \sum_{k=2}^{+\infty} \frac{(-1)^{j}}{j!k!} \cvect{\posl}{\langle J \rangle} I_{\ext}^{\langle K \rangle} (\tcl) \, \partial_{JK} \csphR_{\ext}^{-1}(\tcl) \, ,\label{eq:pottidal2NS}
  \end{empheq}
\end{subequations}
respectively. In these expressions, $\csphR_{\ext} = \vert \xlG \vert = \vert \xb_\ext - \xb_\lune \vert$ and $\partial_{JK}$ is the $(j+k)$-th partial derivative relative to components of $\xb_\ext$, namely
\begin{equation}
  \partial_{JK} \equiv \frac{\partial^{j+k}}{\partial\cvect{\posb_\ext}{i_1}\ldots\partial\cvect{\posb_\ext}{i_j}\partial\cvect{\posb_\ext}{\ell_1}\ldots\partial\cvect{\posb_\ext}{\ell_{k}}} \, .
\end{equation}
We thus infer
\begin{equation}
  \partial_{J} \csphR_{\ext}^{-1} = (-1)^{j} (2j-1)!!\,\cfrac{(\dirlG)_{\langle J\rangle}}{\csphRG^{j+1}} \, ,
\end{equation}
where $\NlG=\xlG/\csphRG$.

To derive estimates of the non-spherical components in equation \eqref{eq:pottidal2NS}, let us assume that solar system bodies are axi-symmetric about their polar axis $\epol_\ext$ (with $\epol_\ext \cdot \epol_\ext = 1$). According to \citet{Poisson2014}, the mass multipole moment of body $\ext$ is thus given by the simple relation:
\begin{equation}
  I_\ext^{\langle J\rangle} = - m_\ext R_\ext^j J_j^\ext \polax_\ext^{\langle J\rangle} \, ,
\end{equation}
for $j \geqslant 2$. In this expression, $J_j^\ext$ is the body $\ext$ zonal harmonic coefficient of degree $j$ and $R_\ext$ is its radius. After inserting it in equations \eqref{eq:pottidal2} and making use of the following STF identities (cf. e.g., \citet{1994CeMDA..60..139H} and \citet{Poisson2014}):
\begin{subequations}
  \begin{align}
    \dirl^{\langle J\rangle}(\dirlG)_{\langle J\rangle} & = \frac{j!}{(2j-1)!!} P_{j}(\Nl\cdot\NlG) \, , \\
    \dirl^J\polax_\ext^K(\dirlG)_{\langle JK\rangle} & = \frac{(j+k)!}{(2j+2k-1)!!} \frac{P_{j+k} (\NlG\cdot\epol_\ext)}{P_{j} (\NlG\cdot\epol_\ext)^2-1} \nonumber\\
    & \times \Big\{\big[ P_{j} (\NlG\cdot\epol_\ext)P_{j+k} (\NlG\cdot\epol_\ext)-P_{k} (\NlG\cdot\epol_\ext) \big] P_j (\Nl\cdot\NlG) \nonumber \\
    & + \big[ P_{j} (\NlG\cdot\epol_\ext)P_{k} (\NlG\cdot\epol_\ext)-P_{j+k} (\NlG\cdot\epol_\ext) \big] P_j (\Nl\cdot\epol_\ext) \Big\} \, ,
  \end{align}
\end{subequations}
we eventually obtain
\begin{subequations}\label{eq:pottideint}
  \begin{align}
    \pottideS(\tcl,\xl) & = \sum_{\ext \neq \lune} \frac{Gm_\ext}{\csphRG} \sum_{j=2}^{+\infty} \left(\frac{\csphR}{\csphRG}\right)^{\!j}  P_{j}(\Nl\cdot\NlG) \, , \label{eq:pottideintS}\\
    \pottideNS(\tcl,\xl) & = \sum_{\ext \neq \lune} \frac{Gm_\ext}{\csphRG} \sum_{j=2}^{+\infty} \sum_{k=2}^{+\infty} (-1)^{k+1} \frac{(j+k)!}{j!k!} \, J_{k}^\ext \left(\frac{\csphR}{\csphRG}\right)^{\!j}\left(\frac{R_\ext}{\csphRG}\right)^{\!k} \frac{P_{j+k} (\NlG\cdot\epol_\ext)}{P_{j} (\NlG\cdot\epol_\ext)^2-1} \nonumber\\
    & \times \Big\{\big[ P_{j} (\NlG\cdot\epol_\ext)P_{j+k} (\NlG\cdot\epol_\ext)-P_{k} (\NlG\cdot\epol_\ext) \big] P_j (\Nl\cdot\NlG) \nonumber \\
    & + \big[ P_{j} (\NlG\cdot\epol_\ext)P_{k} (\NlG\cdot\epol_\ext)-P_{j+k} (\NlG\cdot\epol_\ext) \big] P_j (\Nl\cdot\epol_\ext) \Big\} \, ,\label{eq:pottideintNS}
  \end{align}
\end{subequations}
where $\csphR = \vert \xl \vert$ and $\Nl=\xl/\csphR$. In these expressions, $P_j$ denotes the Legendre polynomial of degree $j$.

Now let us apply equation \eqref{eq:pottideintNS} at the position of clock `$\B$' considering the main term only (i.e., $j=2$ and $k=2$). We thus find that the tides raised by the non-spherical component of body $\ext$ gravitational field at the level of clock `$\B$' scales such as
\begin{equation}
  \frac{1}{c^2}\pottideNS(\tcl,\xlB) \propto -\sum_{\ext \neq \lune} J_{2}^\ext \left(\frac{Gm_\ext}{c^2\csphRG}\right) \left(\frac{\csphRB}{\csphRG}\right)^{\!2}\left(\frac{R_\ext}{\csphRG}\right)^{\!2} \, .
\end{equation}
For tides raised by the Earth's and Sun's quadrupole moments, we find
\begin{equation}
  J_{2}^\terre \left(\frac{Gm_\terre}{c^2\csphR_\terre}\right) \left(\frac{R_\lune}{\csphR_\terre}\right)^{\!2}\left(\frac{R_\terre}{\csphR_\terre}\right)^{\!2} \sim 10^{-23} \, , \quad  J_{2}^\sun \left(\frac{Gm_\sun}{c^2\csphR_\sun}\right) \left(\frac{R_\lune}{\csphR_\sun}\right)^{\!2}\left(\frac{R_\sun}{\csphR_\sun}\right)^{\!2} \sim 10^{-30} \, ,
\end{equation}
considering numerical values: $R_\terre = 6\,378 \, \mathrm{km}$ and $R_\sun = 6.96\times10^5 \, \mathrm{km}$ for radii, and $J_{2}^\terre=1.08\times10^{-3}$ and $J_{2}^\sun=2.11\times10^{-7}$ for quadrupoles (see e.g., \citet{Folkner2014}). We conclude that $\pottideNS$ is completely negligible.

Let us now focus on the monopole component from external bodies, namely $\pottideS$. According to equation \eqref{eq:pottideintS} this term scales such as
\begin{equation}
  \frac{1}{c^2}\pottideS(\tcl,\xlB) \propto \sum_{\ext \neq \lune} \sum_{j=2}^{+\infty} \left(\frac{Gm_\ext}{c^2\csphRG} \right) \left(\frac{\csphRB}{\csphRG}\right)^{\!j} \, ,
\end{equation}
at the level of clock `$\B$'. For Earth, the quadrupole and hexapole terms (i.e., $j=2$ and $j=3$) are of the order of
\begin{equation}
  \left(\frac{Gm_\terre}{c^2\csphR_\terre} \right) \left(\frac{R_\lune}{\csphR_\terre}\right)^{\!2} = 2.37 \times 10^{-16} \, , \qquad \left(\frac{Gm_\terre}{c^2\csphR_\terre} \right) \left(\frac{R_\lune}{\csphR_\terre}\right)^{\!3} = 1.07 \times 10^{-18} \, ,
\end{equation}
while for Sun, the quadrupole term scales such as
\begin{equation}
  \left(\frac{Gm_\sun}{c^2\csphR_\sun} \right) \left(\frac{R_\lune}{\csphR_\sun}\right)^{\!2} = 1.32 \times 10^{-18} \, .
\end{equation}

Therefore, for what concerns us only the Earth's quadrupole tidal potential shall be kept, so that to sufficient accuracy, the tidal potential eventually reads as
\begin{equation}
  \pottide(\tcl,\xlB) = \frac{Gm_\terre}{\csphR_\terre} \left(\frac{R_\lune}{\csphR_\terre}\right)^{\!2} \left( \frac{\csphRB}{R_\lune} \right)^{\!2} P_{2}(\NlB\cdot\NlE) \, .
  \label{eq:pottide3}
\end{equation}
Within the lunar principal axis frame, $\NlE$ is given by
\begin{align}
  \NlE = & +\vect{e}_A \big[ \cos \obqtL \sin(L_\lune-\Omega_\lune) \sin(f_\lune+\omega_\lune)+\cos(L_\lune-\Omega_\lune)\cos(f_\lune+\omega_\lune)\big]\nonumber\\
  &+\vect{e}_B \big[ \cos\obqtL\cos(L_\lune-\Omega_\lune) \sin(f_\lune+\omega_\lune)-\sin(L_\lune-\Omega_\lune)\cos(f_\lune+\omega_\lune)\big] \nonumber\\
  &-\vect{e}_C\sin\obqtL\sin(f_\lune+\omega_\lune)\, ,
\end{align}
where $f_\lune$ is the true anomaly of the lunar orbit and where $\obqtL$ is the lunar obliquity. Given that the Moon orbit is tilted by $\iota_\lune = 0.089\ \mathrm{rad}$ relative to the ecliptic and considering, according to Cassini's laws, that the Moon equator is inclined by $I_\lune = 0.026\ \mathrm{rad}$ with respect to the ecliptic, we infer that $\obqtL = \iota_\lune + I_\lune$, namely $\obqtL = 0.12 \ \mathrm{rad}$.

It is convenient to develop the Legendre polynomial in equation \eqref{eq:pottide3} in a Fourier series of $M_\lune$, the mean longitude of the Moon. This is achieved after expressing the lunar true anomaly in term of the mean anomaly with the following series expansion
\begin{subequations}
  \begin{empheq}[left=\empheqlbrace]{align}
    \cos f_\lune & = -e_\lune + \frac{2(1-e_\lune^2)}{e_\lune} \sum_{s=1}^{+\infty} \mathrm{J}_s(s e_\lune) \cos (s M_\lune) \, , \\
    \sin f_\lune & = (1-e_\lune^2)^{-1/2} \sum_{s=1}^{+\infty} [ \mathrm{J}_{s-1}(s e_\lune) - \mathrm{J}_{s+1}(s e_\lune) ] \sin (s M_\lune) \, ,
  \end{empheq}
\end{subequations}
where $\mathrm{J}_s(x)$ denotes the Bessel function of first kind with $s$-index and $e_\lune$ refers to the eccentricity of the lunar orbit. Only first order in $e_\lune$ is considered in the following discussion since $e_\lune = 0.055$. In addition, we already mentioned that the lunar obliquity is a small angle so only first order in $\obqtL$ is also considered. Couplings terms of order $\obqtL e_\lune$ are neglected for the rest of the discussion. With these simplifications, we find the following expression:
\begin{align}
  \left( \frac{a_\lune}{\csphR_\terre} \right)^{\!3} P_{2}(\NlB\cdot\NlE) & = \frac{1}{2} (3\cos^2\csphTB\cos^2\csphLB-1) (1 + 3 e_\lune \cos M_\lune ) \nonumber \\
  & -\frac{3\obqtL}{2} \sin 2\csphTB \cos\csphLB \sin (M_\lune + \omega_\lune) + 3 e_\lune \cos^2 \csphTB \sin 2\csphLB \sin M_\lune \, .
  \label{eq:FourSerLeg2}
\end{align}
After plugging it into the expression of the tidal potential and then integrating with respect to time, one gets (i) a term linear in time and (ii) periodic variations (at the lunar orbital frequency about the Earth) in the expression of the time difference $[\pptime - \tcl]_{(\tclB,\xlB)}$. The term linear in time scales such as
\begin{equation}
  \left(\frac{Gm_\terre}{c^2a_\lune} \right) \left(\frac{R_\lune}{a_\lune}\right)^{\!2} = 2.37 \times 10^{-16} \, , 
\end{equation}
which represents $7.48 \ \mathrm{ns}/\mathrm{yr}$ and must thus be considered in the modeling. The periodic variations have amplitudes scaling such as
\begin{equation}
  \obqtL \left( \frac{G m_\terre}{c^2 n_\lune a_\lune} \right) \left( \frac{R_\lune}{a_\lune} \right)^{\!2} = 10.7 \ \mathrm{ps} \, , \qquad e_\lune \left( \frac{G m_\terre}{c^2 n_\lune a_\lune} \right) \left( \frac{R_\lune}{a_\lune} \right)^{\!2} = 4.90 \ \mathrm{ps} \, 
\end{equation}
and can then be neglected.

\begin{figure}[t]
  \begin{center}
    \hspace{0.15cm}
    \includegraphics[scale=0.8]{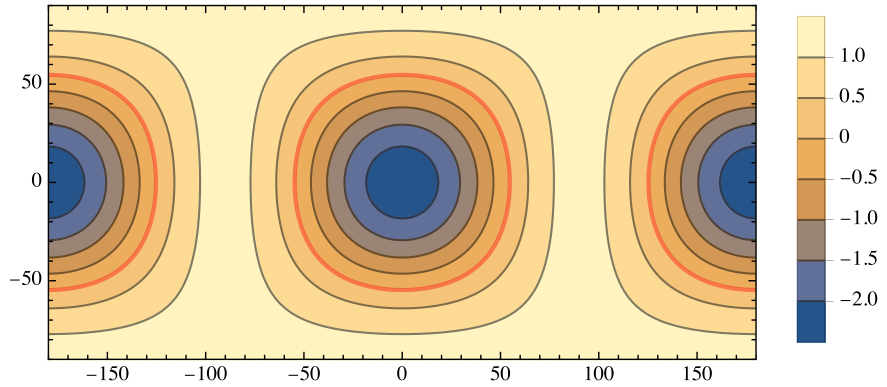}
  \end{center}
  \setlength{\unitlength}{1.0cm}
  \begin{picture}(0,0)(-8,0)
    \put(-1.85,0.25){\rotatebox{0}{longitude $\csphLB$ $[{}^{\circ}]$}}
    \put(-6.7,2.55){\rotatebox{90}{latitude $\csphTB$ $[{}^{\circ}]$}}
    \put(6.4,5.0){\rotatebox{270}{rel. freq. diff.}}
    \put(4.72,6.2){\rotatebox{0}{\footnotesize $[\times 10^{-16}]$}}
    \linethickness{0.4mm}
    \put(4.78,4.24){{\color{red}\line(1,0){0.37}}}
  \end{picture}
  \vspace{0.0cm}
  \caption{Contour map of the contribution to the relative frequency difference in Eq. \eqref{eq:dpptimedtcl} from the permanent tides raised on the Moon by the Earth. The \emph{red contour} corresponds to null time rate. The origin of the longitude represents lunar prime meridian.}
  \label{fig:lunar_tides}
\end{figure}

Therefore, to sufficient accuracy, we have the following expression for the tidal potential:
\begin{equation}
  - \frac{1}{c^2} \int_{\tcl_0}^{\tclB} \dd \tcl' \, \pottide(\tcl',\xlB) = -\frac{1}{2} \left( \frac{G m_\terre}{c^2 a_\lune} \right) \left( \frac{R_\lune}{a_\lune} \right)^{\!2} (3\cos^2\csphTB\cos^2\csphLB-1) (\tclB-\tcl_0) \, .
\end{equation}
It is represented in figure \ref{fig:lunar_tides}. It clearly shows two bulges separated by $180^\circ$ in longitude, with a maximum effect with an amplitude around $-2.5 \times 10^{-16}$ at locations $(\csphTB,\csphLB) = (0^\circ,0^\circ)$ and $(0^\circ,180^\circ)$ .

\subsection{Negligible potentials}
\label{sec:neglpot}

All the terms in equations \eqref{eq:Wscal} and \eqref{eq:Wvect} that have not been discussed yet are actually negligible when relative time keeping is required at the level of $10^{-16}$ in a lunar environment. This is what we intent to demonstrate in this section.

We have already shown that some of the terms proportional to $c^{-4}$ are expected to not contribute in equation~\eqref{eq:dpptimedtcl} when the observer is at rest on the surface of the Moon, and hence, the relative frequency difference shall reduce to
\begin{align}
  \frac{\dd \pptime_\B - \dd \tcl}{\dd \tcl} & = - \frac{1}{2c^2} \left[ \vitlB^2 + 2\poteff(\tcl,\xlB) \right] + \frac{4}{c^4} \vlB \cdot \vect{\poteff}(\tcl,\xlB) \, .
  \label{eq:dpptimedtclsimp}
\end{align}
Let us recall that $\xlB$ and $\vlB$ refer to the trajectory of clock `$\B$', namely $\xlB = \xlB (\tcl)$ and $\vlB = \vlB (\tcl)$.

\subsubsection{Scalar inertial potential.}

Among the $\mathcal{O}(c^{-2})$ terms, is the scalar inertial potential $\potiner$. According to \citet{Soffel2003} it reads such as
\begin{equation}
  \potiner (\tcl,\xlB) = \nongeoacc (\tcl) \cdot \xlB \, ,
\end{equation}
where $\nongeoacc$ represents the non-geodesic acceleration of the Moon. The $a$-th component of $\nongeoacc$ is given, at Newtonian order, by
\begin{equation}
    \cnongeoacc{a} (\tcl) = \delta_{ai} \left[ \partial_i \, \potMnot (\tcb,\xb_\lune) - a_\lune^i (t) \right] \, ;
\end{equation}
note that in this expression, the 4D relations which are used to pass from BCRS
space-time coordinates to LCRS’s, are limited to the Newtonian terms only [cf. Eqs. \eqref{eq:pass_BCRS_LCRS}]. The BCRS acceleration of the Moon is
\begin{equation}
    a_\lune^i (t) = \sum_{j = 0}^{+\infty} \frac{1}{j !} \frac{I_\lune^{\langle J \rangle} (t)}{m_\lune} \, \partial_{iJ} \, \potMnot (\tcb,\xb_\lune) \, .
\end{equation}
Therefore, after invoking \eqref{eq:potMnot}, we deduce the following expression
\begin{align}
  \cnongeoacc{a} (\tcl) & = G\sum_{\ext \neq \lune} \frac{m_\ext}{m_\lune} \sum_{j=2}^{+\infty} \frac{1}{j!} I_\lune^{\langle J \rangle} (\tcl) \, \partial_{aJ} \csphRG^{-1}(\tcl) \nonumber\\
  & + G\sum_{\ext \neq \lune} \sum_{j=2}^{+\infty}\sum_{k=2}^{+\infty} \frac{(-1)^k}{j! k!} \frac{I_\lune^{\langle J \rangle} (\tcl) \, I_\ext^{\langle K \rangle} (\tcl) }{m_\lune} \partial_{aJK} \csphRG^{-1}(\tcl) \, .
  \label{eq:nongeoaccM}
\end{align}
The dominant term, is the first one in the right-hand side. It corresponds to the couplings of the Moon non-spherical potential to the monopole of external bodies. The second term represents the couplings between non-spherical potential of the Moon and non-spherical potential of external bodies; it is expected to be smaller than the first one. Let us thus focus on the first term considering the main effect which is due to the quadrupole moment of the Moon [i.e, $j=2$ in Eq. \eqref{eq:nongeoaccM}]. In this condition, the non-geodesic acceleration of the Moon scales such as
\begin{equation}
  \vert\nongeoacc\vert \propto J_2^\lune\sum_{\ext \neq \lune} \left(\frac{Gm_\ext}{\csphRG^2} \right) \left(\frac{R_\lune}{\csphRG}\right)^{\!2} \, .
\end{equation}
The Earth (resp. the Sun) contribution is of the order of $1.12\times10^{-11}\ \mathrm{m}\,\mathrm{s}^{-2}$ (resp. $1.61\times10^{-16}\ \mathrm{m}\,\mathrm{s}^{-2}$). It is interesting to note that the non-geodesic acceleration of the Moon is thus the same order of magnitude as the Earth's (see discussion in Sect. \ref{subsec:GCRS}). 
Concerning the non-inertial potential, it scales such as
\begin{equation}
  \frac{1}{c^2}\potiner (\tcl,\xlB) \propto J_2^\lune\sum_{\ext \neq \lune} \left(\frac{Gm_\ext}{c^2\csphRG} \right)\left(\frac{\csphRB}{R_\lune}\right) \left(\frac{R_\lune}{\csphRG}\right)^{\!3} \, ,
\end{equation}
when evaluated at the position of the clock on the Moon. The Earth and Sun contributions are of the order of
\begin{equation}
   J_2^\lune \left(\frac{Gm_\terre}{c^2\csphR_\terre} \right) \left(\frac{R_\lune}{\csphR_\terre}\right)^{\!3} = 2.18 \times 10^{-22}\, , \qquad J_2^\lune \left(\frac{Gm_\sun}{c^2\csphR_\sun} \right) \left(\frac{R_\lune}{\csphR_\sun}\right)^{\!3} = 3.11 \times 10^{-35} \, ;
\end{equation}
both are negligible.

\subsubsection{Vector potential.}

Let us now discuss the vector potential contribution in equation \eqref{eq:dpptimedtclsimp}, where $\vect{\poteff}$ is given by equation  \eqref{eq:Wvect}. The vector potential is a combination of three terms whose amplitudes are discussed in turn hereafter. The vector lunar gravitational potential $\vpotM$ is given by
\begin{equation}
  \vpotM(\tcl,\xlB) = \frac{G}{2} \frac{\vspinM(\tcl) \times \xlB}{\csphRB^3} \, ,
\end{equation}
where $\vspinM$ is the spin of Moon, that we approximate by
\begin{equation}
  \vspinM (\tcl) = \frac{2}{5} \, m_\lune R_\lune^2 \, \angvitCass (\tcl)\, .
\end{equation}
By making use of equation \eqref{eq:angvitCass}, we end up with
\begin{equation}
  \frac{4}{c^4}\,\vlB \cdot \vpotM(\tcl,\xlB) = \frac{4}{5}\left(\frac{n_\lune R_\lune}{c}\right)^{\!2} \left(\frac{Gm_\lune}{c^2R_\lune}\right) \left( \frac{R_\lune}{\csphRB} \right) \cos^2 \csphTB \, ,
\end{equation}
which scales such as
\begin{equation}
  \left(\frac{n_\lune R_\lune}{c}\right)^{\!2} \left(\frac{Gm_\lune}{c^2R_\lune}\right) = 7.47 \times 10^{-27}\, ,
\end{equation}
and can be neglected.

Then, let us discuss the tidal vector potential. The dominant part can be obtained from \citet{Damour1991,Damour1992}; we infer the following estimate
\begin{equation}
  \frac{1}{c^4} \, \vlB \cdot\vpottide(\tcl,\xlB) \propto \sum_{\ext\neq\lune} \left( \frac{G m_\ext}{c^2\csphRG} \right) \left(\frac{\vitlG \vitlB}{c^2} \right) \left(\frac{R_\lune}{\csphRG}\right)^{\!2} \left(\frac{\csphRB}{R_\lune}\right)^{\!2} \NlB \, ,
\end{equation}
where $\vitlG$ is the norm of the LCRS velocity of body $\ext$: at the Newtonian level, it is given by $\vitlG = \vert \vb_\ext - \vb_\lune \vert$, where $\vb_\ext$ and $\vb_\lune$ are the BCRS velocities of body $\ext$ and Moon, respectively. For the Earth contribution to the vector tidal potential, we find
\begin{equation}
  \left( \frac{G m_\terre}{c^2\csphR_\terre} \right) \left(\frac{\vitl_\terre \vitlB}{c^2} \right) \left(\frac{R_\lune}{\csphR_\terre}\right)^{\!2} = 1.21 \times 10^{-29} \, ,
\end{equation}
and for the Sun's
\begin{equation}
  \left( \frac{G m_\sun}{c^2\csphR_\sun} \right) \left(\frac{\vitl_\sun \vitlB}{c^2} \right) \left(\frac{R_\lune}{\csphR_\sun}\right)^{\!2} = 1.97 \times 10^{-30} \, ;
\end{equation}
both contributions are negligible in view of our $10^{-16}$ threshold.

Now we examine the effect of the inertial vector potential $\vpotiner$. According to \citet{Soffel2003} it is given by
\begin{equation}
  \vpotiner (\tcl , \xlB) = - \frac{c^2}{4} \, \angvitiner (\tcl) \times \xlB \, ,
\end{equation}
which shows from equation \eqref{eq:dpptimedtclsimp}, that it contributes to the relative frequency difference as follows:
\begin{equation}
  \frac{4}{c^4}\,\vlB \cdot \vpotiner (\tcl , \xlB) = - \frac{1}{c^2} \, \angvitiner (\tcl) \cdot ( \xlB \times \vlB ) \, .
  \label{eq:estimatinert}
\end{equation}
The angular velocity $\angvitiner$ is a combination of the geodetic, Lense-Thirring, and Thomas precessions:
\begin{equation}
  \angvitiner = \angvitGP + \angvitLTP + \angvitTP \, .
\end{equation}
At leading order, these are given by
\begin{subequations}
  \begin{empheq}[left=\empheqlbrace]{align}
  \angvitGP & = -\frac{1}{2c^2} \sum_{\ext \neq \lune} \left( \frac{Gm_\ext}{ \csphRG^2} \right) \NlG \times (3 \vb_\lune - 4 \vb_\ext) \, ,\\
  \angvitLTP & = - \frac{1}{c^2} \sum_{\ext \neq \lune} \left(\frac{G \spinG}{\csphRG^3}\right) \Big[ 3 ( \NlG \cdot \epol_\ext ) \NlG - \epol_\ext \Big] \, , \\
  \angvitTP & = - \frac{3 J_2^\lune}{4c^2} \sum_{\ext \neq \lune} \left( \frac{Gm_\ext}{\csphRG^2} \right) \left( \frac{R_\lune}{\csphRG} \right)^{\!2} \frac{P_3(\NlG \cdot \epol_\lune)}{(\NlG \cdot \epol_\lune)^2-1} \nonumber\\
  & \times \Big\{ \big[ (\NlG \cdot \epol_\lune) P_3(\NlG \cdot \epol_\lune) - P_2(\NlG \cdot \epol_\lune) \big] \NlG \nonumber\\
  & + \big[ (\NlG \cdot \epol_\lune) P_2(\NlG \cdot \epol_\lune) - P_3(\NlG \cdot \epol_\lune) \big] \epol_\lune \Big\} \times \vb_\lune \, ,
\end{empheq}
\end{subequations}
where $\spinG$ and $\epol_\ext$ are respectively the magnitude and direction of $\vspinG$, namely $\vspinG =\spinG \, \epol_\ext$. In the last expression, the non-geodesic acceleration of the Moon was expressed, at lowest order (i.e., considering the quadrupole moment of the Moon only). Let us emphasize that part of the Lense-Thirring effect (term proportional to $\vb_\ext$) was inserted into the geodetic precession for conciseness.

According to expression \eqref{eq:estimatinert}, we deduce that the geodetic precession contribution to the relative frequency difference scales such as
\begin{equation}
  - \frac{1}{c^2} \, \angvitGP \cdot ( \xlB \times \vlB ) \propto \frac{1}{2} \sum_{\ext \neq \lune} \left(\frac{n_\lune R_\lune}{c}\right) \left( \frac{Gm_\ext}{ c^2\csphRG} \right) \left( \frac{\vert 3 \vb_\lune - 4 \vb_\ext \vert}{c} \right) \left( \frac{R_\lune}{\csphRG} \right) \left( \frac{\csphRB}{R_\lune} \right) \, .
\end{equation}
For Earth and Sun contributions, we find $-7.72 \times 10^{-26}$ and $5.46 \times 10^{-25}$, respectively. So the geodetic precession can be omitted considering the $10^{-16}$ threshold.

The Lense-Thirring effect enters the time modeling according to equation \eqref{eq:estimatinert}, that is to say
\begin{equation}
  - \frac{1}{c^2} \, \angvitLTP \cdot ( \xlB \times \vlB ) \propto 2 \sum_{\ext \neq \lune} \left(\frac{n_\lune R_\lune}{c}\right) \left(\frac{G \spinG}{c^3\csphRG^2}\right) \left(\frac{R_\lune}{\csphRG}\right) \left(\frac{\csphRB}{R_\lune}\right) \, .
\end{equation}
For Earth and Sun contributions, we find $8.23 \times 10^{-30}$ and $3.76 \times 10^{-30}$, respectively. For the numerical estimate of $\spin_\sun$ we have considered $\spin_\sun = 1.909 \times 10^{41} \ \mathrm{kg} \, \mathrm{m}^2 \, \mathrm{s}^{-1}$ \citep{1998MNRAS.297L..76P}. Both contributions are negligible.

Finally, let us discuss the Thomas precession. It scales such as
\begin{equation}
  - \frac{1}{c^2} \, \angvitTP \cdot ( \xlB \times \vlB ) \propto \frac{3 J_2^\lune}{4} \sum_{\ext \neq \lune} \left( \frac{\vitb_\lune}{c} \right) \left(\frac{n_\lune R_\lune}{c}\right) \left( \frac{Gm_\ext}{c^2\csphRG} \right) \left( \frac{R_\lune}{\csphRG} \right)^{\!3} \left( \frac{\csphRB}{R_\lune} \right) \, .
\end{equation}
Then, Earth and Sun contributions are $3.43 \times 10^{-34}$ and $4.96 \times 10^{-39}$, respectively. These contributions are perfectly negligible; they are respectively eighteen and twenty-three orders of magnitude smaller than requirement.

\section{Shape of the selenoid}
\label{sec:selenoid}

Let $\scalpotmoy$ be the time average of the equipotential surface representing the sum of the centrifugal potential plus the scalar potential $\poteff$ [cf. Eq. \eqref{eq:equipot}]
\begin{equation}
  \scalpotmoy (\xl) = \lim_{\tau\rightarrow\infty} \frac{1}{\tau} \int_0^\tau \dd \tcl \left[ \frac{\vitl(\tcl)^2}{2} + \poteff(\tcl,\xl) \right] \, ,
\end{equation}

 For the Moon it is the synodic month period. According to previous computations (see \ref{app:comp}), we obtain the following relation:
\begin{align}
  \scalpotmoy (\csphR,\csphT,\csphL) & = \frac{Gm_\lune}{\csphR} \Bigg\{ 1+\frac{J_2^\lune}{2} \! \left(\frac{R_\lune}{\csphR}\right)^{\!2} + \frac{n_\lune^2\csphR^3}{2\,Gm_\lune} \nonumber\\
  & - \left[\frac{3J_2^\lune}{2} \! \left(\frac{R_\lune}{\csphR}\right)^{\!2}+\frac{n_\lune^2\csphR^3}{2\, Gm_\lune}\right] \sin^2\csphT + 3\left(\frac{R_\lune}{\csphR}\right)^{\!2}\left( C_{22}^\lune \cos 2 \csphL + S_{22}^\lune \sin 2 \csphL \right) \cos^2\csphT \nonumber\\
  & + \sum_{\ell=3}^{\ellnum}\sum_{m=0}^\ell\left(\frac{R_\lune}{\csphR}\right)^{\!\ell}P_{\ell m}(\sin \csphT) \left[ C_{\ell m}^\lune \cos (m \csphL) + S_{\ell m}^\lune \sin (m \csphL) \right] \nonumber\\
  & + \frac{1}{2}\left(\frac{m_\terre}{m_\lune}\right) \left(\frac{\csphR}{a_\lune}\right)^{\!3}(3\cos^2\csphTB\cos^2\csphLB-1) \Bigg\} \, .
\end{align}

Let us define $\potsel$, the numerical value of the (mean) potential of the selenoid, as the equipotential $\scalpotmoy$ evaluated on the equatorial radius and at the intersection between the lunar prime meridian and lunar equator, that is to say at the spatial point of coordinates $\xl=(R_\lune,0^\circ,0^\circ)$. Let us emphasize that this choice is purely arbitrary---a common definition should result from a globally admitted convention instead. We thus have $\potsel = \scalpotmoy(R_\lune,0^\circ,0^\circ)$, namely
\begin{align}
  \potsel = \frac{Gm_\lune}{R_\lune} \Bigg\{ 1 & + 3 C_{22}^\lune + \frac{J_2^\lune}{2} + \frac{n_\lune^2R_\lune^3}{2\,Gm_\lune} + \frac{m_\terre}{m_\lune} \! \left(\frac{R_\lune}{a_\lune}\right)^{\!3} + \sum_{\ell=3}^{\ellnum}\sum_{m=0}^\ell C_{\ell m}^\lune P_{\ell m}(0) \Bigg\} \, ,
\end{align}
with numerical value $\potsel=2.82\times10^{6}\ \mathrm{m}^2/\mathrm{s}^2$. From $\potsel$, we can now proceed and solve for $\Rsel$, the shape of the selenoid, from the following relation:
\begin{equation}
  \scalpotmoy (\Rsel,\csphT,\csphL) - \potsel = 0 \, .
  \label{eq:solvRsel}
\end{equation}
This equation can be solved numerically, up to the desired precision, using for instance a simple Newton-Raphson algorithm with $\Rsel=R_\lune$ as first iteration. Instead, hereafter, we provide an approximate algebraic expression.

\begin{figure}[t]
  \begin{center}
    \hspace{0.15cm}
    \includegraphics[scale=0.8]{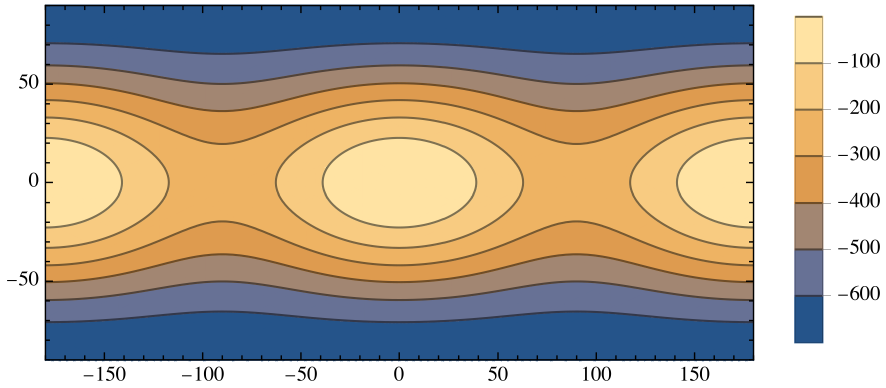}
  \end{center}
  \setlength{\unitlength}{1.0cm}
  \begin{picture}(0,0)(-8,0)
    \put(-1.83,0.25){\rotatebox{0}{longitude $\csphL$ $[{}^{\circ}]$}}
    \put(-6.7,2.58){\rotatebox{90}{latitude $\csphT$ $[{}^{\circ}]$}}
    \put(6.4,5.0){\rotatebox{270}{$\Rsel-R_\lune\ [\mathrm{m}]$}}
  \end{picture}
  \vspace{0.0cm}
  \caption{Contour map of the altitude of the approximate selenoid [cf. Eq. \eqref{eq:Rsel}] with respect to the lunar equatorial radius $R_\lune$.}
  \label{fig:selenoide}
\end{figure}

To this end, let us first recall that the numerical values of the Moon gravitational coefficients are
\begin{equation}
  J_2^\lune = 2.03 \times 10^{-4} \, , \qquad C_{22}^\lune = 2.24\times 10^{-5} \, , \qquad S_{22}^\lune = 5.86\times 10^{-10} \, ,
  \label{eq:deg2}
\end{equation}
and (using \href{https://pds-geosciences.wustl.edu/grail/grail-l-lgrs-5-rdr-v1/grail_1001/shadr/gggrx_1200a_sha.tab}{GRGM1200A} modeling up to $\ellnum=150$)
\begin{equation}
  \sum_{\ell=3}^{\ellnum}\sum_{m=0}^\ell C_{\ell m}^\lune P_{\ell m}(0) = -6.06 \times 10^{-7} \, .
\end{equation}
Moreover, we also have the following numerical estimates:
\begin{equation}
  \frac{n_\lune^2R_\lune^3}{Gm_\lune} = 7.59 \times 10^{-6} \, , \qquad \frac{m_\terre}{m_\lune} \! \left(\frac{R_\lune}{a_\lune}\right)^{\!3} = 7.54 \times 10^{-6} \, .
\end{equation}
These suggest, as a good first approximation, to not consider $S_{22}^\lune$ and $\ell\geqslant 3$ terms while solving for $\Rsel$ from equation~\eqref{eq:solvRsel}. After some algebra we find the first order expression of $\Rsel$; it reads such as
\begin{align}
  \Rsel(\csphT,\csphL) \simeq R_\lune \Bigg\{ 1 & - \left( \frac{3J_2^\lune}{2} + \frac{n_\lune^2 R_\lune^3}{2\, Gm_\lune} \right) \sin^2\csphT + 3 C_{22}^\lune \left( \cos^2\csphT \cos 2\csphL - 1 \right) \nonumber \\
  & + \frac{3}{2}\left(\frac{m_\terre}{m_\lune} \right) \left(\frac{R_\lune}{a_\lune}\right)^{\!3} \left(\cos^2\csphT\cos^2\csphL-1\right) \Bigg\} \, .
  \label{eq:Rsel}
\end{align}
In figure \ref{fig:selenoide}, we represent the altitude of the selenoid with respect to the equatorial radius of the Moon as function of the latitude and longitude on the Moon.

Let us compute $\Rsel$ at some specific locations. For $\csphT = 0$ and $\csphL = 0$ (intersection between lunar prime meridian and equator), we find $\Rsel(0,0) = R_\lune = \Rsel(0,\pi)$, as it is expected from the definition $\potsel = \scalpotmoy(R_\lune,0,0)$. For $\csphT = 0$ and $\csphL = \pi/4$, we obtain
\begin{equation}
  \Rsel(0,\pi/4) = R_\lune \left[ 1 - 3 C_{22}^\lune - \frac{3}{4} \left(\frac{m_\terre}{m_\lune}\right)\left(\frac{R_\lune}{a_\lune}\right)^{\!3} \right] = \Rsel(0,3\pi/4) \, ,
\end{equation}
namely $\Rsel(0,\pi/4) < R_\lune$ since $C_{22}^\lune > 0$. For $\csphT = 0$ and $\csphL = \pi/2$, we find
\begin{equation}
  \Rsel(0,\pi/2) = R_\lune \left[ 1 - 6 C_{22}^\lune - \frac{3}{2} \left(\frac{m_\terre}{m_\lune}\right)\left(\frac{R_\lune}{a_\lune}\right)^{\!3} \right] = \Rsel(0,3\pi/2) \, ,
\end{equation}
so that we eventually have $\Rsel(0,\pi/2) < \Rsel(0,\pi/4) < R_\lune$. Let us now compute $\Rsel$ at the lunar pole, that is to say for $\csphT = \pi/2$ and $\csphL = 0$ (notice that $\csphL$ is not defined when $\csphT = \pi/2$); we obtain
\begin{align}
  \Rsel(\pi/2,0) & = R_\lune \Bigg[ 1 - \frac{3J_2^\lune}{2} - \frac{n_\lune^2 R_\lune^3}{2\, Gm_\lune} - 3 C_{22}^\lune - \frac{3}{2} \left(\frac{m_\terre}{m_\lune}\right)\left(\frac{R_\lune}{a_\lune}\right)^{\!3} \Bigg] \nonumber\\
  & = \Rsel(-\pi/2,0)\, ,
\end{align}
from which we infer $\Rsel(\pi/2,0) < \Rsel(0,\pi/4) < R_\lune$. In order to compare $\Rsel(\pi/2,0)$ and $\Rsel(0,\pi/2)$ let us substitute the numerical values for the lunar (and Earth) physical parameters. We find
\begin{equation}
  1-\frac{\Rsel(0,\pi/2)}{R_\lune} = 1.46 \times 10^{-4} \, , \qquad 1-\frac{\Rsel(\pi/2,0)}{R_\lune} = 3.87 \times 10^{-4} \, ,
  \label{eq:Rshape}
\end{equation}
so that $\Rsel(\pi/2,0)<\Rsel(0,\pi/2)$ as it is expected from the direction of the mass quadrupole moment. Numerical estimates in equation \eqref{eq:Rshape} represent deficits of $253 \ \mathrm{m}$ and $672\ \mathrm{m}$ with respect to $R_\lune$, respectively. In other words, the minimal distance between the surface of the selenoid and the center of mass of the Moon occurs at the poles. The largest distance happens at the intersection between the lunar prime meridian and equator, but also at the symmetric position with respect to the center of mass of the Moon. Within the equatorial plane, the minimal distances are found at $\pm 90^\circ$ from the lunar prime meridian (cf. Fig. \ref{fig:selenoide}).

Let us mention that our approach is just illustrative and would require more careful treatment to provide a comprehensive definition of the selenoid. See for instance results by \citet{Cziraki2023} who considered \href{https://pds-geosciences.wustl.edu/grail/grail-l-lgrs-5-rdr-v1/grail_1001/shadr/gggrx_1200a_sha.tab}{GRGM1200A} modeling but neglected the centrifugal and tidal potentials. Their selenoid might therefore show differences of the order of 10 m with respect to a selenoid built from $\scalpotmoy$ rather than $\potM$ alone. 

\section{Lunar topography}
\label{sec:topo}

\begin{figure}[t]
  \begin{center}
    \includegraphics[scale=0.8]{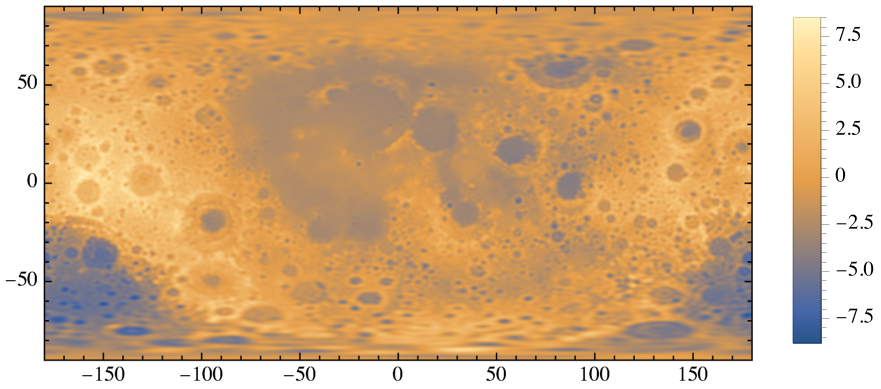}
  \end{center}
  \setlength{\unitlength}{1.0cm}
  \begin{picture}(0,0)(-8,0)
    \put(-1.9,0.25){\rotatebox{0}{longitude $\csphL$ $[{}^{\circ}]$}}
    \put(-6.9,2.6){\rotatebox{90}{latitude $\csphT$ $[{}^{\circ}]$}}
    \put(6.1,5.3){\rotatebox{270}{$\Rtopo-\Rsel$ $[\mathrm{km}]$}}
    \linethickness{0.4mm}
  \end{picture}
  \vspace{0.0cm}
  \caption{Map of the lunar topography reconstructed from LOLA's data considering $\ellnumtopo = 150$ in Eq. \eqref{eq:Rtopo}. The lunar topography [i.e., $\Rtopo(\csphT,\csphL)$ in Eq. \eqref{eq:Rtopo}] is expressed relative to the altitude above the selenoid [i.e., $\Rsel(\csphT,\csphL)$ in Eq. \eqref{eq:Rsel}].}
  \label{fig:topo_sel}
\end{figure}

The topography of the Moon can be assessed from the data obtained by the Lunar Orbiter Laser Altimeter (LOLA) \citep{2009ApOpt..48.3035R,2010SSRv..150..209S}---an instrument onboard the NASA Lunar Reconnaissance Orbiter mission. The lunar topography is conveniently reconstructed using \href{https://imbrium.mit.edu}{LOLA's spherical harmonic coefficients} $(A_{\ell m}^{\lune},B_{\ell m}^{\lune})$ distributed in the principal axis frame. The radial position of a point on the surface of the Moon is thus given by $\csphR = \Rtopo(\csphT,\csphL)$, where $\Rtopo(\csphT,\csphL)$ represents the radial topography of the Moon (radial distance between the center of mass of the Moon and the surface of the Moon) and is given by the following partial sum
\begin{equation}
  \Rtopo(\csphT,\csphL) = \sum_{\ell=0}^{\ellnumtopo} \sum_{m=0}^{\ell} P_{\ell m} (\sin \csphT) \left[ A^\lune_{\ell m} \cos (m \csphL) + B^\lune_{\ell m} \sin (m \csphL) \right] \, ,
  \label{eq:Rtopo}
\end{equation}
with $A^\lune_{\ell m}$ and $B^\lune_{\ell m}$ having length dimension and $\ellnumtopo$ being the degree to which the series is stopped. In figure \ref{fig:topo_sel}, we represent the lunar topography (with $\ellnumtopo = 150$) relative to the selenoid $\Rsel$ derived in equation \eqref{eq:Rsel}. The chosen $\ellnumtopo$ is a compromise between computational time and precision of topographic details of the lunar surface. Roughly speaking, the precision of topographic details goes on like $R_\lune/\ellnumtopo$, that is to say $12 \ \mathrm{km}$ for $\ellnumtopo = 150$.

The maximum of the difference between the mean equatorial radius $R_\lune$ and the topography $R_\topo(\csphT,\csphL)$ is of the order of
\begin{equation}
  \mathrm{max} \left\vert \frac{\Rtopo(\csphT,\csphL)}{R_\lune} - 1 \right\vert \sim 5.2 \times 10^{-3} \, ,
  \label{eq:RtopoRM}
\end{equation}
for $-90^\circ \leqslant \csphTB \leqslant 90^\circ$ and $-180^\circ < \csphLB \leqslant 180^\circ$.

\section{Resolution of the light travel time equation}
\label{sec:lighttt}

Expression \eqref{eq:ttf} being implicit in $\tcbB$, it is usually solved numerically by an iterative process starting with $\tcbB^{(0)}=\tcbA$. After $n$ iterations, we shall have $\tcbB = \tcbB^{(n)}$. This eventually allows one to determine the position of the receiver at time $\tcbB$.

Another possibility is to perform manually the iterations and then the series expansion in the right-hand side of \eqref{eq:lighttime} in order to get a closed-form expression of $\tcbB$ up to the desired order in $c^{-1}$. As an illustration, after three iterations, we find
  \begin{align}
    \tcbB^{(3)} & = \tcbA + \frac{L_{\A\B}}{c} + \frac{L_{\A\B}}{c^2} (\vect{l}_{\B} \cdot \vbB) + \frac{L_{\A\B}}{2c^3} \Bigg\{ 2(\vect{l}_{\B} \cdot \vbB)^2+L_{\A\B} (\vect{l}_{\B}\cdot\abB) \nonumber\\
    &+\frac{L_{\A\B}}{\vert \xb_{\B\A} \vert} \Bigg[ \vitb_\B^2 -\left(\frac{\xb_{\B\A}\cdot\vbB}{\vert\xb_{\B\A}\vert}\right)^{\!2} \Bigg] + L_{\A\B} \, \cvect{\vitb_\B}{a} \cvect{\vitb_\B}{b} \left[\frac{\partial^2\Delta}{\partial \cvect{\posb_\B}{a} \partial \cvect{\posb_\B}{b}} \right]_{(\tcbA,\xb_{\A} (\tcbA),\xb_{\B} (\tcbA))} \Bigg\} \, ,
  \end{align}
where we used $\xb_{\B\A} = \xb_{\B} (\tcbA) - \xb_{\A} (\tcbA)$, $\abB = \abB (\tcbA)$, and  $\vb_\B = \vb_\B (\tcbA)$ with $\vitb_\B = \vert \vb_\B \vert$. The instantaneous distance $L_{\A\B}$ is defined by
\begin{equation}
  L_{\A\B} = \vert \xb_{\B\A} \vert + \Delta(\tcbA,\xbA(\tcbA),\xbB(\tcbA)) \, ,
\end{equation}
and $\vect{l}_\B$, the tangent vector to the ray at position $\xbB (\tcbA)$, is given by (see \citet{2004CQGra..21.4463L} and \citet{PhysRevD.101.064035})
\begin{equation}
  \vect{l}_{\B} = \frac{\xb_{\B\A}}{\vert\xb_{\B\A}\vert} + \left[\frac{\partial \Delta}{\partial \xb_{\B}}\right]_{(\tcbA,\xb_{\A} (\tcbA),\xb_{\B} (\tcbA))} \, .
\end{equation}

In most applications, we have $\vert \xb_{\B\A} \vert \gg \Delta(\tcbA,\xbA(\tcbA),\xbB(\tcbA))$. Therefore, it is convenient to apply another series expansion in $\Delta/\vert\xb_{\B\A}\vert$ around $\Delta/\vert\xb_{\B\A}\vert = 0$. At zeroth-order in $\Delta/\vert\xb_{\B\A}\vert$, we therefore obtain (see also \citet{1994A&A...286..971P}):
\begin{align}
  \tcbB^{(3)} & = \tcbA + \frac{\vert\xb_{\B\A}\vert}{c} + \frac{\xb_{\B\A}\cdot \vbB }{c^2} + \frac{\vert\xb_{\B\A}\vert}{2c^3} \Bigg[ \vitb_\B^2 +\left(\frac{\xb_{\B\A}\cdot\vbB}{\vert\xb_{\B\A}\vert}\right)^{\!2}+\xb_{\B\A}\cdot\abB \Bigg] \, .
\end{align}
This approximate time of reception can be used eventually to estimate, in the right-hand side of equation \eqref{eq:lighttime}, the position of the receiver $\xbB = \xbB(\tcbB)$ with $\tcbB = \tcbB^{(3)}$.

For the one-way time or frequency transfer between a clock on the Earth and a clock on the Moon, $\vert \xb_{\B\A} \vert \sim 4 \times 10^{8} \ \mathrm{m}$ while the Shapiro delay caused by the Sun, the Earth, and the Moon, are of the order of $10\ \mathrm{ns}$, $50\ \mathrm{ps}$, and $0.8\ \mathrm{ps}$, respectively (i.e., $4\ \mathrm{m}$, $2\ \mathrm{cm}$, and $0.3\ \mathrm{mm}$, resp. in terms of range). The delay caused by the atmosphere is typically the same order of magnitude as the Shapiro delay caused by the Sun. These estimates justify the approximation $\vert \xb_{\B\A} \vert \gg \Delta(\tcbA,\xbA(\tcbA),\xbB(\tcbA))$.

\section{Demonstration of range covariance using either TCB or TDB}
\label{sec:TDBVSTCB}

When relying on TDB (labelled $\tdb$) rather than TCB (labelled $\tcb$), the light travel time in equation \eqref{eq:range} can be decomposed such as
\begin{equation}
  [\tcbB - \tcbA]_{(\tcbA,\xbA,\xbB)} = [\tdb - \tcb]_{(\tcbA)} + [\tdbB - \tdbA]_{(\tdbA,\xbdA,\xbdB)} - [\tdb - \tcb]_{(\tcbB)} \, , 
  \label{eq:TCBlt2TDBlt}
\end{equation}
where $[\tdbB - \tdbA]_{(\tdbA,\xbdA,\xbdB)}$ is now the light travel time expressed in TDB. The function $[\tdb - \tcb]_{(\tcb)}$ appearing in equation \eqref{eq:TCBlt2TDBlt} represents the difference between TDB and TCG and is given by the right-hand side of equation \eqref{eq:TDB-TCB}.

According to the general covariance principle, making use, in equation \eqref{eq:range}, of either the left-hand side or the right-hand side of \eqref{eq:TCBlt2TDBlt},  shall not affect the proper time difference $\pptimeB - \pptimeA$. Let us thus demonstrate that the right-hand side of equation \eqref{eq:TCBlt2TDBlt} do indeed reduce to the left-hand side.

On one side, according to the definition in equation \eqref{eq:TDB-TCB}, we have
\begin{equation}
  [\tdb - \tcb]_{(\tcbA)}- [\tdb - \tcb]_{(\tcbB)} = L_B ( \tcbB - \tcbA ) \, .
  \label{eq:TDB-TCBenAB}
\end{equation}
On the other side, the expression for the light travel time expressed in TDB reads
\begin{equation}
  [\tdbB - \tdbA]_{(\tdbA,\xbdA,\xbdB)} = \frac{\vert \xbdB - \xbdA \vert}{c} + \frac{\Delta^{**}(\tdbA,\xbdA,\xbdB)}{c} \, .
  \label{eq:lighttimeTDB}
\end{equation}
As discussed in \citet{Klioner2008}, in order for the light travel time expression to be the same in equations \eqref{eq:lighttime} and \eqref{eq:lighttimeTDB}, the TDB-compatible distances and mass parameters have to be rescaled by the same factor than the one used to rescale TDB in equation \eqref{eq:TDB-TCB}, namely $\tdb = (1-L_B)\tcb+\tdb_0$, and hence, $\xbd = (1-L_B) \xb$ and $\Delta^{**} = (1-L_B) \Delta$. Inserting these two last relations into the right-hand side of equation \eqref{eq:lighttimeTDB}, we find
\begin{equation}
  [\tdbB - \tdbA]_{(\tdbA,\xbdA,\xbdB)} = (1-L_B)(\tcbB - \tcbA) \, ,
  \label{eq:lighttimeTCBVSTDB}
\end{equation}
after invoking equations \eqref{eq:lighttime} and then \eqref{eq:ttf}. Equation \eqref{eq:TCBlt2TDBlt} is finally recovered by adding together equations \eqref{eq:lighttimeTCBVSTDB} and \eqref{eq:TDB-TCBenAB}, and then making use of \eqref{eq:ttf}, which achieves the demonstration.

\section{Demonstration of range covariance using either BCRS or GCRS as intermediate system}
\label{sec:GCRSVSBCRS}

In order to show that the proper time difference $\pptimeB - \pptimeA$ is independent of the intermediate system being used for the light travel time computation, we shall demonstrate the equality in equation \eqref{eq:temp1}. Following \citet{Kopeikin2024}, let us start by rewriting the first term in the left-hand side of equation \eqref{eq:temp1} as
\begin{equation}
  [\tcl - \tcg]_{(\tcgB,\xgB)} = [ \tcl - \tcb]_{(\tcbB,\xbB)} - [\tcg - \tcb]_{(\tcbB,\xbB)} \, ,
\end{equation}
where $[\tcl - \tcb]_{(\tcb,\xb)}$ is given by equation \eqref{eq:TCB-TCL} and $[\tcg - \tcb]_{(\tcb,\xb)}$ by equation \eqref{eq:TCB-TCG}. Equation \eqref{eq:temp1}, thus reduces to
\begin{equation}
  [\tcgB - \tcgA]_{(\tcgA,\xgA,\xgB)} - [\tcbB - \tcbA]_{(\tcbA,\xbA,\xbB)} = [\tcg - \tcb]_{(\tcbB,\xbB)} - [\tcg - \tcb]_{(\tcbA,\xbA)} \, .
  \label{eq:temp2}
\end{equation}
The equality between the left-hand side---i.e., the difference between the light travel times between clock `$\B$' and clock `$\A$' in GCRS and BCRS---and the right-hand side---i.e., the difference between the time transformation from TCB to TCG at clock `$\B$' and at clock `$\A$'---can be easily demonstrated at the order $c^{-2}$.

As a matter of fact, the time transfer function expressed in TCG and at second order in $c^{-2}$ is given as follows (see \ref{sec:lighttt} for more details):
\begin{equation}
  [\tcgB - \tcgA]_{(\tcgA,\xgA,\xgB)} = \frac{\vert \xgB - \xgA \vert}{c} + \frac{\vgB \cdot (\xgB - \xgA )}{c^2} + \mathcal{O}(c^{-3})\, ,
  \label{eq:lighttimeTCG}
\end{equation}
where all quantities in the right-hand side are expressed at $\tcgA$. Let us emphasize that we neglected the delay function for simplicity. Then, according to resolution B1.3 of IAU 2000 \citep{Soffel2003}, the geocentric and barycentric coordinates of a given space-time event are related, up to second order terms in $c^{-2}$, by
\begin{subequations}
\begin{empheq}[left=\empheqlbrace]{align}
  \xg (\tcg) & = \xb (\tcb) - \xb_\terre (\tcb) + \mathcal{O}(c^{-2}) \, , \\
  \vg (\tcg) & = \vb (\tcb) - \vb_\terre (\tcb) + \mathcal{O}(c^{-2}) \, .
\end{empheq}
\end{subequations}
Applying these to equation \eqref{eq:lighttimeTCG}, we thus deduce
\begin{equation}
  [\tcgB - \tcgA]_{(\tcgA,\xgA,\xgB)} - [\tcbB - \tcbA]_{(\tcbA,\xbA,\xbB)} = -\frac{ \vb_{\terre} \cdot (\xbB - \xbA )}{c^2} + \mathcal{O}(c^{-3})\, ,
  \label{eq:lighttimeGCRSandBCRS}
\end{equation}
where $[\tcbB - \tcbA]_{(\tcbA,\xbA,\xbB)}$ is given, at second order in $c^{-1}$, by
\begin{equation}
  [\tcbB - \tcbA]_{(\tcbA,\xbA,\xbB)} = \frac{\vert \xbB - \xbA \vert}{c} + \frac{\vbB \cdot (\xbB - \xbA )}{c^2} + \mathcal{O}(c^{-3})\, ,
  \label{eq:lighttimeBCRSapprox}
\end{equation}
where all quantities in the right-hand sides of equations \eqref{eq:lighttimeGCRSandBCRS} and \eqref{eq:lighttimeBCRSapprox} are expressed at $\tcbA$. 

Let us now focus on the right-hand side of equation \eqref{eq:temp2}. According to equation \eqref{eq:TCB-TCG}, the difference between TCG and TCB at clock `$\B$' is 
given by
\begin{equation}
  [\tcg - \tcb]_{(\tcbB,\xbB)} = -\frac{1}{c^{2}} \left\{\int_{\tcb_0}^{\tcbB}\left[\frac{\vitb^2_\terre(\tcb)}{2}+\overline w_{\terre}(\xb_\terre(\tcb))\right] \dd \tcb + \vb_\terre (\tcbB) \cdot [\xbB(\tcbB) -\xb_\terre(\tcbB)] \right\}+\mathcal{O}(c^{-4}) \, .
  \label{eq:TCG-TCBl}
\end{equation}
Substituting for $\tcbB$ from $\tcbB = \tcbA + [\tcbB - \tcbA]_{(\tcbA,\xbA,\xbB)}$ into \eqref{eq:TCG-TCBl}, while making use of \eqref{eq:lighttimeBCRSapprox}, shall return
\begin{equation}
  [\tcg - \tcb]_{(\tcbB,\xbB)} - [\tcg - \tcb]_{(\tcbA,\xbA)} = -\frac{\vb_\terre \cdot (\xbB -\xb_\terre)}{c^{2}} +\mathcal{O}(c^{-3}) \, ,
  \label{eq:TCG-TCBldev}
\end{equation}
after Taylor expanding the expression around $\tcbB = \tcbA$ and after keeping only terms of order $c^{-2}$.

The equality between equations \eqref{eq:lighttimeGCRSandBCRS} and \eqref{eq:TCG-TCBldev} verifies the correctness of equation \eqref{eq:temp2}, which achieves the demonstration.

\bibliographystyle{plainnat}
\bibliography{lunar_time}

@Book{1991ercm.book.....B,
  Title                    = {Essential relativistic celestial mechanics},
  Author                   = {Brumberg, V.~A.},
  Year                     = {1991},
  publisher                = {Adam Hilger},
  Booktitle                = {Bristol, England and New York, Adam Hilger, 1991, 271 p.},
  Url                      = {http://adsabs.harvard.edu/abs/1991ercm.book.....B}
}

@ARTICLE{2010ITN....36....1P,
       author = {{Petit}, G. and {Luzum}, B.},
        title = "{IERS Conventions (2010)}",
      journal = {IERS Technical Note},
         year = 2010,
        month = jan,
       volume = {36},
        pages = {1},
       url = {https://ui.adsabs.harvard.edu/abs/2010ITN....36....1P}
}

@Article{2004CQGra..21.4463L,
  Title                    = {World function and time transfer: general post-Minkowskian expansions},
  Author                   = {Le Poncin-Lafitte, C. and Linet, B. and Teyssandier, P.},
  Journal                  = {Classical and Quantum Gravity},
  Year                     = {2004},

  Month                    = sep,
  Pages                    = {4463-4483},
  Volume                   = {21},
  Doi                      = {10.1088/0264-9381/21/18/012},
  Eprint                   = {arXiv:gr-qc/0403094},
  Url                      = {http://adsabs.harvard.edu/abs/2004CQGra..21.4463L}
}

@ARTICLE{2019NSTIM.109.....F,
    author = {{Fienga}, A. and {Deram}, P. and {Viswanathan}, V. and {Di Ruscio}, A. and {Bernus}, L. and {Durante}, D. and {Gastineau}, M. and {Laskar}, J.},
    title = "{INPOP19a planetary ephemerides}",
    journal = {Notes Scientifiques et Techniques de l'Institut de Mecanique Celeste},
    year = 2019,
    month = dec,
    volume = {109},
    url = {https://ui.adsabs.harvard.edu/abs/2019NSTIM.109.....F}
}

@Article{2021AJ....161..105P,
  author   = {{Park}, R.~S. and {Folkner}, W.~M. and {Williams}, J.~G. and {Boggs}, D.~H.},
  title    = {{The JPL Planetary and Lunar Ephemerides {DE440} and {DE441}}},
  doi      = {10.3847/1538-3881/abd414},
  eid      = {105},
  number   = {3},
  pages    = {105},
  volume   = {161},
  url   = {https://ui.adsabs.harvard.edu/abs/2021AJ....161..105P},
  journal  = {\aj},
  month    = mar,
  year     = {2021},
}

@Article{1994A&A...286..971P,
  Title                    = {Relativistic theory for picosecond time transfer in the vicinity of the {E}arth},
  Author                   = {Petit, G. and Wolf, P.},
  Journal                  = {Astronomy and Astrophysics},
  Year                     = {1994},

  Month                    = jun,
  Pages                    = {971-977},
  Volume                   = {286},
  Url                      = {http://adsabs.harvard.edu/abs/1994A%26A...286..971P}
}

@Article{PhysRevD.93.044028,
  author    = {Linet, B. and Teyssandier, P.},
  title     = {Time transfer functions in Schwarzschild-like metrics in the weak-field limit: A unified description of Shapiro and lensing effects},
  journal   = {Phys. Rev. D},
  year      = {2016},
  volume    = {93},
  issue     = {4},
  month     = {Feb},
  pages     = {044028},
  doi       = {10.1103/PhysRevD.93.044028},
  url       = {https://link.aps.org/doi/10.1103/PhysRevD.93.044028},
  numpages  = {12},
  publisher = {American Physical Society},
}

@Article{PhysRevD.101.064035,
  author    = {Bourgoin, A.},
  title     = {General expansion of time transfer functions in optical spacetime},
  journal   = {Phys. Rev. D},
  year      = {2020},
  volume    = {101},
  issue     = {6},
  month     = {Mar},
  pages     = {064035},
  doi       = {10.1103/PhysRevD.101.064035},
  url       = {https://link.aps.org/doi/10.1103/PhysRevD.101.064035},
  numpages  = {28},
  publisher = {American Physical Society},
}

@Article{2008CQGra..25n5020T,
  author        = {Teyssandier, P. and Le Poncin-Lafitte, C.},
  title         = {General post-Minkowskian expansion of time transfer functions},
  doi           = {10.1088/0264-9381/25/14/145020},
  eprint        = {0803.0277},
  number        = {14},
  pages         = {145020},
  url           = {http://adsabs.harvard.edu/abs/2008CQGra..25n5020T},
  volume        = {25},
  archiveprefix = {arXiv},
  comment       = {Modeling most of the tests of general relativity requires us to know the function relating light travel time to the coordinate time of reception and to the spatial coordinates of the emitter and the receiver. We call such a function the reception time transfer function. Of course, an emission time transfer function may as well be considered. We present here a recursive procedure enabling us to expand each time transfer function into a perturbative series of ascending powers of the Newtonian gravitational constant G (general post-Minkowskian expansion). Our method is self-sufficient in the sense that neither the integration of null geodesic equations nor the determination of Synge's world function is necessary. To illustrate the method, the time transfer function of a three-parameter family of static, spherically symmetric metrics is derived within the post-linear approximation.},
  journal       = {Classical and Quantum Gravity},
  month         = jul,
  year          = {2008},
}

@ARTICLE{1998MNRAS.297L..76P,
       author = {{Pijpers}, F.~P.},
        title = "{Helioseismic determination of the solar gravitational quadrupole moment}",
      journal = {\mnras},
         year = 1998,
        month = jul,
       volume = {297},
       number = {3},
        pages = {L76-L80},
          doi = {10.1046/j.1365-8711.1998.01801.x},
archivePrefix = {arXiv},
       eprint = {astro-ph/9804258},
 primaryClass = {astro-ph},
       url = {https://ui.adsabs.harvard.edu/abs/1998MNRAS.297L..76P}
}

@Proceedings{1989ASSL..154.....K,
  title     = {{Reference frames in astronomy and geophysics}},
  doi       = {10.1007/978-94-009-0933-5},
  editor    = {{Kovalevsky}, J. and {Mueller}, I.~I. and {Kolaczek}, B.},
  url    = {https://ui.adsabs.harvard.edu/abs/1989ASSL..154.....K},
  booktitle = {Reference Frames},
  month     = jan,
  year      = {1989},
}

@Article{1990CeMDA..48...23B,
  author   = {{Brumberg}, V.~A. and {Kopeikin}, S.~M.},
  title    = {{Relativistic time scales in the solar system}},
  doi      = {10.1007/BF00050674},
  number   = {1},
  pages    = {23-44},
  volume   = {48},
  url   = {https://ui.adsabs.harvard.edu/abs/1990CeMDA..48...23B},
  journal  = {Celestial Mechanics and Dynamical Astronomy},
  month    = mar,
  year     = {1990},
}

@Article{1993PhRvD..48.1451K,
  Title                    = {Relativistic theory of astronomical reference systems in closed form},
  Author                   = {Klioner, S.~A. and Voinov, A.~V.},
  Journal                  = {Physical Review D},
  Year                     = {1993},

  Month                    = aug,
  Pages                    = {1451-1461},
  Volume                   = {48},
  Doi                      = {10.1103/PhysRevD.48.1451},
  Url                      = {http://adsabs.harvard.edu/abs/1993PhRvD..48.1451K}
}

@Article{Soffel2003,
  author   = {Soffel, M. and Klioner, S.~A. and Petit, G. and Wolf, P. and Kopeikin, S.~M. and Bretagnon, P. and Brumberg, V.~A. and Capitaine, N. and Damour, T. and Fukushima, T. and Guinot, B. and Huang, T.-Y. and Lindegren, L. and Ma, C. and Nordtvedt, K. and Ries, J.~C. and Seidelmann, P.~K. and Vokrouhlick{\'y}, D. and Will, C.~M. and Xu, C.},
  journal  = {Astronomical Journal},
  title    = {The {IAU} 2000 Resolutions for Astrometry, Celestial Mechanics, and Metrology in the Relativistic Framework: Explanatory Supplement},
  year     = {2003},
  month    = dec,
  pages    = {2687-2706},
  volume   = {126},
  doi      = {10.1086/378162},
  eprint   = {arXiv:astro-ph/0303376},
  url      = {http://adsabs.harvard.edu/abs/2003AJ....126.2687S},
}

@Article{Damour1991,
  author    = {Damour, T. and Soffel, M. and Xu, C.},
  journal   = {Physical Review D},
  title     = {General relativistic celestial mechanics. {I.} Method and definition of reference systems},
  year      = {1991},
  month     = oct,
  pages     = {3273-3307},
  volume    = {43},
  url       = {http://adsabs.harvard.edu/abs/1991PhRvD..43.3273D},
}

@Article{Damour1992,
  author    = {Damour, T. and Soffel, M. and Xu, C.},
  journal   = {Physical Review D},
  title     = {General-relativistic celestial mechanics. {II.} Translational equations of motion},
  year      = {1992},
  month     = feb,
  pages     = {1017-1044},
  volume    = {45},
  doi       = {10.1103/PhysRevD.45.1017},
  url       = {http://adsabs.harvard.edu/abs/1992PhRvD..45.1017D},
}

@Article{1993PhRvD..47.3124D,
  Title                    = {General-relativistic celestial mechanics. III. Rotational equations of motion},
  Author                   = {Damour, T. and Soffel, M. and Xu, C.},
  Journal                  = {Physical Review D},
  Year                     = {1993},

  Month                    = apr,
  Pages                    = {3124-3135},
  Volume                   = {47},
  Doi                      = {10.1103/PhysRevD.47.3124},
  Url                      = {http://adsabs.harvard.edu/abs/1993PhRvD..47.3124D}
}

@Article{1994PhRvD..49..618D,
  Title                    = {General-relativistic celestial mechanics. IV. Theory of satellite motion},
  Author                   = {Damour, T. and Soffel, M. and Xu, C.},
  Journal                  = {Physical Review D},
  Year                     = {1994},

  Month                    = jan,
  Pages                    = {618-635},
  Volume                   = {49},
  Doi                      = {10.1103/PhysRevD.49.618},
  Url                      = {http://adsabs.harvard.edu/abs/1994PhRvD..49..618D}
}

@Article{1993A&A...279..273K,
  Title                    = {On the hierarchy of relativistic kinematically nonrotating reference systems},
  Author                   = {Klioner, S.~A.},
  Journal                  = {Astronomy and Astrophysics},
  Year                     = {1993},

  Month                    = nov,
  Pages                    = {273-277},
  Volume                   = {279},
  Url                      = {http://adsabs.harvard.edu/abs/1993A%26A...279..273K}
}

@book{Handbook_GNSS,
title = "Springer Handbook of Global Navigation Satellite Systems",
editor = "Teunissen, P. and Montenbruck, O.",
year = "2017",
language = "English",
isbn = "978-3-319-42926-7",
volume = "XXXI",
publisher = "Springer",
doi = "10.1007/978-3-319-42928-1",
url = "https://doi.org/10.1007/978-3-319-42928-1"
}

@article{petit2015,
title = {International atomic time: Status and future challenges},
journal = {Comptes Rendus Physique},
volume = {16},
number = {5},
pages = {480-488},
year = {2015},
note = {The measurement of time / La mesure du temps},
issn = {1631-0705},
doi = {https://doi.org/10.1016/j.crhy.2015.03.002},
url = {https://www.sciencedirect.com/science/article/pii/S1631070515000535},
author = {Petit, G. and Arias, F. and Panfilo, G.}
}

@ARTICLE{2025ApJ...985..140T,
       author = {{Turyshev}, S.~G. and {Williams}, J.~G. and {Boggs}, D.~H. and {Park}, R.~S.},
        title = "{Relativistic Time Transformations between the Solar System Barycenter, Earth, and Moon}",
      journal = {\apj},
         year = 2025,
        month = may,
       volume = {985},
       number = {1},
          eid = {140},
        pages = {140},
          doi = {10.3847/1538-4357/adcc18},
archivePrefix = {arXiv},
       eprint = {2406.16147},
 primaryClass = {astro-ph.EP},
       url = {https://ui.adsabs.harvard.edu/abs/2025ApJ...985..140T}
}

@Article{Kopeikin2024,
  author        = {{Kopeikin}, S.~M. and {Kaplan}, G.~H.},
  journal       = {\prd},
  title         = {{Lunar time in general relativity}},
  year          = {2024},
  month         = oct,
  number        = {8},
  pages         = {084047},
  volume        = {110},
  url           = {https://ui.adsabs.harvard.edu/abs/2024PhRvD.110h4047K},
  archiveprefix = {arXiv},
  doi           = {10.1103/PhysRevD.110.084047},
  eid           = {084047},
  eprint        = {2407.04862},
  primaryclass  = {gr-qc},
}

@Article{Newhall1996,
  author   = {{Newhall}, X.~X. and {Williams}, J.~G.},
  journal  = {Celestial Mechanics and Dynamical Astronomy},
  title    = {{Estimation of the Lunar Physical Librations}},
  year     = {1996},
  month    = mar,
  number   = {1},
  pages    = {21-30},
  volume   = {66},
  url   = {https://ui.adsabs.harvard.edu/abs/1996CeMDA..66...21N},
  doi      = {10.1007/BF00048820},
}

@Article{Eckhardt1981,
  author    = {{Eckhardt}, D.~H.},
  journal   = {Moon and Planets},
  title     = {{Theory of the libration of the moon}},
  year      = {1981},
  month     = aug,
  pages     = {3-49},
  volume    = {25},
  doi       = {10.1007/BF00911807},
  url       = {http://adsabs.harvard.edu/abs/1981M%26P....25....3E},
}

@ARTICLE{Simon1994,
       author = {{Simon}, J.~L. and {Bretagnon}, P. and {Chapront}, J. and {Chapront-Touze}, M. and {Francou}, G. and {Laskar}, J.},
        title = "{Numerical expressions for precession formulae and mean elements for the Moon and the planets.}",
      journal = {\aap},
         year = 1994,
        month = feb,
       volume = {282},
        pages = {663},
          url = {https://ui.adsabs.harvard.edu/abs/1994A&A...282..663S}
}

@Article{Folkner2014,
  author    = {{Folkner}, W.~M. and {Williams}, J.~G. and {Boggs}, D.~H. and {Park}, R.~S. and {Kuchynka}, P.},
  journal   = {Interplanetary Network Progress Report},
  title     = {The Planetary and Lunar Ephemerides {DE430} and {DE431}},
  year      = {2014},
  month     = feb,
  pages     = {1-81},
  volume    = {42-196},
  url       = {http://adsabs.harvard.edu/abs/2014IPNPR.196C...1F},
}

@Book{Poisson2014,
  author    = {{Poisson}, E. and {Will}, C.~M.},
  title     = {{Gravity}},
  year      = {2014},
  month     = may,
  booktitle = {Gravity},
  publisher = {Cambridge University Press },
  url       = {http://adsabs.harvard.edu/abs/2014grav.book.....P},
}

@Article{Lemoine2014,
  author    = {{Lemoine}, F.~G. and {Goossens}, S. and {Sabaka}, T.~J. and {Nicholas}, J.~B. and {Mazarico}, E. and {Rowlands}, D.~D. and {Loomis}, B.~D. and {Chinn}, D.~S. and {Neumann}, G.~A. and {Smith}, D.~E. and {Zuber}, M.~T.},
  journal   = {\grl},
  title     = {{GRGM900C: A degree 900 lunar gravity model from GRAIL primary and extended mission data}},
  year      = {2014},
  month     = may,
  pages     = {3382-3389},
  volume    = {41},
  doi       = {10.1002/2014GL060027},
  url       = {http://adsabs.harvard.edu/abs/2014GeoRL..41.3382L},
}

@article{NovaMoon,
  title={NovaMoon: A New Paradigm in Lunar Exploration },
  author={Ventura-Traveset, J. and Swinden, R. and  Melman, F. and  Psychas, D. and  Audet, Y.},
  journal={Inside GNSS},
  year={2025},
  volume={20},
  number={2},
  url={https://insidegnss.com/novamoon-a-new-paradigm-in-lunar-exploration/},
}

@article{Topo,
    author ={Barker, M.~K. and Mazarico, E. and Neumann, G.~A. and Zuber, M.T. and Haruyama, J. and Smith, D.~E. } ,
    title ={A new lunar digital elevation model from the {L}unar {O}rbiter {L}aser {A}ltimeter and {SELENE} Terrain Camera } ,
    journal ={Icarus } ,
  volume    = {273},
    pages = {346-355},
    year = {2015},
    url = {http://dx.doi.org/10.1016/j.icarus.2015.07.039 }
}

@Article{Konopliv2014,
  author    = {{Konopliv}, A.~S. and {Park}, R.~S. and {Yuan}, D.-N. and {Asmar}, S.~W. and {Watkins}, M.~M. and {Williams}, J.~G. and {Fahnestock}, E. and {Kruizinga}, G. and {Paik}, M. and {Strekalov}, D. and {Harvey}, N. and {Smith}, D.~E. and {Zuber}, M.~T.},
  journal   = {\grl},
  title     = {{High-resolution lunar gravity fields from the GRAIL Primary and Extended Missions}},
  year      = {2014},
  month     = mar,
  pages     = {1452-1458},
  volume    = {41},
  doi       = {10.1002/2013GL059066},
  url       = {http://adsabs.harvard.edu/abs/2014GeoRL..41.1452K},
}

@Article{1994CeMDA..60..139H,
  Title                    = {On the use of {STF}-tensors in celestial mechanics},
  Author                   = {Hartmann, T. and Soffel, M.~H. and Kioustelidis, T.},
  Journal                  = {Celestial Mechanics and Dynamical Astronomy},
  Year                     = {1994},
  Month                    = {sep},
  Pages                    = {139-159},
  Volume                   = {60},
  Doi                      = {10.1007/BF00693097},
  Url                      = {http://adsabs.harvard.edu/abs/1994CeMDA..60..139H}
}

@article{Blanchet2001,
    author = {Blanchet, L. and Salomon, C. and  Teyssandier, P. and Wolf, P.},
    title = {Relativistic theory for time and frequency transfer to order $c^{-3}$ },
    journal = {Astronomy and Astrophysics},
    year = {2001},
    pages = {320-329},
    volume = {370},
    url =  {https://doi.org/10.1051/0004-6361:20010233},
    doi =  {10.1051/0004-6361:20010233},
}

@article{PavlisWeiss2017,
    author = {Pavlis, N. and Weiss, M.},
    title ={ Re-evaluation of the relativistic redshift on frequency standards at {NIST}, {B}oulder, {C}olorado, {USA}},
  journal = {Metrologia},
  volume = {54},
  year = {2017},
  pages = {535–548},
  doi = {10.1088/1681-7575/aa765c},
  url = {https://doi.org/10.1088/1681-7575/aa765c},
}

@article{Cziraki2023,
author={Cziráki, T. and Kamilla, G.},
year={2023},
title={Parameters of the best fitting lunar ellipsoid based on {GRAIL}’s selenoid model},
journal={Acta Geodaetica et Geophysica},
pages={139-147},
volume={58},
doi = {10.1007/s40328-023-00415-w},
url = {https://doi.org/10.1007/s40328-023-00415-w},
}

@article{Ashby2024,
author={Ashby, N.  and Patla, B.},
year={2024},
title={A Relativistic Framework to Estimate Clock Rates on the Moon},
journal={The Astronomical Journal},
pages={112},
volume={168},
doi = {10.48550/ARXIV.2402.11150},
url = {https://doi.org/10.48550/ARXIV.2402.11150},
}

@Proceedings{IAUGA1991,
  year         = {1991},
  title        = {IAU, Proceedings of the Twenty-First General Assembly, Buenos Aires},
  eventdate    = {1991},
  organization = {IAU},
  venue        = {Buenos Aires},
}

@Proceedings{IAUGA2000,
  year         = {2000},
  title        = {IAU, Proceedings of the Twenty-Fourth General Assembly, Manchester},
  eventdate    = {2000},
  organization = {IAU},
  venue        = {Manchester},
}

@Proceedings{IAUGA2024,
  year         = {2024},
  title        = {IAU, Proceedings of the Thirty-Second General Assembly, Cape Town},
  eventdate    = {2024},
  organization = {IAU},
  venue        = {Cape Town},
}

@Proceedings{IAUGA2006,
  year         = {2006},
  title        = {IAU, Proceedings of the Twenty-Sixth General Assembly, Prague},
  eventdate    = {2006},
  organization = {IAU},
  venue        = {Prague},
}

@Proceedings{CGPM2018res2,
  date         = {2019},
  year         = {2019},
  title        = {Proceedings of the 26th CGPM (2018), 2019, p475 },
  eventdate    = {2018},
  key = {CGPM(2018)},
  organization = {CGPM},
  doi          = {10.59161/CGPM2018RES2E},
  url          = {https://doi.org/10.59161/CGPM2018RES2E},
}

@article{Klioner2008,
    author = {Klioner, S.~A.},
    title ={Relativistic scaling of astronomical quantities and the system of astronomical units },
    journal = {Astronomy and Astrophysics},
    volume={478},
    page={951-958},
    year = {2008},
    url = {https://doi.org/10.1051/0004-6361:20077786},
    doi = {10.1051/0004-6361:20077786},
}

@Proceedings{Klioner2010,
    author = {Klioner, S.~A. and Capitaine, N. and  Folkner, W.~M. and Guinot, B. and Huang, T.~Y. and Kopeikin, S.~M. and Pitjeva, E.~V. and Seidelmann, P.~K. and  Soffel, M.~H.},
    title ={Units of relativistic time scales
and associated quantities },
    journal = {Relativity in Fundamental Astronomy},
    event={ IAU Symposium No. 261},
editor={S.~A. Klioner and P.~K. Seidelman and M.~H. Soffel},
    page={951-958},
    year = {2009},
    url = {https://core.ac.uk/download/pdf/191829735.pdf},
}

@Article{Panfilo2019,
  author    = {Panfilo, G. and Arias, F.},
  title     = {The {C}oordinated {U}niversal {T}ime ({UTC})},
  doi       = {10.1088/1681-7575/ab1e68},
  url       = {https://doi.org/10.1088/1681-7575/ab1e68},
  issn      = {0026-1394},
  language  = {en},
  number    = {4},
  pages     = {042001},
  urldate   = {2019-06-18},
  volume    = {56},
  journal   = {Metrologia},
  month     = {6},
  publisher = {{IOP} Publishing},
  year      = {2019},
}

@InProceedings{Parker2022,
  author     = {Parker, J.J.K. and Dovis, F. and Anderson, B. and Ansalone, L. and Ashman, B. and Bauer, F.~H. and D’Amore, G. and Facchinetti, C. and Fantinato, S. and Impresario, G. and McKim, S.~A. and Miotti, E. and Miller, J.~J. and Musmeci, M. and Pozzobon, O. and Schlenker, L. and Tuozzi, A. and Valencia, L.},
  booktitle  = {Proceedings of the 2022 International Technical Meeting of The Institute of Navigation},
  date       = {2022-02},
  year       = {2022},
  title      = {The {L}unar {GNSS} {R}eceiver {E}xperiment ({LuGRE})},
  doi        = {10.33012/2022.18199},
  url        = {https://doi.org/10.33012/2022.18199},
  pages      = {420--437},
  publisher  = {Institute of Navigation},
  series     = {ITM 2022},
  collection = {ITM 2022},
  issn       = {2330-3646},
}

@ARTICLE{2009ApOpt..48.3035R,
       author = {{Ramos-Izquierdo}, L. and {Scott III}, V.~S. and {Connelly}, J. and {Schmidt}, S. and {Mamakos}, W. and {Guzek}, J. and {Peters}, C. and {Liiva}, P. and {Rodriguez}, M. and {Cavanaugh}, J. and {Riris}, H.},
        title = "{Optical system design and integration of the Lunar Orbiter Laser Altimeter}",
      journal = {\ao},
         year = {2009},
        month = {may},
       volume = {48},
       number = {16},
        pages = {3035},
          doi = {10.1364/AO.48.003035},
       url = {https://ui.adsabs.harvard.edu/abs/2009ApOpt..48.3035R}
}

@ARTICLE{2010SSRv..150..209S,
       author = {{Smith}, D.~E. and {Zuber}, M.T. and {Jackson}, G.~B. and {Cavanaugh}, J.~F. and {Neumann}, G.~A. and {Riris}, H. and {Sun}, X. and {Zellar}, R.~S. and {Coltharp}, C. and {Connelly}, J. and {Katz}, R.~B. and {Kleyner}, I. and {Liiva}, P. and {Matuszeski}, A. and {Mazarico}, E.~M. and {McGarry}, J.~F. and {Novo-Gradac}, A.-M. and {Ott}, M.~N. and {Peters}, C. and {Ramos-Izquierdo}, L.~A. and {Ramsey}, L. and {Rowlands}, D.~D. and {Schmidt}, S. and {Scott}, V.~S. and {Shaw}, G.~B. and {Smith}, J.~C. and {Swinski}, J.-P. and {Torrence}, M.~H. and {Unger}, G. and {Yu}, A.~W. and {Zagwodzki}, T.~W.},
        title = "{The Lunar Orbiter Laser Altimeter Investigation on the Lunar Reconnaissance Orbiter Mission}",
      journal = {\ssr},
         year = {2010},
        month = {jan},
       volume = {150},
       number = {1-4},
        pages = {209-241},
          doi = {10.1007/s11214-009-9512-y},
          url = {https://doi.org/10.1007/s11214-009-9512-y},
       adsurl = {https://ui.adsabs.harvard.edu/abs/2010SSRv..150..209S}
}

@INPROCEEDINGS{2016LPI....47.1484G,
       author = {{Goossens}, S. and {Lemoine}, F.~G. and {Sabaka}, T.~J. and {Nicholas}, J.~B. and {Mazarico}, E. and {Rowlands}, D.~D. and {Loomis}, B.~D. and {Chinn}, D.~S. and {Neumann}, G.~A. and {Smith}, D.~E. and {Zuber}, M.~T.},
        title = "{A Global Degree and Order 1200 Model of the Lunar Gravity Field Using GRAIL Mission Data}",
    booktitle = {47th Annual Lunar and Planetary Science Conference},
         year = {2016},
       series = {Lunar and Planetary Science Conference},
        month = {mar},
        pages = {1484},
       url = {https://ui.adsabs.harvard.edu/abs/2016LPI....47.1484G}
}

@INPROCEEDINGS{Israel2020,
author={Israel, D.~J. and Mauldin, K.~D. and Roberts, C.~J. and Mitchell, J.~W. and Pulkkinen, A.~A. and Cooper, L.~V.~D. and Johnson, M.~A. and Christe, S.~D. and Gramling, C.~J.},
year={2020},
title={ {LunaNet}: a Flexible and Extensible Lunar Exploration Communications and Navigation Infrastructure},
booktitle={ IEEE Aerospace Conference },
pages={1-14},
publisher={ Big Sky, MT, USA},
doi={10.1109/AERO47225.2020.9172509},
url={https://doi.org/10.1109/AERO47225.2020.9172509 },
}

@INPROCEEDINGS{Moonlight2022,
title={The Lunar Pathfinder {PNT} Experiment and {M}oonlight Navigation Service: The Future of Lunar Position, Navigation and Timing},
author={Giordano, P. and  Malman, F. and  Swinden, R. and   Zoccarato, P. and  Ventura-Traveset, J.},
year={2022},
booktitle={International Technical Meeting of The Institute of Navigation, Long Beach, California},
pages={632-642},
url={https://doi.org/10.33012/2022.18225},
doi={10.33012/2022.18225},
}

@article{Wolf-Petit1995,
author = {Wolf, P. and Petit, G.},
title = {Relativistic theory for clock syntonization and the realization of geocentric coordinate times},
journal = {\aap},
volume = {304},
number = {653-661},
year = {1995},
url = {https://ui.adsabs.harvard.edu/abs/1995A&A...304..653W},
}

@article{Ashby2003,
author ={Ashby, N.},
title = {Relativity in the Global Positioning System},
journal = { Living Rev. Relativ.},
volume={6,1},
year={2003},
doi={10.12942/lrr-2003-1},
url={https://doi.org/10.12942/lrr-2003-1}
}

@article{Petit-Wolf2005,
author = {  Petit, G. and Wolf, P.},
title = {Relativistic theory for time comparisons:
a review},
journal = {Metrologia},
volume = {52},
number = {S138-S144},
year = {2005},
doi={10.1088/0026-1394/42/3/S14},
url={https://doi.org/10.1088/0026-1394/42/3/S14},
}

@article{China2019,
author = {Ma, L. and Xie, P. and Liu, D. and Wu, Y.},
title = {Research on the Influence of China's Commercial Spaceflight on the Economic and Social Development of the Regions Along the Belt and Road},
journal = {New Space},
volume = {7},
number = {4},
pages = {223-234},
year = {2019},
url = {https://doi.org/10.1089/space.2019.0012},
doi = {10.1089/space.2019.0012},
}

@misc{CircularT,
  author = {BIPM},
  title = { {Circular T} },
  year = {2025},
  url = {https://www.bipm.org/en/time-ftp/circular-t},
  urldate = {2025}
}

@misc{SOFA,
 author       = {{IAU SOFA Board}},
 title        = {{IAU} {SOFA} Software Collection},
 edition      = {Issue 2021-01-25},
 url          = {http://www.iausofa.org},
 year         = {2025},
 }

@misc{Fienga2024,
      title={Lunar References Systems, Frames and Time-scales in the context of the {ESA} Programme {M}oonlight}, 
      author={Fienga, A. and Rambaux, N. and Sosnica, K.},
      year={2024},
      eprint={2409.10043},
      archivePrefix={arXiv},
      primaryClass={astro-ph.EP},
      url={https://arxiv.org/abs/2409.10043}, 
}

\vspace{0.8cm}

\end{document}